\documentclass[aps,pre,amsmath,amsfonts,amssymb,floatfix,superscriptaddress,nofootinbib]{revtex4}
\usepackage{mathrsfs} \usepackage{graphicx,color} \usepackage{bbm} \usepackage{multirow}

\let\a=\alpha \let\b=\beta \let\g=\gamma \let\d=\delta
\let\e=\varepsilon  \let\h=\eta \let\k=\kappa
\let\l=\lambda  \let\n=\nu  
 \let\t=\tau \let\f=\varphi 
   \let\G=\Gamma
\let\D=\Delta \let\Th=\Theta\let\L=\Lambda  
\let\Si=\Sigma   
\let\ee=\epsilon \let\r=\rho \let\th=\theta \let\io=\infty

\def\PP{{\cal P}} 
\def\FF{{\cal F}} 
 \def\BB{{\cal B}}\def\II{{\cal I}}
\def\LL{{\cal L}}  \def\OO{{\cal O}}
\def\DD{{\cal D}}\def\GG{{\cal G}} \def\SS{{\cal S}}

  \def\erf{\text{erf}}

\def\de{\mathrm d}

\def\to{\rightarrow} \def\la{\left\langle} \def\ra{\right\rangle}

\newcommand{\beq}{\begin{equation}} \newcommand{\eeq}{\end{equation}}
\newcommand{\wh}{\widehat} \newcommand{\wt}{\widetilde}
\newcommand{\Tr}{\text{Tr}}

\newcommand{\afunc}[1]{\operatorname{\mathsf{#1}}}
\def\DE{\afunc{D}}

\allowdisplaybreaks

\begin{document}

\title{
Exact theory of dense amorphous hard spheres in high dimension \\
III. The full replica symmetry breaking solution
} 

\author{Patrick Charbonneau}
\affiliation{Department of Chemistry, Duke University, Durham,
North Carolina 27708, USA}
\affiliation{Department of Physics, Duke University, Durham,
North Carolina 27708, USA}

\author{Jorge Kurchan}
\affiliation{
LPS,
\'Ecole Normale Sup\'erieure, UMR 8550 CNRS, 24 Rue Lhomond, 75005 Paris, France
}

\author{Giorgio Parisi}
\affiliation{Dipartimento di Fisica,
Sapienza Universit\`a di Roma,
P.le A. Moro 2, I-00185 Roma, Italy}
\affiliation{
INFN, Sezione di Roma I, IPFC -- CNR,
P.le A. Moro 2, I-00185 Roma, Italy
}

\author{Pierfrancesco Urbani}
\affiliation{
IPhT, CEA/DSM-CNRS/URA 2306, CEA Saclay, F-91191 Gif-sur-Yvette Cedex, France
}

\author{Francesco Zamponi}
\affiliation{LPT,
\'Ecole Normale Sup\'erieure, UMR 8549 CNRS, 24 Rue Lhomond, 75005 Paris, France}

\begin{abstract}
In the first part of this paper,
we derive the general replica equations that describe infinite-dimensional hard spheres at any level of replica symmetry breaking (RSB)
and in particular in the fullRSB scheme.
We show that these equations are formally very similar to the ones that have been derived for spin glass models, thus showing that
the analogy between spin glasses and structural glasses conjectured by Kirkpatrick, Thirumalai, and Wolynes
is realized in a strong sense in the mean field limit. 
We also suggest how the computation could be generalized in an approximate way to finite dimensional hard spheres.
In the second part of the paper,
we discuss the solution of these equations and we derive from it a number of physical predictions. 
We show that, below the Gardner transition where the 1RSB solution
becomes unstable, a fullRSB phase exists and we locate the boundary of the fullRSB phase. Most importantly, we show that the fullRSB solution
predicts correctly that jammed packings are isostatic, and allows one to compute analytically
the critical exponents associated with the jamming transition, which are missed by the 1RSB solution.
We show that these predictions compare very well with numerical results.
\end{abstract}

\maketitle

\tableofcontents

\clearpage

\section{Introduction}

In the two previous papers of this series~\cite{KPZ12,KPUZ13} (on which the present work relies heavily), we have considered a system of
identical hard spheres in spatial dimension $d\to\io$, and we have obtained an exact expression of its replicated partition function. 
Within the Random First Order Transition (RFOT) general scenario for the glass 
transition~\cite{KW87b,KT88,KTW89,WL12}, which relies on the conjecture that structural glasses behave in the same way as a certain class of
mean field spin glass models, the replicated partition function describes the amorphous arrested states of the system, i.e. its glasses~\cite{KT89,Mo95}.

Previous work on structural glasses in the RFOT context always focused on the simplest replica scheme, 
the one-step replica symmetry breaking (1RSB). The 1RSB scheme indeed already predicts the most important phase transitions
that happen upon approaching the glass phase, namely the dynamical and Kauzmann transitions. An approximate 1RSB treatment of
finite-dimensional glasses was developed in~\cite{MP99,MP09}, and applied to hard spheres in~\cite{PZ10}; these works have shown
that the 1RSB approach gives quite accurate predictions of thermodynamic and structural quantities in three-dimensional glasses.
Moreover, within the 1RSB scheme, 
the exact expression obtained in~\cite{KPZ12,KPUZ13} coincides with the infinite-dimensional limit of the finite-dimensional approach of~\cite{PZ10}
while a series of numerical works~\cite{CIPZ11,CIPZ12} in $d = 3, \cdots, 12$ have shown that the main qualitative
properties of the system evolve smoothly with dimensions 
(although the investigated dimensions are quite far from the asymptotic $d\to\io$ regime~\cite{CIPZ11}).
One could therefore think that the 1RSB scheme is sufficient to describe the properties of glasses within RFOT theory in all the range of physical parameters.

However, when one tries to apply the 1RSB scheme to study the properties of hard or soft-sphere glasses at very high pressures and low temperatures,
close to the jamming transition that marks the transition from a mechanically loose to a mechanically rigid glass state,
one finds contradictory results. On the one hand, the behavior of the main thermodynamic quantities (pressure, energy, ...) is quite well reproduced~\cite{PZ10,BJZ11}.
On the other hand, the scaling properties of other observables are not. It is therefore natural to search for the origin of this discrepancy.

A first step in this direction was made through the investigation of nearly jammed sphere packings.
It was shown that hard sphere glasses at high pressures are close to a mechanical instability, 
the so-called ``isostatic'' point where the number of mechanical constraints
exactly equals the number of degrees of freedom~\cite{Mo98,TW99,Ro00}. 
Due to this proximity, anomalous low-frequency modes appear in the vibrational 
spectrum~\cite{OLLN02,OSLN03,WSNW05,BW06,BW07}.
Based on this observation, in a series of papers Wyart and coworkers~\cite{WSNW05,BW09b,LNSW10,Wy12}
assumed that hard sphere glasses at high pressure are
{\it marginally stable}, and under this assumption they derived a scaling theory of the jamming transition 
that is able to describe most of its basic phenomenology. In particular, it was shown both analytically and 
numerically~\cite{WSNW05,BW09b,IBB12} that marginality is associated
with a particular scaling of the mean square displacement $\D$ in the hard sphere glass, which vanishes when the pressure $p\to\io$
as $\D \sim p^{-\k}$ with $\k \sim 3/2$ (naive free-volume considerations would suggest $\D \sim p^{-2}$).
It was confirmed in~\cite{IBB12} that the exponent $\k$ plays an important role, and in fact controls all the criticality of the jamming transition.

This analysis suggests that a 1RSB description is incorrect in this regime, because 1RSB states are perfectly stable and do not show
any sign of marginality. In fact, the 1RSB solution further wrongly\footnote{Still, as noted in~\cite{IBB12}, the 1RSB prediction 
shows a sign of an anomalous behavior of the mean square displacement with respect to the naive expectation.}
 predicts $\D \sim p^{-1}$~\cite{PZ10}, and furthermore misses other critical
exponents associated to the structure and the force distribution~\cite{CCPZ12,Wy12,LDW13}.
In the context of spin glasses, it is well known that fullRSB phases are always associated with marginal stability and anomalous
low-frequency modes~\cite{BM79,MPV87}, which affect the low-temperature scaling of physical quantities~\cite{MPV87}.
It appears therefore that a fullRSB phase is a natural candidate for explaining the marginal stability of low-temperature glasses.

Additional insights came from analyzing the out-of-equilibrium dynamical behavior of one of the simplest spin glass models that
are at the basis of the RFOT scenario, the Ising $p$-spin model~\cite{MR03,MR04,Ri13,KZ13}. 
It was shown that, although the 1RSB solution correctly describes the approach
to the glass phase at equilibrium, when the system is instantaneously quenched out-of-equilibrium it evolves towards 
a region of phase space where the 1RSB solution
is unstable, the so-called Gardner phase\footnote{
The existence of a transition at low temperature from a 1RSB
to a fullRSB solution was discovered independently in~\cite{Ga85,GKS85} and
its precise location was found in~\cite{Ga85}.
}~\cite{Ga85,GKS85}, in which a fullRSB solution is present~\cite{Ri13}. The behavior of the Ising $p$-spin
model is not yet fully understood, 
but this model nonetheless strongly suggests that fullRSB effects may be observable in low-temperature glasses.

Many other observations hint to the presence of a fullRSB phase in low temperature or high pressure glasses
(see e.g.~\cite{FM13}), 
which we reviewed in~\cite{KPUZ13}.
Based on these observations, in the previous paper of this series~\cite{KPUZ13} we used our exact expression for the replicated partition function
of infinite-dimensional hard spheres to investigate the stability of the 1RSB phase, a computation that could not be done previously because
the finite-dimensional approximations of~\cite{MP99,MP09,PZ10} have been formulated only at the 1RSB level. Our analysis showed that the
1RSB phase is stable around the dynamical glass transition, but becomes unstable at high pressures close to the jamming transition.
This result gives additional indications in favor of the presence of a different phase that could better describe the properties of high pressure
(or low temperature) glasses.

The aim of this work is to write explicitly the replica equations in the $k$RSB scheme and send $k\to\io$ to describe the fullRSB solution, 
in order to check if this solution predicts correctly the scaling properties of the jamming transition.
In the first part of this work we derive the set of equations that give the replicated entropy at the level of $k$RSB. 
Interestingly,
the equations we obtain are remarkably similar to those describing the Ising $p$-spin model with, however, some crucial
differences. The proof of this formal similarity between the equations that describe a disordered mean field
spin glass model and a system of interacting particles without any quenched disorder
somehow completes the program initiated by Kirkpatrick, Thirumalai
and Wolynes, who constructed the RFOT scenario by assuming that such a similarity existed and proved some arguments supporting it, 
see e.g.~\cite{KW87,KT89}.

In the second part of the paper, we extract physical quantities from the equations. Our main results are that {\it (i)}~a fullRSB phase 
always exists at high pressures, hence the jamming transition always lies in the fullRSB region; 
{\it (ii)}~as in the SK model~\cite{TAK80,DK83,Go83,KD86},
the fullRSB phase is {\it marginally stable}, because one of the eigenvalues of the stability matrix of the replicated entropy
is identically vanishing;
{\it (iii)} jammed packings are {\it predicted} to be isostatic at all densities;
{\it (iv)} the fullRSB solution
predicts a different scaling of the mean square displacement in the glass, namely $\D \sim p^{-\k}$ with an exponent $\k$ very close
to $3/2$; {\it (v)} the fullRSB solution predicts a power-law divergence of the pair correlation function of jammed packings at contact,
$g(r) \sim (D-r)^{-\a}$; {\it (vi)} the force distribution vanishes at small forces as $P(f) \sim f^\th$; 
{\it (vii)} we obtain analytical predictions for the exponents $\k,\a,\th$.

In~\cite{Wy12,LDW13,DLBW14}, 
the exponents $\theta$ and $\alpha$ were argued to control the stability of packings and the presence of avalanches, 
and a scaling relation between them was derived assuming marginal stability. 
The value of one exponent was lacking however to have a complete scaling picture.
Our predicted values of $\k,\a,\th$ are perfectly compatible with the scaling relations derived in~\cite{Wy12,LDW13,DLBW14}
and with previous numerical investigations~\cite{DTS05,SLN06,TS10,Wy12,CCPZ12,LDW13}
(except for $\theta$, where the situation remains somehow unclear).
By means of additional numerical simulations, we show that the prediction for $\k$ is correct in all spatial dimensions.

These results open the way to many different studies. For example, one could now hope to compute 
the shear modulus~\cite{BW06,YM10,Yo12,Yo12b}, the distribution of avalanche 
sizes~\cite{LMW10}, the complete scaling on both sides of the jamming transition~\cite{IBB12,BJZ11}, and so on.
Moreover, the results could be extended to finite dimensional systems within the effective potential approximation scheme of~\cite{PZ10,BJZ11}, 
which would allow one to directly compare the predictions with numerical simulations and experiments.
We discuss briefly these possible developments in the conclusions. An important open question is how the 
fullRSB structure affects the off-equilibrium dynamics~\cite{MR03,MR04,Ri13}.

Because the paper is quite long, we do not present a detailed plan here. At the beginning of each section, we explain what
is the aim of the section and detail the structure of subsections.
Reading the technical part of the paper requires familiarity with the general concepts of 
spin glass theory (see e.g.~\cite{MPV87,CC05} for reviews), with its application to structural glasses~\cite{MP09} and jamming~\cite{PZ10},
and with the previous papers of this series~\cite{KPZ12,KPUZ13}.
A short account of this work, where the main physical ideas behind it and the main results are presented in a more accessible way for the
reader not interested in technical details,
has been presented in~\cite{nature}.

\clearpage

\part{General equations}

\section{Generic expression of the entropy}
\label{generalities}

In the following, we consider a system of $N$ identical spheres of diameter $D$ in a $d$-dimensional volume $V$,
in the thermodynamic limit $N = \r V$ and $V\to\io$.
At sufficiently high density, this system exhibits a glass phase, i.e., its dynamics arrests and an amorphous solid
phase forms due to self-induced frustration.
A strategy to obtain a thermodynamic description of the glass states has been derived in~\cite{KT88,KW87,Mo95}.
The idea behind these works is the following. In the dense supercooled regime the liquid phase splits
in a collection of glass basins that can be defined as amorphous minima of an appropriate density functional~\cite{KT88}. 
Hence if one extracts one liquid configuration in equilibrium, 
this configuration falls into one of the multiple glass basins. To describe this glass basin thermodynamically, one has to consider 
a second configuration that is coupled to the first one. Its partition function gives the properties of the glass. 
The coupling to the first configuration acts as an external quenched disorder, and that disorder 
can be treated using replicas, in analogy with spin glasses~\cite{KT87b}. This construction was made precise
through the introduction of the so-called Franz-Parisi potential~\cite{FP95,FP97,CFP98b}. 
In this way one can easily follow the {\it adiabatic evolution} with density and temperature of glassy states that originate from
an equilibrium liquid configuration (also known as the ``state following'' procedure~\cite{KZ10}). Dynamically,
this corresponds to preparing glassy states through a slow annealing.

A slightly different (and computationally simpler)
strategy was proposed by Monasson~\cite{Mo95}. Here $m$ replicas are coupled to an external random field that
selects glassy states. After averaging out the external field one is left with $m$ coupled replicas. 
Because one uses here a completely random field, this procedure gives the properties of the ``typical'' states that exist
at a given temperature or density~\cite{Mo95,PZ10}. In this way one can compute the dynamical glass transition density,
the Kauzmann density (when it exists), and the so-called dynamical line (or threshold line) that delimits the region of
existence of glassy states. Dynamically, it is expected that after a {\it fast quench} the system becomes arrested in the glassy
states that lie on this dynamical/threshold line~\cite{CK93,CC05}.
This strategy is very efficient and was used in many previous studies, see e.g.~\cite{Me99,MP00} for pedagogical introductions
and~\cite{MP09,PZ10} for reviews of previous results.
In this paper,
as in the previous papers of this series~\cite{KPZ12,KPUZ13}, 
 we follow this simpler approach and consider an
$m$-times replicated system in order to describe glassy states. We will see that this treatment is sufficient to describe the marginal stability 
of jammed packings and to
 obtain the critical properties
around the jamming transition.
The state-following (or Franz-Parisi) computation is also possible, but is left for a future publication.
In the second part of this paper,
we will explain more precisely what kind of information on the phase diagram
can be obtained from this procedure.

We consider $m$ identical copies of the original hard sphere system,
assuming that the spheres are arranged
in molecules, each molecule containing an atom of each replica. A molecule is thus described by $m$ vectors
$\bar x = \{x_1\cdots x_m\}$, each $x_a$ being $d$-dimensional. The molecular liquid is translationally invariant, so each atom
in a molecule fluctuates around the center of mass of the molecule $X=m^{-1}\sum_a x_a$. We call the displacement
of one atom $u_a = x_a - X$.
In the following, $m \times m$ matrices in replica spaces are indicated by a hat, e.g. the matrix $\hat\a$ has elements $\a_{ab}$, 
and we denote by $\hat\a^{a,b}$ the $m-1 \times m-1$ matrix obtained from $\hat\a$ by removing line $a$ and column $b$.
Wide hats are used to denote quantities that have been properly scaled to be finite in the limit $d\to\io$; for
example, $\wh\f = 2^d \f/d$ is the scaled packing fraction, with the usual packing fraction $\f = \r V_d$, where
$V_d$ is the volume of a sphere of unit radius~\cite{KPZ12}.

The aim of this section is to write the entropy of the replicated system in a convenient way.
With the notations introduced above,
we consider the general form of the replicated
entropy that has been derived in paper II of this series~\cite{KPUZ13}:
\beq\label{eq:gauss_r}
s[\hat \a]  = 1 - \log\r + d \log m + \frac{d}{2}(m-1) \log(2 \pi e D^2/d^2) + \frac{d}2 \log \det(\hat \a^{m,m}) -  \frac{d}2 \wh \f \,
\FF\left( 2 \hat \a \right) \ ,
\eeq
where the matrix $\hat\a$ with elements $\a_{ab} = d \la u_a \cdot u_b \ra / D^2$ encodes the fluctuations of the replica displacement vectors $u_a$.
The term proportional to $\wh\f$ comes from the density-density interaction and we refer to it below as the ``interaction term''. The other terms
encode the entropic contributions and we refer to them all as the ``entropic term''.
It is useful to define the scaled mean square displacement between two replicas
\beq\label{eq:Dabdef}
\D_{ab}= \frac{d}{D^2} \left<(u_a-u_b)^2\right>=\alpha_{aa} + \alpha_{bb} -2\alpha_{ab}\:.
\eeq
It has been shown in~\cite{KPZ12,KPUZ13} that the elements of the matrices $\hat\D$ and $\hat\a$ are finite in the limit $d\to\io$.
Because the vector $u_a$ has $d$ components, Eq.~\eqref{eq:Dabdef} 
implies that the variance of the norm of the vector $u_a-u_b$ is of order $1/d$,
while the variance of each one of its components is of order $1/d^2$.
In order to find the thermodynamic entropy of the replicated liquid,
the replicated entropy \eqref{eq:gauss_r} must be optimized\footnote{For integer $m>1$, the optimum is a maximum, but for real $m<1$ analytic continuation
changes the sign of some eigenvalues and the optimum becomes a saddle point~\cite{MPV87}.} with respect to the matrices $\hat\a$ or
$\hat \D$.

In~\cite{KPUZ13} a generic expression for the interaction term $\mathcal F(2\hat \alpha)$ has been derived, but we now want to write
it in a more convenient form. 
It reads:
\beq
\mathcal F(2\hat \a)=\lim_{n\to 0}\sum_{n_1,\ldots,n_m: \sum_a n_a=n}\frac{n!}{n_1!\ldots n_m!}\exp\left[- \sum_{a=1}^m \a_{aa}\frac{n_a}{n}+ 
\sum_{a,b}^m \a_{ab}\frac{n_a n_b}{n^2}\right]\:.
\eeq
We will now suppose that the diagonal part of the matrix $\a$ is constant and $\a_{aa}=\a_d$ for all $a$, 
because this is true for matrices with a general $k$RSB structure. 
This implies that
\beq
2\a_{ab}=2\a_d- \Delta_{ab} \ ,
\eeq
and therefore the function $\FF$ can be rewritten in terms of the matrix $\hat\D$ as
\beq
\begin{split}
\mathcal F(\hat \Delta)&=\lim_{n\to 0}\sum_{n_1,\ldots,n_m: \sum_a n_a=n}\frac{n!}{n_1!\ldots n_m!}\exp\left[-\a_d +\frac 12 \sum_{a,b}^m\left(2\a_d- \Delta_{ab}\right)\frac{n_a n_b}{n^2}\right] \\
&=\lim_{n\to 0}\sum_{n_1,\ldots,n_m: \sum_a n_a=n}\frac{n!}{n_1!\ldots n_m!}\exp\left[-\frac 12 \sum_{a,b}^m \Delta_{ab}\frac{n_a n_b}{n^2}\right] \ .
\end{split}
\eeq
Note that the diagonal elements of the displacement matrix $\hat \Delta$ are all zero. Moreover, we temporarily define a matrix
\beq\label{eq:deltastar}
\Delta_{ab}^*=\Delta_{ab}+\L \delta_{ab}
\eeq
and we write the interaction term as
\beq\label{gen_simp}
\mathcal F(\hat \Delta)= \lim_{n\to 0}\sum_{n_1,\ldots,n_m: \sum_a n_a=n}\frac{n!}{n_1!\ldots n_m!}\exp\left[-\frac 12 \sum_{a,b}^m \Delta^*_{ab}\frac{n_a n_b}{n^2}+\frac{\L}{2} \sum_{a=1}^m\frac{n_a^2}{n^2}\right] \ .
\eeq
We assume that the parameter $\L$ is positive. 
Using the identity
\beq
e^{-\frac 12 \sum_{a,b}^m \Delta^*_{ab}\frac{n_a n_b}{n^2}} =\left. e^{-\frac{1}{2}\sum_{a,b=1}^m  \Delta^*_{ab}\frac{\partial^2}{\partial h_a\partial h_b}} 
e^{-\sum_{a=1}^mh_a\frac{n_a}{n}}\right|_{\{h_a=0\}} \ ,
\eeq
and introducing $\mathcal D\l_a = \de \l_a \, e^{-\l_a^2/2}/\sqrt{2\pi}$ as a Gaussian measure with zero mean and unit variance,
$\mathcal F(\hat \Delta)$ can be rewritten as
\beq
\begin{split}
\mathcal F(\hat \Delta) = & \left.\lim_{n\to 0}\sum_{n_1,\ldots,n_m: \sum_a n_a=n}\frac{n!}{n_1!\ldots n_m!}\exp\left[-\frac{1}{2}\sum_{a,b=1}^m  \Delta^*_{ab}\frac{\partial^2}{\partial h_a\partial h_b}\right]\exp\left[ \frac{\L}{2}\sum_{a=1}^m\frac{n_a^2}{n^2}-\sum_{a=1}^mh_a\frac{n_a}{n}\right]\right|_{\{h_a=0\}} \\
= & \lim_{n\to 0}\sum_{n_1,\ldots,n_m: \sum_a n_a=n}\frac{n!}{n_1!\ldots n_m!}\exp\left[-\frac{1}{2}\sum_{a,b=1}^m \Delta^*_{ab}\frac{\partial^2}{\partial h_a\partial h_b}\right]\times\\
&\times \left.\int \left(\prod_{a=1}^m\mathcal D\l_a\right) \exp\left[-\sqrt{\L}\sum_{a=1}^m\frac{n_a}{n}\l_a-\sum_{a=1}^m\frac{n_a}{n}h_a\right]\right|_{\{h_a=0\}} \\
= & \left.\exp\left[-\frac{1}{2}\sum_{a,b=1}^m \Delta^*_{ab}\frac{\partial^2}{\partial h_a\partial h_b}\right]\int\left(\prod_{a=1}^m\mathcal D\l_a\right)\exp\left[-\min_a\left(\sqrt\L \l_a+h_a\right)\right]\right|_{\{h_a=0\}} \ .
\end{split}
\eeq
In the above expression, the integration measure of the $\l_a$ corresponds to independent random variables with a Gaussian probability distribution. 
Then, the probability distribution of the random
variable $h = -\min_a\left(\sqrt\L \l_a+h_a\right) = \max_a\left(-\sqrt\L \l_a- h_a\right)$ 
is given by
\beq\begin{split}
\mu_{\rm min}(h) 
&= \frac{\de}{\de h}  \prod_{a=1}^m\int_{-(h + h_a)/\sqrt{\L} }^\infty \DD \l_a = \frac{\de}{\de h}  \prod_{a=1}^m \Th\left(  \frac{h + h_a}{\sqrt{2\L}}     \right)
\ ,
\end{split}\eeq
because the product of integrals is the probability that $h \geq -\sqrt{\L}\l_a - h_a , \ \forall a$, 
hence it is the probability that $h \geq \max_a\left(-\sqrt\L \l_a- h_a\right)$,
which is nothing but the cumulative distribution of $h$.
Here we introduced the function $\Th(z) = (1+\erf(z))/2$ and $\int_{-h}^\io \DD \l = \Th(h/\sqrt{2})$.
Therefore
\beq
\int\left(\prod_{a=1}^m\mathcal D\l_a\right)\exp\left[-\min_a\left(\sqrt\L \l_a + h_a\right)\right]
= \int_{-\io}^\io \de h \, e^{h }\, \mu_{\rm min}(h) =
\int_{-\io}^\io \de h \, e^{h }\, \frac{\de}{\de h}  \prod_{a=1}^m \Th\left(  \frac{h + h_a}{\sqrt{2\L}}     \right)
\eeq
and collecting these results we obtain
\beq
\begin{split}
\mathcal F(\hat\Delta) 
= & \left.\exp\left[-\frac{1}{2}\sum_{a,b=1}^m \Delta^*_{ab}\frac{\partial^2}{\partial h_a\partial h_b}\right]\int_{-\infty}^\infty \de h\, e^{h} 
 \frac{\de}{\de h}  \prod_{a=1}^m \Th\left(  \frac{h + h_a}{\sqrt{2\L}}     \right)
\right|_{\{h_a=0\}} \\
= & \int_{-\infty}^\infty \de h\, e^{h} 
 \frac{\de}{\de h} \left\{
\exp\left[-\frac{1}{2}\sum_{a,b=1}^m \Delta^*_{ab}\frac{\partial^2}{\partial h_a\partial h_b}\right] \prod_{a=1}^m \Th\left(  \frac{ h_a}{\sqrt{2\L}}     \right)
 \right\}_{\{h_a=h\}} \ .
 \end{split} 
\eeq
By defining the Gaussian kernel $\g_a(z) = e^{-z^2/(2a)}/\sqrt{2\pi a}$, one obtains 
the relation
\beq\label{eq:derint}
\exp\left[ \frac a2 \frac{\de^2}{\de h^2}\right]f(h) = \int_{-\infty}^\infty\frac{\de z}{\sqrt{2\pi a}}e^{-\frac{z^2}{2a}}f\left( h- z\right)
= \g_a \star f(h) \ ,
\eeq
where we defined the $\g_a \star f$ operation as the convolution of the Gaussian kernel with the function $f(h)$.
Then
\beq
\Th(h/\sqrt{2a}) = \int_{-\infty}^\infty\frac{\de z}{\sqrt{2\pi a}}e^{-\frac{z^2}{2a}}\theta(h-z) = \exp\left[ \frac a2 \frac{\de^2}{\de h^2}\right] \th(h) =  \g_{a} \star \th(h)    \ .
\eeq
Using this relation we obtain the compact result
\beq\label{int_gen}
\begin{split}
\mathcal F(\hat\Delta) 
= & \int_{-\infty}^\infty \de h\, e^{h} 
 \frac{\de}{\de h} \left\{
\exp\left[-\frac{1}{2}\sum_{a,b=1}^m \Delta^*_{ab}\frac{\partial^2}{\partial h_a\partial h_b}\right] \prod_{a=1}^m e^{ \frac12 \L \frac{\partial^2}{\partial h_a^2}}  \th(h_a)  
 \right\}_{\{h_a=h\}} \\
= & \int_{-\infty}^\infty \de h\, e^{h} 
 \frac{\de}{\de h} \left\{
\exp\left[-\frac{1}{2}\sum_{a,b=1}^m \Delta_{ab}\frac{\partial^2}{\partial h_a\partial h_b}\right] \prod_{a=1}^m  \th(h_a)  
 \right\}_{\{h_a=h\}} \ . 
\end{split} 
\eeq
The replicated entropy is therefore given by Eq.~\eqref{eq:gauss_r}, with the function $\FF$ given in Eq.~\eqref{int_gen}.
This expression is quite general and will be used to study $k$-step replica symmetry breaking schemes in the following. 
Note that it has been derived under the only assumption that
the diagonal elements of $\a_{ab}$ are all equal.

\section{The replicated entropy for hierarchical RSB matrices}

In spin glasses, it has been shown that a correct description of the system can be achieved by considering a special
class of matrices, known as hierarchical $k$RSB matrices~\cite{MPV87}.
Therefore, we want to specialize the general expression of the replicated entropy given by Eqs.~\eqref{eq:gauss_r} and~\eqref{int_gen} 
to these matrices.
In Sec.~\ref{sec:IIIA}, we introduce hierarchical $k$RSB matrices and discuss some of their properties.
In Sec.~\ref{sec:entropic}, we compute the entropic term of the replicated entropy 
for a $k$RSB matrix, and in Sec.~\ref{sec:interactionterm} we compute
the interaction term. In Sec.~\ref{sec:IIID}, we present the final explicit expressions for the 1RSB, 2RSB, $k$RSB and fullRSB case.

\subsection{Parametrization of hierarchical matrices}
\label{sec:IIIA}

The structure of hierarchical $k$RSB matrices is well-known and here we just summarize it briefly.
Remember that with respect to the formalism of~\cite{MPV87} we have an important difference, in that the diagonal elements of the matrices
we consider are determined by the condition that $\sum_b \a_{ab} = 0, \ \forall b$.
The simplest class is that of 1RSB matrices\footnote{In the standard notation of~\cite{MPV87} this would be called a replica-symmetric
matrix, but remember that here we are using the Monasson's real replica scheme~\cite{Mo95} where we consider $m$ coupled replicas and treat
$m$ as a parameter to select different metastable states. It is a standard convention to denote a RS matrix in the Monasson's scheme as a ``1RSB matrix'':
the reason is that in models with quenched disorder the two schemes are indeed equivalent. 
},
that has been studied in the first~\cite{KPZ12} and second~\cite{KPUZ13} paper of this series. It corresponds to $\a_{ab} = -\wh \a_1, \ \forall a\neq b$:
\beq\label{alpha1RSB}
\a_{ab}^{\rm 1RSB}= \wh \a_1 \left(m \delta_{ab}- 1 \right) \ .
\eeq
Here we used a slightly different but equivalent notation with respect to previous papers~\cite{PZ10,KPZ12,KPUZ13},
to be consistent with the general $k$RSB notation\footnote{Note that in previous papers~\cite{PZ10,KPZ12,KPUZ13} 
we dropped the subscript, $\wh\a_1 = \wh\a$ because
we only considered the 1RSB case, and we used $\wh A = m \wh\a$.
}.
A 2RSB matrix is parametrized as follows. Replicas are arranged in $m/m_1$ groups, and the matrix
elements $\a_{ab} = -\wh \a_2$ if they belong to the same group, or $\a_{ab} = -\wh \a_1$ if they do not. 
If we say that $b\sim a$ when $b$ is in the same subgroup of $a$, and $b\nsim a$ otherwise, and
we define a function $\mathrm{I}(\text{e})=1$ if the event ``e'' is true and $\mathrm{I}(\text{e})=0$ otherwise,
a 2RSB matrix has the form:
\beq\label{alpha2RSB}
\a^{\rm 2RSB}_{ab}=\delta_{ab}\left[(m_1-1) \wh \a_2+(m-m_1) \wh \a_1 \right]-(1-\delta_{ab})\left[ \wh \a_2\mathrm I(b \sim a)+ \wh \a_1\mathrm I (b \nsim a)    \right] \ .
\eeq
At the 3RSB level, the blocks of $m_1$ replicas are each divided in $m_1/m_2$ sub-blocks of $m_2$ replicas, and the construction can be
iterated to any desired level of RSB (which we denote $k$RSB). Note that diagonal elements of hierarchical matrices are all equal.

For each $k$, these ``hierarchical'' matrices form a closed algebra. Moreover, in the interesting case where one performs an analytic continuation to $m<1$,
a generic hierarchical matrix $\hat Q$ can be parametrized by its diagonal element $q_d$ and a single function $q(x)$.
This is done as follows.  If we formally define $m_0 = m$ and $m_k = 1$,
we can observe that for integer $m>1$, one has $m_k \equiv 1 < m_{k-1} < m_{k-2} < \cdots < m_1 < m_0 \equiv m$.
When $m<1$, these inequalities are reversed~\cite{MPV87} and one has $1 > m_{k-1} > m_{k-2} > \cdots > m_1 > m > 0$.
Then we can define $q(x)$ to be a piecewise constant function for $x\in [0,1]$, which in the interval $x \in [m_{i-1} , m_{i}]$ takes the value of the elements $q_{i}$ 
in the corresponding sub-block (with $i=1 \cdots k$), and $q(x)=0$ for $x \in [0,m_0]$. We write this parametrization as $\hat Q \leftrightarrow \{ q_d , q(x) \}$.

\begin{figure}
\includegraphics[width=.45\textwidth]{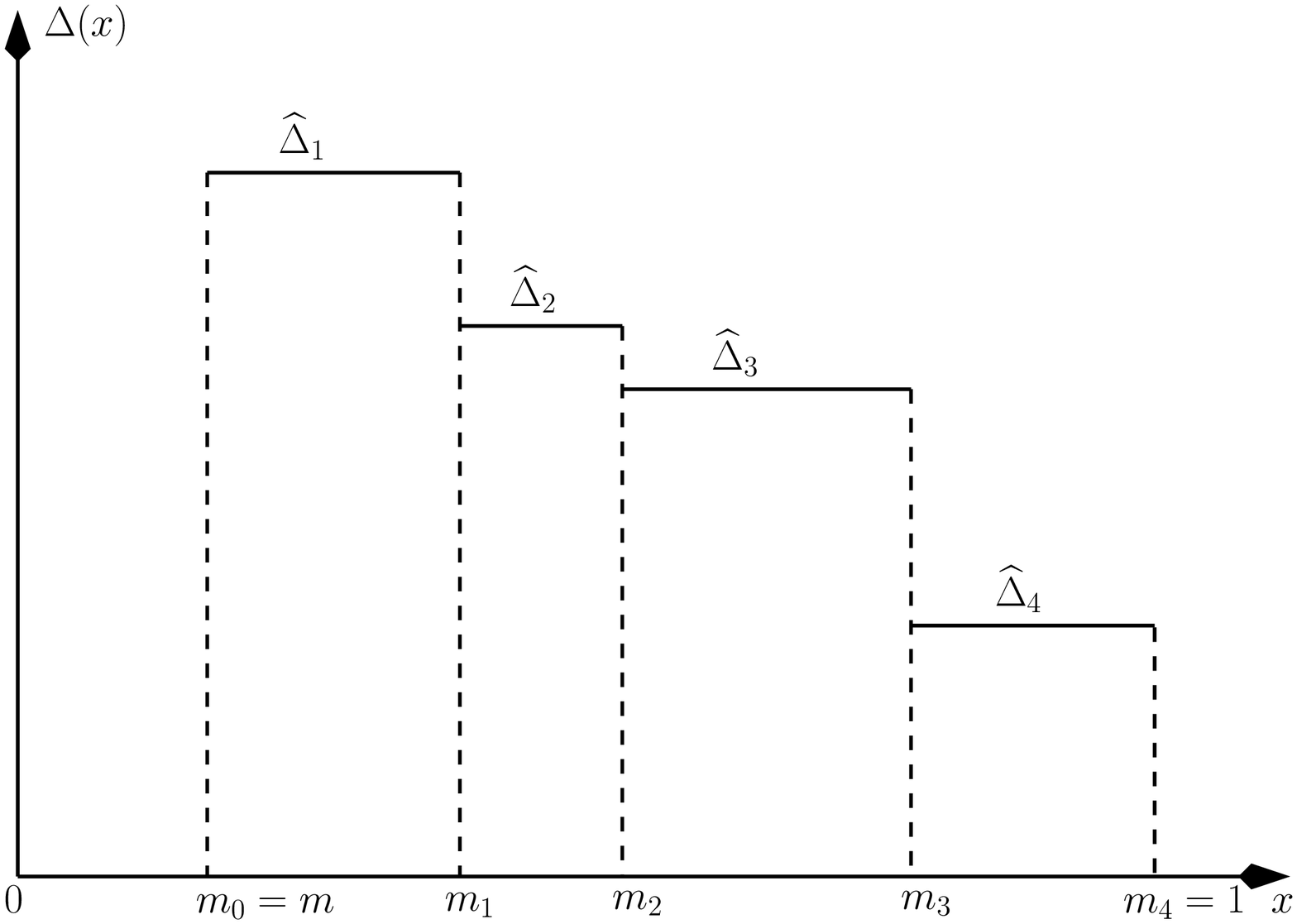}
\includegraphics[width=.45\textwidth]{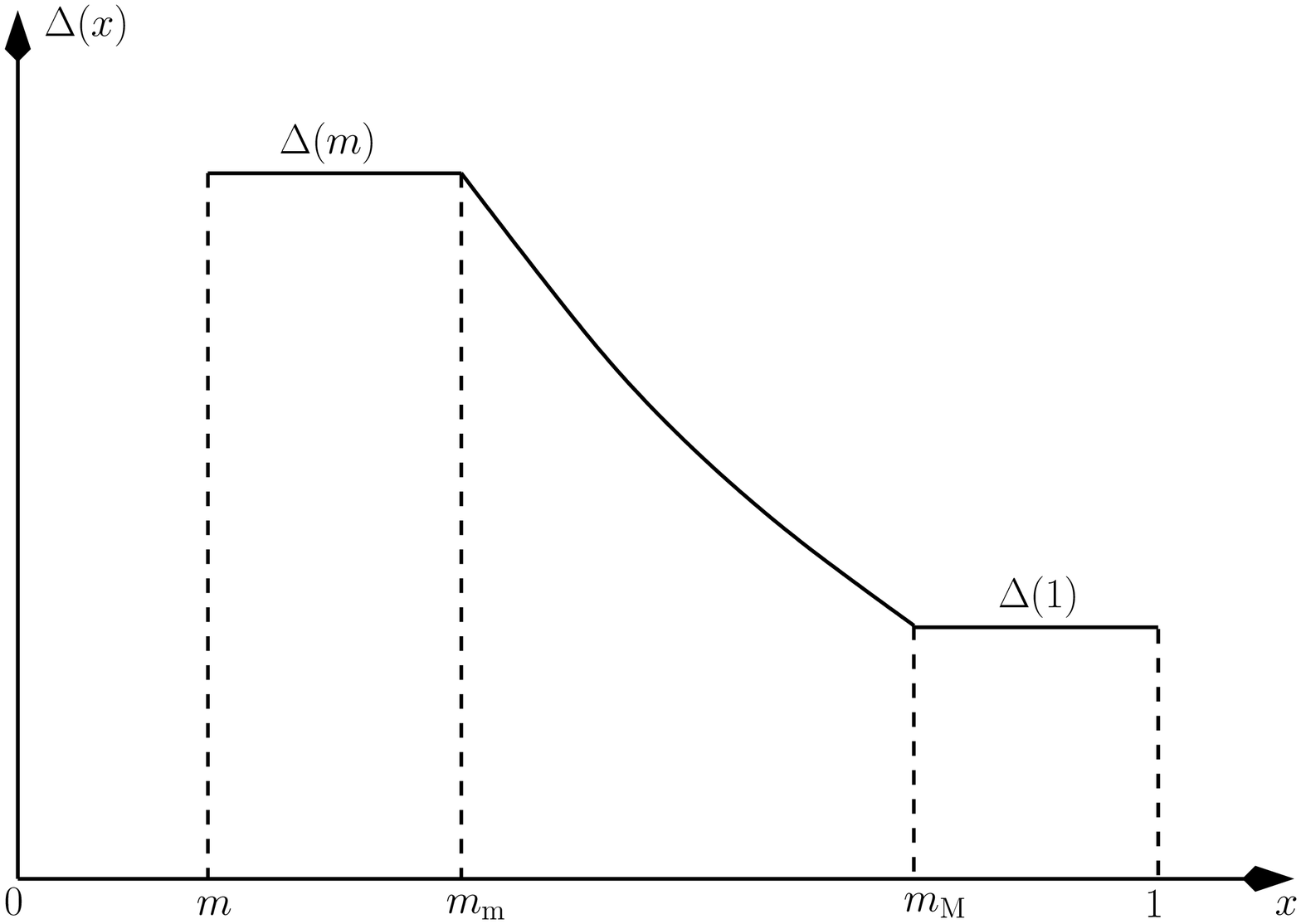}
\caption{(Left) An example of the parametrization of the matrix $\D_{ab}$ for a $4$RSB case. When needed for notational purposes,
we use the convention $\wh\D_0 \equiv 0$.
(Right) The expected form of the function $\D(x)$ in the fullRSB limit.
}
\label{fig:hier}
\end{figure}

The matrices $\hat\a$ are therefore defined by $\hat\a \leftrightarrow \{\a_d,-\a(x)\}$ where $\a(x)$ is a piecewise constant function given by the $\wh\a_i$ in each
block,
while $\a_d$ is fixed by the condition $\sum_b \a_{ab}=0$ and is given by
\beq
\a_d =  (m_{k-1}-1)\wh\a_{k}+(m_{k-2}-m_{k-1})\wh\a_{k-1}+\ldots +(m-m_1)\wh\a_1  =- \int_m^1	\de x\, \alpha(x) \ .
\eeq
For a given matrix $\a_{ab}$, one has a corresponding matrix $\D_{ab} = 2 \a_{aa} - 2 \a_{ab}$, which is therefore parametrized as
$\hat \D \leftrightarrow \{ 0,  \D(x) \}$ with
\beq
\Delta(x)=2\a_d+2\a(x)=-2\int_m^1\de y\, \a(y)+2\a(x) \ .
\eeq
By integrating this relation we obtain
$\int_m^1\de x\,\a(x)=\frac{1}{2m}\int_m^1\de x\,\Delta(x)$, therefore we can invert the relation to obtain
\beq\label{eq:aD}
\a(x)=\frac 12\Delta(x)+\frac{1}{2m}\int_m^1\de y\,\Delta(y) \ .
\eeq
The seemingly strange conventions that we adopted for the signs of the above parametrization are justified by the fact that with this choice both 
$\a(x)$ and $\D(x)$ are positive functions. Clearly this is the case by definition for $\D(x)$, and from Eq.~\eqref{eq:aD} we deduce that the same must be
true for $\a(x)$. Moreover, $\D(x)$ must be a {\it decreasing} function of $x$ (contrary to the usual overlap function~\cite{MPV87}). This is because
larger values of $x$ correspond to inner blocks of the matrix $\D_{ab}$, hence to replicas that are ``closer'' to each other in the usual interpretation, and
therefore must have a smaller value of $\D$.
An example is given in Fig.~\ref{fig:hier}. 
Of course, the above construction can be generalized to the case where the function $\D(x)$ is allowed to have a continuous part: this can
be thought as an appropriate limit of the $k$RSB construction when $k\to\io$ and is therefore called ``fullRSB" or ``$\io$RSB''.

\subsection{The algebra of hierarchical matrices and the entropic term}
\label{sec:entropic}

\subsubsection{General expression of the entropic term}

In order to compute the entropic term, we need to compute $\log \det \hat \alpha^{m,m}$. For this we need to recall some standard
results for the algebra of hierarchical matrices. Let us consider a generic hierarchical matrix $q_{ab}$, parametrized by the corresponding
function $q(x)$ and with diagonal element $q_d$. 
Although our matrices $\hat Q_m$ are $m\times m$ matrices with fixed $m<1$, we can think to embed them in a $n \times n$ dimensional
matrix $\hat Q_n$ with $n\to 0$, and just set to zero the element of the outermost block. This is consistent with the fact that $q(x)$ is defined in $x\in [0,1]$,
and it corresponds to the special choice that $q(x)$ vanishes for $0 \leq x < m$. We can therefore use standard results for the algebra of 
$n \times n$ hierarchical matrices, in the limit $n\to 0$, see e.g.~\cite{MP91}. 
The important result of~\cite{MP91} that we need is that,
introducing the notations
\beq
[q](x)=x q(x) - \int_0^x\de y\, q(y) \ , 
\hskip2cm
 \langle q \rangle = \int_0^1\de x\, q(x) \ ,
\eeq
we have for the determinant~\cite[Eq.(AII.11)]{MP91}
\beq
\lim_{n\to 0} \frac1n \log\det \hat Q_n = \log(q_d - \la q \ra) + \frac{q(0)}{q_d - \la q \ra} - \int_0^1 \frac{\de y}{y^2} \log\left(
\frac{ q_d - \la q \ra - [q](y)}{q_d - \la q\ra}
\right) \ ,
\eeq
and the inverse matrix is parametrized by~\cite[Eq.(AII.7)]{MP91}
\beq\begin{split}
(q^{-1})_d &= \frac{1}{q_d - \la q \ra} \left( 1 - \int_0^1  \frac{\de y}{y^2}  \frac{[q](y)}{q_d - \la q \ra - [q](y)} - \frac{q(0)}{q_d - \la q\ra} \right) \ , \\
(q^{-1})(x)&=-\frac{1}{q_d-\langle q\rangle}\left[\frac{q(0)}{q_d-\langle q\rangle}+\frac{[q](x)}{x(q_d -\langle q\rangle-[q](x))}+
\int_0^x\frac{\de y}{y^2}\frac{[q](y)}{q_d-\langle q\rangle-[q](y)}\right] \ . 
\end{split}\eeq
To adapt these results to our case, we note that if the outermost blocks vanish, we have $q(0)=0$. Furthermore, $q(x) = [q](x) = 0$ for $x<m$.
Finally, $\det \hat Q_n = (\det \hat Q_m)^{n/m}$, hence $\log \det \hat Q_m = (m/n) \log \det \hat Q_n$, 
and the diagonal element of the inverse of $\hat Q_n$ and $\hat Q_m$ are identical. We conclude
that in the $m\times m$ space
\beq\label{eq:hieralg}
\begin{split}
\log\det \hat Q_m &= m \log(q_d - \la q \ra)  - m \int_m^1 \frac{\de y}{y^2} \log\left(
\frac{ q_d - \la q \ra - [q](y)}{q_d - \la q\ra}
\right) \\
&=  \log(q_d - \la q \ra)  - m \int_m^1 \frac{\de y}{y^2} \log\left(
q_d - \la q \ra - [q](y)
\right)
\ , \\
(q^{-1})_d &= \frac{1}{q_d - \la q \ra} \left( 1 - \int_m^1  \frac{\de y}{y^2}  \frac{[q](y)}{q_d - \la q \ra - [q](y)}  \right) \ , \\
(q^{-1})(x) 
&=-\frac{1}{q_d-\langle q\rangle}\left[\frac{[q](x)}{x(q_d-\langle q\rangle-[q](x))}+\int_m^x\frac{\de y}{y^2}\frac{[q](y)}{q_d-\langle q\rangle-[q](y)}\right] \ , \\
[q](x) &=x q(x) - \int_m^x\de y\, q(y) \ , \\
 \langle q \rangle & = \int_m^1\de x\, q(x) \ .
\end{split}\eeq

In order to compute $\log \det \hat \alpha^{m,m}$, 
we introduce a matrix $\hat \b$ which is ``regularized''
in such a way that $\sum_b \b_{ab} = \e$, and
$\hat\a = \lim_{\e\to 0}\hat \b$. 
For instance we can choose 
\beq
\hat \b \leftrightarrow \{ \b_d, \b(x) \} = \{\a_d ,  -\a(x) +\e \d(x-x_0) \} \ ,
\eeq
for any $x_0 \in [m,1]$.
The matrix $\hat \b$ is invertible, hence
we can use the relation $\det \hat \b^{m,m} = (\hat \b^{-1})_{mm} \det \hat \b $.
Using Eq.~\eqref{eq:hieralg} we get
\beq\begin{split}
\log \det \hat\a^{m,m} &= \lim_{\e\to 0}\left\{ \log \det \hat \b +\log (\hat \b^{-1})_{mm}\right\}
= \lim_{\e\to 0}\left\{ \log \det \hat \b +\log \b^{-1}_d \right\}
 \\
&=
\lim_{\e\to 0}\left\{
- m \int_m^1 \frac{\de y}{y^2} \log\left(
\b_d - \la \b \ra - [\b](y)
\right)
+ \log \left( 1 - \int_m^1  \frac{\de y}{y^2}  \frac{[\b](y)}{\b_d - \la \b \ra - [\b](y)}  \right) 
\right\} \ .
\end{split}\eeq
Note that for $\e\to 0$ we have $\la \b \ra = -\int_m^1 \de x \, \a(x) = \a_d = \b_d$. However, the singular terms $\log(\b_d - \la \b \ra)$ cancel and we
obtain a finite result
\beq\label{eq:entropicA}
\log \det \hat\a^{m,m} =
- m \int_m^1 \frac{\de y}{y^2} \log\left(
 [\a](y)
\right)
+ \log \left( 1 + \int_m^1  \frac{\de y}{y^2}  \right) 
= -\log m - m \int_m^1 \frac{\de y}{y^2} \log\left(
 [\a](y)
\right)
 \ .
\eeq
Expressed in terms of the function $\D(x)$ by means of Eq.~\eqref{eq:aD}, 
we have that
\beq
-m\int_m^1\frac{\de y}{y^2}\log\left( [\a](y) \right)=-(m-1) \log 2 -m\int_m^1\frac{\de y}{y^2}\log\left[y\D(y) +  \int_y^1 \de z \D(z) \right] \ ,
\eeq
therefore
\beq\label{eq:entropicD}
\log\det\hat \a^{m,m}=-\log m-(m-1) \log 2-m\int_m^1\frac{\de y}{y^2}\log\left[ y\D(y)+ \int_y^1 \de z \D(z)\right] \ .
\eeq

\subsubsection{1RSB and 2RSB}

When specialized to the 1RSB case Eq.~\eqref{alpha1RSB}, which corresponds to $ \a(x) = \wh\a_1$ and $\D(x) = \wh \D_1 = 2m \wh\a_1$, 
we get
\beq\label{eq:entropic1RSB}
\log\det\hat \a_{\rm 1RSB}^{m,m}=-\log m + (m-1) \log(m \wh \a_1) =-\log m + (m-1) \log(\wh \D_1/2) \ ,
\eeq
which reproduces the results of~\cite{KPZ12,KPUZ13}.
The 2RSB solution is parametrized by a step function
\beq
\a(x)=
\begin{cases}
\wh\a_1 & m \leq x < m_1\\
\wh\a_2 & m_1 \leq x \leq 1
\end{cases}
\hskip2cm
[\a](x)=
\begin{cases}
m \wh\a_1 & m \leq x < m_1\\
m_1 \wh\a_2 + (m-m_1) \wh\a_1 & m_1 \leq x \leq 1
\end{cases}
\eeq
and therefore
\beq
\log\det\hat\a_{\rm 2RSB}^{m,m}=-\log m +\left(\frac{m}{m_1}-1\right)\log(m\wh\a_1)+\left(m-\frac{m}{m_1}\right)\log\left(m_1\wh\a_2+(m-m_1)\wh\a_1\right) \ .
\eeq
Using the relations
\beq\label{eq:Da2RSB}
\begin{split}
\wh\D_2 &= 2 m_1 \wh \a_2 + 2\left( m - m_1 \right)\wh \a_1 \ , \\
\wh\D_1 &= \wh\D_2 + 2 (\wh \a_1 - \wh \a_2) \ , 
\end{split}\eeq
or directly Eq.~\eqref{eq:entropicD},
we have
\beq\label{eq:entropic2RSB}
\log\det\hat\a_{\rm 2RSB}^{m,m}=-\log m -(m-1)\log 2 +\left(\frac{m}{m_1}-1\right)\log[  m_1 \wh\D_1 + (1 -m_1) \wh\D_2 ]+\left(m-\frac{m}{m_1}\right)\log \wh\D_2 \ .
\eeq

\subsubsection{$k$RSB}

Writing explicitly the $k$RSB expression requires the introduction of a function
\beq\label{eq:GxDx}
G(x) = x \D(x) + \int_x^1 \de z \D(z) \ .
\eeq
In fact, if $\D(x)$ is a $k$RSB piecewise constant function and using the notations of Fig.~\ref{fig:hier}, 
it is easy to check that $G(x)$ is also a piecewise constant function parametrized by $\wh G_i$, with
\beq
\wh G_i = m_i \wh\D_i + \sum_{j=i+1}^k (m_j - m_{j-1}) \wh \D_j \ .
\eeq
Note that $\frac{\de G(x)}{\de x} = x \frac{\de\D(x)}{\de x}$, and $G(1) = \D(1)$, therefore from $G(x)$ one can reconstruct $\D(x)$ as 
\beq\label{eq:DxGx}
\D(x) = G(1) - \int_x^1 \frac{\de z}z \frac{\de G(z)}{\de z} = \frac{G(x)}x - \int_x^1 \frac{\de z}{z^2} G(z) \ ,
\eeq
or for a $k$RSB function
\beq\label{eq:DiGi}
\wh\D_i = \frac{\wh G_i}{m_i} + \sum_{j=i+1}^k \left(\frac1{m_j} - \frac1{m_{j-1}} \right) \wh G_j \ .
\eeq
Then, from Eq.~\eqref{eq:entropicD}, we have
\beq\label{eq:entropickRSB}
\begin{split}
\log\det\hat \a^{m,m}_{k{\rm RSB}} &= -\log m-m\int_m^1\frac{\de y}{y^2}\log\left[ G(y)/2\right]  \\
& =  -\log m + \sum_{i=1}^k \left( \frac{m}{m_{i}} - \frac{m}{m_{i-1}}  \right) \log ( \wh G_i/2 )
\end{split}\eeq
Note that from this result we can recover the 1RSB and 2RSB results obtained above.

\subsection{The interaction term}
\label{sec:interactionterm}

\subsubsection{$k$RSB derivation}

We now compute the interaction term of the replicated entropy for a generic hierarchical matrix $\D_{ab}$ parametrized by $\D(x)$. 
We start from Eq.~(\ref{int_gen}),
and we need to compute the function
\beq\label{eq:gmhder}
g(m,h) =
\left\{
\exp\left[-\frac{1}{2}\sum_{a,b=1}^m \Delta_{ab}\frac{\partial^2}{\partial h_a\partial h_b}\right] \prod_{a=1}^m \th\left(  h_a     \right)
 \right\}_{\{h_a=h\}} \ .
 \eeq
Because $\D_{ab}$ is a hierarchical matrix, this computation can be done by taking 
the derivative with respect to the external fields in a hierarchical way~\cite{Du81}. 
Let us define the matrix $I^{m_i}_{ab}$, which has elements equal to 1 in blocks of size $m_i$ around the diagonal, and zero otherwise. In other
words, the matrix $I^{m_i}_{ab}$ is parametrized by a function $I^{m_i}(x) = 1$ for $m_i \leq x \leq 1$ and zero otherwise. 
Then, recalling the notations of Fig.~\ref{fig:hier}, and noting that $I^{m_k=1}_{ab} = \d_{ab}$, one can easily check that
\beq
\D_{ab} = \sum_{i=1}^k \wh\D_i  (  I^{m_{i-1}}_{ab} - I^{m_i}_{ab}     )   =  \sum_{i=0}^{k-1} (\wh \D_{i+1} - \wh \D_{i} ) I^{m_{i}}_{ab}  - \wh\D_k \d_{ab} \ .
\eeq
Inserting this form in Eq.~\eqref{eq:gmhder}, one obtains a sequence of differential operators acting on the product of theta functions,
each of them being the sum of partial derivatives inside a block. This sequence of operations can be written as a recursion;
since the procedure is very well explained in~\cite{Du81}, we
only report the main results here.
When acting with the term containing $\wh\D_k$ we obtain, recalling Eq.~\eqref{eq:derint}:
\beq\label{eq:g1}
g(1,h)= \exp\left[\frac{\wh \D_{k}}{2}\frac{\de^2}{\de h^2}\right] \th(h) =  \g_{\wh\D_k} \star \th(h) = \Th\left(\frac{h}{\sqrt{2\wh\D_k}}\right) \ .
\eeq
Then, the action of each of the terms $i= k-1, \cdots, 0$ induces a recursion of the form
\beq\label{eq:gi}
g(m_{i},h)=\exp\left[\frac{\wh \D_{i}-\wh \D_{i+1}}{2}\frac{\de^2}{\de h^2}\right]g(m_{i+1},h)^{\frac{m_{i}}{m_{i+1}}}
= \g_{\wh\D_i - \wh\D_{i+1}} \star  g(m_{i+1},h)^{\frac{m_{i}}{m_{i+1}}}
 \ .
\eeq
Note that the last of these iterations can be written explicitly as
\beq\label{eq:gm}
g(m,h) = \g_{-\wh\D_1} \star g(m_1,h)^{\frac{m}{m_1}} \ .
\eeq
This recursive procedure
allows us to compute easily $g(m,h)$ for any $k$, and, according to Eq.~\eqref{int_gen}, the function $\FF(\D)$.

The last iteration might seem problematic, because the kernel $\g_a$ has a negative parameter $a = -\wh\D_1$ and therefore cannot be
represented as in Eq.~\eqref{eq:derint}. Luckily enough, this last iteration can be eliminated. In fact,
we have
\beq\label{eq:FDapp}
\begin{split}
\mathcal F(\Delta) 
&= \int_{-\infty}^\infty \de h\, e^{h} 
 \frac{\de}{\de h} g(m,h)
= \int_{-\infty}^\infty \de h\, e^{h}  e^{- \frac{\wh \D_1}{2}\frac{\de^2}{\de h^2}}    \frac{\de}{\de h}  g(m_1,h)^{\frac{m}{m_1}}  \\
& = \int_{-\infty}^\infty \de h\,     \left[ e^{- \frac{\wh \D_1}{2}\frac{\de^2}{\de h^2}}  e^{h}   \right]  \frac{\de}{\de h}  g(m_1,h)^{\frac{m}{m_1}}
 = e^{- \frac{\wh \D_1}{2}} \int_{-\infty}^\infty \de h\,     e^{h}   \frac{\de}{\de h}  g(m_1,h)^{\frac{m}{m_1}} \ ,  \\
\end{split}\eeq
where we performed an integration by parts and assumed that all the boundary terms vanish because all the derivatives of
the function $\frac{\de}{\de h}  g(m_1,h)^{\frac{m}{m_1}}$ vanish for $h\to\io$.
This is justified by the fact that $g(m_i,h)$ behaves, for large $h$, similarly to a $\Th(h)$ function, as it can be seen from the recurrence equations \eqref{eq:g1} and \eqref{eq:gi}.
Assuming, for the same reason, that $1-g(m_1,h)^{\frac{m}{m_1}}$ decays at $h\to -\infty$ faster than a simple exponential, in order to perform another integration by parts,
we have
\beq\label{eq:FD}
\mathcal F(\Delta) 
=e^{- \frac{\wh \D_1}{2}}  \int_{-\infty}^\infty \de h\, e^{h} 
 \frac{\de}{\de h} g(m_1,h)^{\frac{m}{m_1}}
=e^{- \frac{\wh \D_1}{2}}  \int_{-\infty}^\infty \de h\, e^{h} 
  \left\{ 1- g(m_1,h)^{\frac{m}{m_1}} \right\} \ .
\eeq

\subsubsection{$1$RSB and $2$RSB}

Using Eq.~\eqref{eq:g1} we obtain, at the 1RSB level,
\beq\label{eq:int1RSB}
\FF_{\rm 1RSB}(\D) =e^{- \frac{\wh \D_1}{2}}   \int_{-\infty}^\infty \de h\, e^{h} 
  \left\{ 1- [  \g_{\wh\D_{1}} \star \th(h)]^m \right\} \ ,
\eeq
at the 2RSB level
\beq\label{eq:int2RSB}
\begin{split}
 g(m_1,h)^{\frac{m}{m_1}} & =   \left[  \g_{\wh\D_{1} - \wh\D_2 } \star g(1,h)^{m_{1}} \right]^{\frac{m}{m_1}} =
 \left[  \g_{\wh\D_{1} - \wh\D_2 } \star \left(  \g_{\wh\D_{2}} \star \th(h)     \right)^{m_{1}} \right]^{\frac{m}{m_1}} 
\ , \\
\FF_{\rm 2RSB}(\D) &=e^{- \frac{\wh \D_1}{2}}  \int_{-\infty}^\infty \de h\, e^{h} 
  \left\{ 1-  \left[  \g_{\wh\D_{1} - \wh\D_2 } \star \left(  \g_{\wh\D_{2}} \star \th(h)     \right)^{m_{1}} \right]^{\frac{m}{m_1}} 
  \right\} \ ,
\end{split}\eeq
and so on. At the 1RSB level, we reproduce the result of~\cite{KPZ12, KPUZ13}.

\subsubsection{Continuum limit}
\label{sec:continuum_g}

The above strategy is not very practical when $\D(x)$ is a continuous function. In this case, to
take the continuum limit it is convenient to follow again~\cite{Du81}. We have $m_i = x-\de x$, $m_{i+1} = x$, and
$\wh \D_{i}-\wh \D_{i+1} =- \dot\D(x) \de x$, where a dot denotes derivative with respect to $x$.
Hence Eq.~\eqref{eq:gi} becomes in the region where $\D(x)$ is a continuous function
\beq
g(x-\de x,h)=\exp\left[-\de x\frac{\dot \D(x)}{2}\frac{\de^2}{\de h^2}\right]g(x,h)^{\frac{x-\de x}{x}}
\eeq
and then expanding in powers of $\de x$, at linear order we get the Parisi equation
\beq
\frac{\partial g(x,h)}{\partial x}=\frac 12 \dot\D(x)\frac{\partial^2 g(x,h)}{\partial h^2}+\frac 1x g(x,h)\log g(x,h)
\eeq
that has to be solved with the initial condition~\eqref{eq:g1}.
Note that the partial differential equation is well defined because $\dot \D(x)\leq 0$. 
This equation has to be integrated from $x=1$ down to $x=m_1$. The resulting $g(m_1,h)$ can be inserted in Eq.~\eqref{eq:FD}
to obtain $\FF(\D)$.

The partial differential equation written above can be also put in a more convenient form by introducing the function	
\beq\label{eq:fdef}
f(x,h)=\frac{1}{x}\log g(x,h)
\eeq
which obeys the equation
\beq\label{eq:Parisi}
\frac{\partial f(x,h)}{\partial x}=\frac 12\dot\D(x)\left[ \frac{\partial^2 f(x, h)}{\partial h^2}+x\left(\frac{\partial f(x, h)}{\partial h}\right)^2    \right]
\eeq
with initial condition
\beq\label{eq:initialf}
f(1,h)=\log\Th\left[\frac{h}{\sqrt{2\D(1)}}\right]\:.
\eeq
Remarkably enough, the equations written above are identical to those of the SK model~\cite{MPV87,Du81}. 
The only difference is in the initial condition for the Parisi equation.

\subsection{1RSB, 2RSB, $k$RSB, fullRSB expressions of the replicated entropy}
\label{sec:IIID}

We therefore obtained the expression of the replicated entropy, under the assumption that the matrices $\hat\a$ and $\hat\D$ are hierarchical
$k$RSB matrices, and
we now summarize the results that we obtained at different levels of RSB for the replicated entropy~\eqref{eq:gauss_r}.
At any level of $k$RSB, the replicated entropy has the form
\beq\label{eq:SSKRSBdef}
\begin{split}
s[\hat\a_{k{\rm RSB}}] &= 1 - \log\r + \frac{d}2 m \log m + \frac{d}{2}(m-1) \log( \pi e D^2/d^2) + \frac{d}2 \SS_{k{\rm RSB}} \ , \\
\SS_{k{\rm RSB}} &= (2-m) \log m + (m-1) \log 2 + \log \det(\hat \a^{m,m}_{k{\rm RSB}}) -   \wh \f \,
\FF\left( 2 \hat \a_{k{\rm RSB}} \right) 
\end{split}\eeq
where $\SS_{k{\rm RSB}}$ has been defined in such a way that it contains the non-trivial dependence on $\hat \a$ and
it has a good limit for $m\to 0$ (as we will discuss later).

At the 1RSB level, using Eqs.~\eqref{eq:entropic1RSB}, \eqref{eq:int1RSB} and \eqref{eq:FD}, we obtain
\beq\begin{split}
\SS_{\rm 1RSB}  &=  
  (m-1) \log(\wh \D_1/m) - \wh \f \,
 e^{-\wh\D_1/2}\int_{-\infty}^\infty \de h\, e^{h} 
  \{ 1-  [  \g_{\wh\D_{1}} \star \th(h)]^m  \} 
\end{split}\eeq
which coincides with the previously derived results~\cite{PZ10,KPZ12,KPUZ13}.
At the 2RSB level, we have, using Eqs.~\eqref{eq:entropic2RSB}, \eqref{eq:int2RSB} and \eqref{eq:FD}, 
\beq\label{eq:sD2RSB}
\begin{split}
\SS_{\rm 2RSB} &= 
\left( \frac{m}{m_1}-1 \right) \log[ (m_1 \wh\D_1 + (1-m_1) \wh\D_2)/m ] +\left( m - \frac{m}{m_1} \right)  \log (\wh\D_2/m) \\
&- \wh \varphi
  e^{-\frac{\wh \D_1}{2}}\int_{-\infty}^\infty \de h\, e^{h}\left[ 1-\left[ \g_{\wh\D_1-\wh\D_2}\star\left(\g_{\wh\D_2}\star\theta(h) \right)^{m_1} \right]^{m/m_1} \right]
\end{split}
\eeq
At the generic $k$RSB level we obtain from Eq.~\eqref{eq:entropickRSB}, \eqref{eq:FD}, \eqref{eq:g1}, \eqref{eq:gi}, \eqref{eq:gm}:
\beq\label{eq:sDkRSB}
\begin{split}
\SS_{k{\rm RSB}} &=  \sum_{i=1}^k \left( \frac{m}{m_{i}} - \frac{m}{m_{i-1}}  \right) \log ( \wh G_i/m ) - \wh\f \, e^{-\wh\D_1/2}\int_{-\infty}^\infty \de h\, e^{h} 
  \left\{ 1- g(m_1,h)^{\frac{m}{m_1}} \right\}  \ , \\
\wh G_i &= m_i \wh\D_i + \sum_{j=i+1}^k (m_j - m_{j-1}) \wh \D_j \ , \\
g(1,h) & = \g_{\wh\D_k} \star \th(h) \ ,  \\
g(m_i,h) &= \g_{\wh\D_i - \wh\D_{i+1}} \star g(m_{i+1},h)^{\frac{m_i}{m_{i+1}}} \ , \hskip20pt i = 1 \cdots k-1 \ . \\
\end{split}
\eeq
Finally, the fullRSB expression is, from Eqs.~\eqref{eq:entropicD}, \eqref{eq:FD}, \eqref{eq:fdef}, \eqref{eq:Parisi}, \eqref{eq:initialf}: 
\beq\label{eq:sDfullRSB}
\begin{split}
\SS_{\io{\rm RSB}} & = 
-  m \int_m^1\frac{\de y}{y^2}\log\left[ \frac{y\D(y)}m+ \int_y^1 \de z \frac{ \D(z) }m \right] 
-   \wh \f \,
e^{-\D(m)/2}\int_{-\infty}^\infty \de h\, e^{h} 
  [1- e^{m f(m,h)}] \ , \\
  \frac{\partial f(x,h)}{\partial x}&=\frac 12\dot\D(x)\left[ \frac{\partial^2 f(x, h)}{\partial h^2}+x\left(\frac{\partial f(x, h)}{\partial h}\right)^2    \right] \ , \\
  f(1,h)&=\log\Th\left[\frac{h}{\sqrt{2\D(1)}}\right]\:.
\end{split}\eeq

\section{Variational equations}
\label{sec:variational}

The replicated entropy depends on the function $\D(x)$ that parametrizes the hierarchical matrix $\hat\D$. This function
is determined by optimization of the replicated entropy through a variational principle~\cite{MPV87}.
In this section we derive the variational equations for the function $\D(x)$ that are obtained by optimization of the free energy,
i.e. by imposing the equation $\frac{\partial s[\hat \a]}{\partial \D_{ab}}=0$.
It is well known in the context of spin glasses~\cite{MPV87} and structural glasses~\cite{MP09,PZ10} that,
for integer $m>1$, this corresponds to the usual maximization of the entropy or minimization of the free energy with respect to the
variational parameters $\a_{ab}$, but when the problem is analytically continued to real $m < 1$,
the solution of these equations does not correspond to a maximum of the entropy, but to a saddle point.
This not very important because to really characterize the stability of the saddle point one has first to compute the matrix
of second derivatives $\frac{\partial^2 s[\hat \a]}{\partial \D_{ab} \partial \D_{cd}}$, and then perform the analytic continuation of its
eigenvalues to $m<1$. The continued eigenvalues are required to be positive. 

In this section we derive the variational equations for $\D(x)$, and we postpone a partial analysis of the stability matrix
to the following sections.
In Sec.~\ref{sec:vark} we derive the equation for $\wh\D_i$ in the case of a $k$RSB structure. In Sec.~\ref{sec:varfull}
we derive the fullRSB equations in two equivalent ways: first by using Lagrange multipliers, and then by taking the $k\to\io$
limit of the $k$RSB equations.

\subsection{Variational equations for the $k$RSB solution}
\label{sec:vark}

We consider first the $k$RSB solution for fixed $k$. We start from Eq.~\eqref{eq:sDkRSB} and we want to impose
the condition $\frac{\partial \SS_{k{\rm RSB}}}{\partial \wh\D_i} =0 $.
To do this, we consider Eq.~\eqref{eq:gmhder} and \eqref{eq:FD}.
We have, without taking into account that the matrix $\D_{ab}$ is symmetric and for $a\neq b$:
\beq\begin{split}
\frac{\partial g(m,h)}{\partial \D_{ab}} &= 
\left\{
\exp\left[-\frac{1}{2}\sum_{c,d=1}^m \D_{cd}\frac{\partial^2}{\partial h_c\partial h_d}\right]  \left( - \frac12 \frac{\partial^2}{\partial h_a\partial h_b} \right) 
 \prod_{f=1}^m \th\left( h_f    \right)
 \right\}_{\{h_a=h\}} \\
 &=-\frac12 \left\{
\exp\left[-\frac{1}{2}\sum_{c,d=1}^m \D_{cd}\frac{\partial^2}{\partial h_c\partial h_d}\right]  \d(h_a) \d(h_b)
\prod_{f\neq a,b}^{1,m} \th(h_f)
 \right\}_{\{h_a=h\}} 
\end{split}\eeq
Note that there are $m ( m_{\ell-1} - m_\ell )$ elements in the block $\ell = 1 \cdots k$ and they all give the same contribution, therefore
\beq\begin{split}
\frac{\partial g(m,h)}{\partial \wh\D_{\ell}}  &=-\frac{m ( m_{\ell-1} - m_\ell )}2 \left\{
\exp\left[-\frac{1}{2}\sum_{c,d=1}^m \D_{cd}\frac{\partial^2}{\partial h_c\partial h_d}\right]  \d(h_a) \d(h_b)
\prod_{f\neq a,b}^{1,m} \th(h_f)
 \right\}_{\{h_a=h\}} \ .
\end{split}\eeq
For a hierarchical matrix, we can use again the strategy of~\cite{Du81}. The difference is that at the beginning of the iteration, 
there is a special block that contains the delta instead of the theta function. At some level, the two blocks that contain the delta are merged.
If the two replicas $ab$ are in the $\ell$-th block, then we have the following iteration equations:
\beq\label{recN}
\begin{split}
N_\ell(1,h) & =  \g_{\wh\D_k} \star \d(h) = \frac{e^{-h^2/(2\wh\D_k)}}{\sqrt{2\pi\wh\D_k}}  \\
N_\ell(m_i,h) & = \g_{\wh\D_{i} - \wh\D_{i+1}} \star [ N_\ell(m_{i+1},h) g(m_{i+1},h)^{m_i/m_{i+1}-1} ] \hskip20pt i=k-1,\cdots,\ell \\
N_\ell(m_{\ell-1},h) &= \g_{\wh\D_{\ell-1} - \wh\D_{\ell}} \star [ N_\ell(m_{\ell},h)^2 g(m_{\ell},h)^{m_{\ell-1}/m_{\ell}-2} ] \\
N_\ell(m_i,h) & = \g_{\wh\D_{i} - \wh\D_{i+1}} \star [ N_\ell(m_{i+1},h) g(m_{i+1},h)^{m_i/m_{i+1}-1} ] \hskip20pt i=\ell-2,\cdots,0 \\
\frac{\partial g(m,h)}{\partial \wh\D_{\ell}}  &=-\frac{m ( m_{\ell-1} - m_\ell )}2 N_\ell(m,h) 
\end{split}\eeq
Note that these equations hold for all $\ell = k, \cdots, 1$.
In the discrete $k$RSB case, Eq.~\eqref{eq:fdef} reads
\beq
f(m_i,h) = \frac{1}{m_i} \log g(m_i,h) \ ,
\eeq
and we note that for $i > \ell$, namely before the two blocks are merged, we can show by recurrence that
(denoting with primes the derivatives with respect to $h$)
\beq
g'(m_i,h) =  m_i N_\ell(m_i,h) \hskip20pt \Leftrightarrow \hskip20pt
f'(m_i,h) = N_\ell(m_i,h) / g(m_i,h)
\ .
\eeq
This allows for an important simplification because we do not need $N_\ell$ for $i \geq \ell$ anymore. We can rewrite Eqs.~\eqref{recN} equivalently
as
\beq
\begin{split}
N_\ell(m_{\ell-1},h) &= \g_{\wh\D_{\ell-1} - \wh\D_{\ell}} \star [ f'(m_{\ell},h)^2 g(m_{\ell},h)^{m_{\ell-1}/m_{\ell}} ] \\
N_\ell(m_i,h) & = \g_{\wh\D_{i} - \wh\D_{i+1}} \star [ N_\ell(m_{i+1},h) g(m_{i+1},h)^{m_i/m_{i+1}-1} ] \hskip20pt i=\ell-2,\cdots,0 \\
\frac{\partial g(m,h)}{\partial \wh\D_{\ell}}  &=-\frac{m ( m_{\ell-1} - m_\ell )}2 N_\ell(m,h) 
\end{split}\eeq
Now, let us introduce a series of operators $\G_\ell$ that are defined by the following equations for an arbitrary test function $t(h)$:
\beq
\begin{split}
\G_1 \star t(h) &= g(m_1,h)^{\frac{m}{m_1}} t(h) \ , \\
\G_i \star t(h) &= \G_{i-1} \star \left[ \frac{1}{g(m_{i-1},h)} \, \g_{\wh\D_{i-1} - \wh\D_i} \star \, g(m_i,h)^{\frac{m_{i-1}}{m_i}} t(h) \right]
\hskip20pt i = 2 , \cdots , k
\ .
\end{split}
\eeq
Then we have
\beq
N_\ell(m,h) = \g_{-\wh\D_1} \star \G_\ell \star f'(m_\ell,h)^2 \ .
\eeq

The next step is to take the derivative with respect to $\wh G_i$ of the $k$RSB free energy, which we write in the form
\beq
\SS_{k{\rm RSB}} =   \sum_{i=1}^k \left( \frac{m}{m_{i}} - \frac{m}{m_{i-1}}  \right) \log ( \wh G_i/m ) - \wh\f \, \int_{-\infty}^\infty \de h\, e^{h} \frac{\de}{\de h} g(m,h) \ ,
 \eeq
 and make use of the relation~\eqref{eq:DiGi}.
We get
\beq\begin{split}
\left( \frac{1}{m_{i}} - \frac{1}{m_{i-1}}  \right) \frac{m}{\wh G_i} & = \wh\f \int_{-\infty}^\infty \de h\, e^{h} \frac{\de}{\de h} \sum_{j=1}^k \frac{\partial g(m,h)}{\partial\wh\D_j} 
\frac{\partial\wh\D_j}{\partial\wh G_i} \\
& = \wh\f \int_{-\infty}^\infty \de h\, e^{h} \frac{\de}{\de h} \left\{
\frac{1}{m_i} \frac{\partial g(m,h)}{\partial \wh\D_i} + 
\sum_{j=1}^{i-1} \frac{\partial g(m,h)}{\partial \wh\D_j} \left( \frac{1}{m_{i}} - \frac{1}{m_{i-1}}  \right)
\right\} \\
& = \frac{m \wh\f}2 \int_{-\infty}^\infty \de h\, e^{h} \frac{\de}{\de h} \left\{
\frac{1}{m_i}(m_i - m_{i-1}) N_i(m,h) +
\sum_{j=1}^{i-1} ( m_j - m_{j-1}) N_j(m,h) \left( \frac{1}{m_{i}} - \frac{1}{m_{i-1}}  \right)
\right\}
\ .
\end{split}\eeq
We observe from Eq.~\eqref{recN} that $N_i(m,h)$ behaves like a Gaussian for $h\to\pm \io$, therefore we can safely integrate by parts,
and the last expression can be written as
\beq\begin{split}
 \frac{1}{\wh G_i} & = -\frac{ \wh\f}2 \int_{-\infty}^\infty \de h\, e^{h} \frac{\de}{\de h} \left\{
  m_{i-1} N_i(m,h) -
\sum_{j=1}^{i-1} ( m_j - m_{j-1}) N_j(m,h) 
\right\} \\
& = -\frac{ \wh\f}2 \int_{-\infty}^\infty \de h\, e^{h} \g_{-\wh\D_1} \star \frac{\de}{\de h} \left\{
  m_{i-1} \G_i \star f'(m_i,h)^2 -
\sum_{j=1}^{i-1} ( m_j - m_{j-1}) \G_j \star f'(m_j,h)^2 
\right\} \\
& = -\frac{ \wh\f}2 e^{-\wh\D_1 / 2} \int_{-\infty}^\infty \de h\, e^{h} \, \frac{\de}{\de h} \left\{
  m_{i-1} \G_i \star f'(m_i,h)^2 -
\sum_{j=1}^{i-1} ( m_j - m_{j-1}) \G_j \star f'(m_j,h)^2 
\right\} \\
& = \frac{ \wh\f}2 e^{-\wh\D_1 / 2} \int_{-\infty}^\infty \de h\, e^{h} \, \left\{
  m_{i-1} \G_i \star f'(m_i,h)^2 -
\sum_{j=1}^{i-1} ( m_j - m_{j-1}) \G_j \star f'(m_j,h)^2 
\right\} \ , 
\end{split}\eeq
using the same trick as in Eq.~\eqref{eq:FDapp}.
Finally, we can show that
\beq
 \int_{-\infty}^\infty \de h\, e^{h} \,  \G_i \star f'(m_i,h)^2 =  \int_{-\infty}^\infty \de h P(m_i,h) f'(m_i,h)^2 \ ,
\eeq
provided $P(m_i,h)$ satisfies the following recurrence equations:
\begin{eqnarray}
\label{eq:recP_b}
P(m_1,h) &=& e^h \, g(m_1,h)^{\frac{m}{m_1}} \ , \\
\label{eq:recP}
P(m_i,h) &=& \int \de z  \frac{P(m_{i-1},z)}{g(m_{i-1},z)} \g_{\wh\D_{i-1}-\wh\D_{i}}(h-z) g(m_i,h)^{\frac{m_{i-1}}{m_i}} 
\hskip20pt i = 2 , \cdots , k
\ .
\end{eqnarray}
We obtain therefore the final result
\beq\label{eq:varGfin}
 \frac{1}{\wh G_i}  = \frac{ \wh\f}2 e^{-\wh\D_1 / 2} \int_{-\infty}^\infty \de h \, \left\{
  m_{i-1} P(m_i,h) f'(m_i,h)^2 -
\sum_{j=1}^{i-1} ( m_j - m_{j-1}) P(m_j,h) f'(m_j,h)^2
\right\} \ .
\eeq
Eqs.~\eqref{eq:varGfin}, \eqref{eq:recP_b}-\eqref{eq:recP}, \eqref{eq:g1}-\eqref{eq:gi}, \eqref{eq:DiGi} constitute a set of closed equations for the $\wh G_i$, or
equivalently the $\wh\D_i$. They can be solved by the following iteration: starting from a guess for $\wh\D_i$, one can solve first
the recurrence \eqref{eq:g1}-\eqref{eq:gi} and then the recurrence \eqref{eq:recP_b}-\eqref{eq:recP}. From the solutions one can compute the new $\wh G_i$
using Eq.~\eqref{eq:varGfin} and from these the new $\wh\D_i$ using Eq.~\eqref{eq:DiGi}.

\subsection{Variational equations for the fullRSB solution}
\label{sec:varfull}

In this section we will derive the saddle point equations for this function $\D(x)$ in the
continuum limit of fullRSB. 
We will do this in two different ways.
The first derivation starts from Eq.~\eqref{eq:sDfullRSB}, and it imposes the variational condition
$\frac{\de \SS_{\io{\rm RSB}}}{\de \D(x)}=0$
making use of Lagrange multipliers to enforce the Parisi equation~\cite{SD84}.
The second derivation simply consists of taking the continuum limit of the variational equations
obtained in the previous section. We obtain equivalent results which confirms that the calculation
is correct. Given the complexity of the computation, this is a very useful check.

\subsubsection{Lagrange multipliers}

We need to introduce Lagrange multipliers because
the entropy is written in terms of the function $g(x,h)$ 
or, equivalently, of the function $f(x,h)$, 
and this function must satisfy the Parisi equation~\eqref{eq:Parisi}. 
In order to do properly the optimization, one can introduce two Lagrange multipliers: 
$P(x,h)$ is the one needed to enforce the partial differential equation and 
$P(1,h)$ is the one needed to enforce the initial condition.
Let us start from the complete fullRSB expression of the entropy, Eq.~\eqref{eq:sDfullRSB}
(where terms that are not important to derive the saddle point equations are omitted), 
to which we add these Lagrange multipliers. 
As a reminder,
we indicate the derivative with respect to $x$ with a dot and the derivative with respect to $h$ with a prime.
With these notations, and using Eqs.~\eqref{eq:DxGx}, \eqref{eq:GxDx} with $\D(m) = \frac{G(m)}{m} - \int_m^1 \frac{\de z}{z^2} G(z)$,
we have to impose stationarity of the function
\beq
\begin{split}
\SS_{\io{\rm RSB}}&= - m \int_m^1\frac{\de x}{x^2}\log[ G(x) / m] 
-   \wh \f \,
e^{-\frac{\D(m)}{2} }\int_{-\infty}^\infty \de h\, e^{h} 
  [1- e^{m f(m,h)}]
  \\
&+ m  \wh \f \,
e^{-\frac{\D(m)}{2} } \int_{-\infty}^\infty \de h \int_m^1\de x\,P(x,h)\left\{\dot f(x,h) - \frac 12 \frac{\dot G(x)}x \left[ f''(x, h) + x  f'(x, h)^2    \right]\right\} \\
&- m  \wh \f \,
e^{-\frac{\D(m)}{2} }\int_{-\infty}^\infty\de h\,P(1,h)\left\{f(1,h)-\log\Th\left(\frac{h}{\sqrt{2 G(1)}}\right)\right\}
\end{split}
\eeq
over $\D(x)$, $f(x,h)$, $P(x,h)$, $f(m,h)$ and $P(1,h)$.
The first two equations can be obtained by taking the variation with respect to $P(x,h)$ and $f(x,h)$ 
\begin{eqnarray}\label{eq:full1}
\dot f(x,h) &=& \frac 12\frac{\dot G(x)}x\left[  f''(x, h) + x f'(x, h)^2    \right]  \ , \\
\dot P(x,h)&=&-\frac12 \frac{\dot G(x)}x\left[P''(x,h)-2x(P(x,h)f'(x,h))'\right] \ .
\label{eq:full2}
\end{eqnarray}
Taking the variation over $P(1,h)$ and $f(m,h)$ we obtain
\begin{eqnarray}
\label{eq:full4}
f(1,h) &=&  \log\Th\left(\frac{h}{\sqrt{2 G(1)}}\right) \ , \\
\label{eq:full5}
P(m,h) &=& e^{ mf(m,h) + h} \ .
\end{eqnarray}
Finally, 
taking the variation of $G(x)$ (for $x\neq 1$ and $x\neq m$) we obtain the following equation
\beq\nonumber
\begin{split}
\frac{m}{G(x)} &= - \frac{m \wh\f}2
e^{-\frac{\D(m)}{2} } \int_{-\io}^\io \de h P(x,h) f''(x,h) - \frac{\wh\f}2 e^{-\frac{\D(m)}{2}  } \int_{-\infty}^\infty \de h\, e^{h} 
  [1- e^{m f(m,h)}] \\
  &= - \frac{m\wh\f}2 
e^{-\frac{\D(m)}{2} }\int_{-\io}^\io \de h P(x,h) f''(x,h) - \frac{\wh\f}2 e^{-\frac{\D(m)}{2}   } \int_{-\infty}^\infty \de h\, e^{h} 
   e^{m f(m,h)} m f'(m,h) \\
  &= - \frac{m \wh \f} 2 
e^{-\frac{\D(m)}{2} } \left\{ \int_{-\io}^\io \de h P(x,h) f''(x,h) +  \int_{-\infty}^\infty \de h\, P(m,h)  f'(m,h) \right\} \ ,
\end{split}\eeq
which is more conveniently rewritten as
\beq\label{eq:full3}
\frac{1}{G(x)}
  = - \frac{\wh \f} 2 
e^{-\frac{\D(m)}{2} } \int_{-\io}^\io \de h \big[  P(x,h) f''(x,h) +   P(m,h)  f'(m,h) \big] \ .
\eeq
The system of Eqs.~\eqref{eq:full1}-\eqref{eq:full3} can be in principle solved numerically, with the following procedure:
\begin{itemize}
\item one starts with a guess for $G(x)$;
\item from this one solves Eq.~\eqref{eq:full1} with boundary condition \eqref{eq:full4} to get $f(x,h)$;
\item then one can solve Eq.~\eqref{eq:full2} with boundary condition \eqref{eq:full5} to
obtain $P(x,h)$;
\item from Eq.~\eqref{eq:full3} one obtains the new $G(x)$
\end{itemize}

\subsubsection{Continuum limit of the $k$RSB variational equations}

We now take the continuum limit of the discrete $k$RSB equations following the strategy
of section~\ref{sec:continuum_g}. It was already shown in that section that in this limit 
the recurrence
equations for $g(x,h)$, Eqs.~\eqref{eq:g1} and \eqref{eq:gi}, become the Parisi equation
\eqref{eq:full1} with boundary condition \eqref{eq:full4}.
The boundary condition for $P(x,h)$ in the discrete, Eq.~\eqref{eq:recP_b}, is 
clearly equivalent to the one in the continuum, Eq.~\eqref{eq:full5}.
It is quite simple, following the lines as in section~\ref{sec:continuum_g}, to show
that Eq.~\eqref{eq:recP} becomes, in the continuum limit, Eq.~\eqref{eq:full2}, so we do
not report the derivation. 

It remains to derive Eq.~\eqref{eq:full3}.
We start from Eq.~\eqref{eq:varGfin} which becomes in the continuum limit
\beq
\label{eq:varGfincont}
 \frac{1}{ G(x)}  = \frac{ \wh\f}2 e^{-\frac{\D(m)}2} \int_{-\infty}^\infty \de h \, \left\{
  x P(x,h) f'(x,h)^2 -
\int_{m}^{x} \de z P(z,h) f'(z,h)^2
\right\} \ .
\eeq
We now show that Eq.~\eqref{eq:full3} and Eq.~\eqref{eq:varGfincont} are equivalent.
This amounts to showing that
\beq\label{eq:app1}
\int_{-\io}^\io \de h \big[  P(x,h) f''(x,h) +   P(m,h)  f'(m,h) \big]
=  - \int_{-\infty}^\infty \de h \, \left\{
  x P(x,h) f'(x,h)^2 -
\int_{m}^{x} \de z P(z,h) f'(z,h)^2
\right\} \ .
\eeq
The strategy to prove the equality \eqref{eq:app1} is the following. 
We first see that the two sides are the same at $x=m$ 
and then we prove that their derivatives with 
respect to $x$ are the same.
In the following, we say that $a(x,h)\sim b(x,h)$ if
$\int_{-\infty}^\infty \de h a(x,h)=\int_{-\infty}^\infty \de h b(x,h)$ to simplify the notations.
Doing integration by parts, and noting that 
from the initial condition \eqref{eq:full5} it follows that $P'(m,h)=[1+mf'(m,h)]P(m,h)$,
we find
\beq
P(m,h)f''(m,h)+P(m,h)f'(m,h)\sim f'(m,h) [P(m,h)-P'(m,h)] \sim -m P(m,h) f'(m,h)^2
\eeq
which shows that Eq.~\eqref{eq:app1} holds at $x=m$.
Next, we compute the derivative with respect to $x$ of the arguments of the integrals that appear in 
Eq.~\eqref{eq:app1}. We have, using Eqs.~\eqref{eq:full1} and \eqref{eq:full2}, that
\beq
\dot P f''+P\dot f''\sim \dot Pf''+P''\dot f\sim \frac{1}{2}\frac{\dot G}{x}\left[ \left(2x(Pf')'-P''\right)f''+P''\left(f''+xf'^2\right)    \right]\sim \dot G P (f'')^2 \ ,
\eeq
and
\beq
x\dot P f'^2+2xPf'\dot f'\sim\frac12 {\dot G}\left[\left(2x(Pf')'-P''\right)f'^2 +2Pf'\left(f'''+2xf'f''\right)   \right]\sim -\dot G P (f'')^2 \ .
\eeq
This proves that the derivatives of the two sides of Eq.~\eqref{eq:app1} with respect to $x$ coincide
and therefore completes the proof of Eq.~\eqref{eq:app1}, and of the equivalence of
Eq.~\eqref{eq:full3} and Eq.~\eqref{eq:varGfincont}.
We have therefore derived the set of fullRSB equations
Eqs.~\eqref{eq:full1}-\eqref{eq:full3} in two independent ways.

\section{Derivation within the Gaussian ansatz}
\label{sec:Gaussian}

Although the above results have been derived from an exact evaluation of the replicated entropy
following~\cite{KPZ12,KPUZ13}, they could be equivalently obtained from a suitable Gaussian
ansatz in replica space~\cite{KPZ12}. Here we discuss the appropriate form of this ansatz. This approach is interesting for two 
reasons:
it sheds some light
on the physical interpretation of both the $k$RSB ansatz and the function $P(m_i,h)$ introduced in Sec.~\ref{sec:variational},
and it opens the way to extend the result 
(in an approximate way) to finite dimensions, following the approach of~\cite{PZ10,BJZ11}.

In general, the replicated entropy is a functional of the
single molecule density $\r(\bar x)$, where $\bar x = \{x_1,\cdots,x_m\}$
and $x_a$ are the $d$-dimensional vectors corresponding to the positions of particles in the molecule.
In terms of this object, the replicated entropy for $d\to\io$ is given in~\cite[Eqs.(2)]{KPZ12}:
\beq\label{eq:SGauss}
\begin{split}
\SS[\r(\bar x)] & = \int \de\bar x \r(\bar x) [ 1 - \log \r(\bar x) ] + \frac12 
\int \de\bar x \de\bar y \r(\bar x) \r(\bar y) f(\bar x - \bar y) \  , \\
f(\bar x - \bar y) &= \prod_{a=1}^m e^{-\b v(x_a-y_a)} - 1 \ .
\end{split}
\eeq
In Eq.~\eqref{eq:SGauss}, we introduced a generic interparticle potential $v(r)$ at inverse temperature $\b$.  
In this paper we restrict ourselves to the hard
sphere potential, where
\beq
e^{-\b v(x-y)} = \th(|x - y|-D) \ ,
\eeq
but since the Gaussian derivation allows one to consider a generic potential, it will be useful to write the expressions
for a generic $v(r)$ because this will be surely useful for future applications, e.g. to the soft sphere case following~\cite{BJZ11}.
Note that in finite dimensions, the replicated entropy can be expressed as an infinite sum of diagrams~\cite{PZ10}, but 
for $d\to\io$, and for potentials that have a hard core or a properly scaled soft core, one can truncate the series at the lowest
order~\cite{FP99,PZ10}, hence obtaining Eq.~\eqref{eq:SGauss}.

The Gaussian ansatz consists in making an appropriate Gaussian assumption on the function $\r(\bar x)$, thus introducing a set
of variational parameters that are related to the matrix $\hat\a$ considered above.
In Sec.~\ref{sec:VA} we discuss the proper Gaussian parametrization of the density function, we compute the entropic
term and we show that it has the same form as the one we found before. Finally, in Sec.~\ref{sec:VB} we compute the interaction
term, we discuss the connection with the correlation function, and we show that in the limit $d\to\io$ we recover the results
obtained above.

\subsection{Gaussian parametrization and the entropic term}
\label{sec:VA}

For the entropic term, we just need to recall some results already discussed in~\cite{KPZ12,KPUZ13}, to which we refer for details.
Thanks to translational invariance, 
we can choose a parametrization of $\r(\bar x)$ in terms of vectors $\bar u = \{u_1,\cdots,u_m\}$ such that
$\sum_{a=1}^m u_a =0$, and
in terms of a $m \times m$ symmetric matrix $\hat A$ such that $\sum_{a=1}^m A_{ab} = 0$
for all $b$. 
Calling $\hat A^{m,m}$ the $(m-1) \times (m-1)$ matrix obtained from $\hat A$ by removing the last line and column,
the most general Gaussian form of $\r(\bar u)$ is
\beq\label{Gaussparam}
\r(\bar u) = \frac{\r \, m^{-d}}{(2 \pi)^{(m-1)d/2} \det(\hat A^{m,m})^{d/2}} e^{-\frac12 \sum_{ab}^{1,m-1} (\hat A^{m,m})^{-1}_{ab} u_a \cdot u_b}
\eeq
which is normalized according to $\r = \int \DD \bar u \r(\bar u)$
and $\DD \bar u = m^d \d(\sum_a u_a) du_1 \cdots du_m$.
The parameters $A_{ab}$ give the average of the squared replica displacements
\beq\label{eq:alpha}
\la u_a \cdot u_b \ra = \frac{1}\r \int \DD \bar u \r(\bar u) u_a \cdot u_b = d \, A_{ab} \ , 
\eeq
for $a,b \in [1,m-1]$, while $\la u_a \cdot u_m \ra = -\sum_{b=1}^{m-1} \la u_a \cdot u_b \ra = A_{am}$ and
$\la u_m \cdot u_m \ra = \sum_{ab}^{1,m-1} \la u_a \cdot u_b \ra = A_{mm}$.
Hence, the relative replica displacements $\DE_{ab}$ are given by
\beq
 \la (u_a - u_b)^2 \ra = d \, \DE_{ab} =  d \, (A_{aa} + A_{bb} - 2 A_{ab} ) \ .
\eeq
These corresponds to the physical ``overlaps'', i.e. the mean square displacements between different states.
By comparison with Eq.~\eqref{eq:Dabdef}, we see that $\D_{ab} = d^2 \DE_{ab}/D^2$ and $\a_{ab} = d^2 A_{ab}/D^2$. 

The entropic term is better computed by using Eq.~\eqref{Gaussparam}.
Using the fact that $\de \bar x = \de X \DD \bar u$ with $X$ a vector in the physical volume $V$~\cite{KPZ12}, and using
Eq.~\eqref{eq:alpha} and Eq.~\eqref{Gaussparam}, the entropy per particle is
\beq\begin{split}
\frac{1}N  \int \de \bar x \r(\bar x) [ 1 - \log \r(\bar x) ] &=
\frac1\r \int \DD\bar u \r(\bar u) [1 - \log\r(\bar u)] = 1 - \la \log \r(\bar u) \ra \\ &= 1 -\log\r + d\log m + (m-1)\frac{d}2 \log(2\pi e)
+ \frac{d}2 \log\det \hat A^{m,m} \ ,
\end{split}\eeq
which corresponds indeed to the entropic term in Eq.~\eqref{eq:gauss_r}, with the rescaling $\a_{ab} = d^2 A_{ab}/D^2$.
Apart from this rescaling, all the derivation of section~\ref{sec:entropic} can therefore be repeated within the Gaussian ansatz and
we arrive to exactly the same results for the entropic term, which therefore has the same form both in infinite dimensions
and in finite dimensions within the Gaussian ansatz.

\subsection{$k$RSB Gaussian parametrization and the interaction term}
\label{sec:VB}

\subsubsection{Finite dimensions}

To compute the interaction term, we need to find a simpler parametrization of the $k$RSB form of the Gaussian single molecule density. 
Recall first that in the 1RSB case one has $A_{ab} = A_1 (\d_{ab} - 1/m)$, hence $\DE_{aa}=0$ and for $a\neq b$, 
$\DE_{ab} = \DE_1 = 2 A_1$.
In that case Eq.~\eqref{Gaussparam} can be written as~\cite{MP99,MP09,PZ10}:
\beq\label{Gauss1RSB}
\r(\bar x) = \r \int \de X^1 \prod_{a=1}^m \g^d_{\DE_1/2}(X^1 - x_a) \ ,
\eeq
where the superscript on $X^1$ is here added for the purpose of later generalization, and
\beq
\g^d_A(x) = \frac{e^{- \frac{x^2}{2 A} } }{(2\pi A)^{d/2} }
\eeq
is a $d$-dimensional centered Gaussian. The simplest proof of the equivalence of Eq.~\eqref{Gaussparam} 
and Eq.~\eqref{Gauss1RSB} for 1RSB matrices
is obtained by computing for $a\neq b$
\beq\label{app1RSBG}
\begin{split}
\la (u_a - u_b)^2 \ra &= \la (x_a - x_b)^2 \ra = \frac1{\r V} \int \de \bar x \r(\bar x) (x_a - x_b)^2 \\
&= \frac1V \int \de X^1 \de x_a \de x_b (x_a - x_b)^2 \g^d_{\DE_1/2}(X^1 - x_a) \g^d_{\DE_1/2}(X^1 - x_b) \\& = 
\int \de u_a \de u_b (u_a - u_b)^2 \g^d_{\DE_1/2}(u_a) \g^d_{\DE_1/2}(u_b) = d \, \DE_1 \ .
\end{split}\eeq
This shows that Eqs.~\eqref{Gaussparam} and \eqref{Gauss1RSB} have the same (vanishing) first and (finite) second moments,
and therefore they coincide because a Gaussian is only specified by its first two moments.

Let us consider next a matrix $A_{ab}$ with a 2RSB
structure. In this case we have two physical overlaps, the Edwards-Anderson $\DE_2$ and the inter-state $\DE_1> \DE_2$.
Let us call $B_i = \{ 1 + (i-1) m_1, \cdots,  i m_1  \}$, with $i=1\cdots m/m_1$ the $i$-th block of the 2RSB matrix.
Then we can write
\beq\label{Gauss2RSB}
\r(\bar x) = \r  \int \de X^1  \prod_{i = 1}^{m/m_1} \left[  \int \de X^2_i \g^d_{(\DE_1 - \DE_2)/2}(X^1 - X^2_i)  \left(  \prod_{a \in B_i} 
\gamma^d_{\DE_2/2}(X^2_i - x_a)  \right) \right]
\eeq
It is easy to check, through a computation very similar to Eq.~\eqref{app1RSBG}, that 
$\la (u_a - u_b)^2 \ra = \la (x_a - x_b)^2 \ra$ is equal to $\DE_2$ if $a,b$ belong to the same block $B_i$, and $\DE_1$ otherwise
(and of course zero if $a=b$), hence we conclude that Eq.~\eqref{Gauss2RSB} is identical to Eq.~\eqref{Gaussparam} for a 2RSB matrix.

\begin{figure}[t]
\includegraphics[width=.4\textwidth]{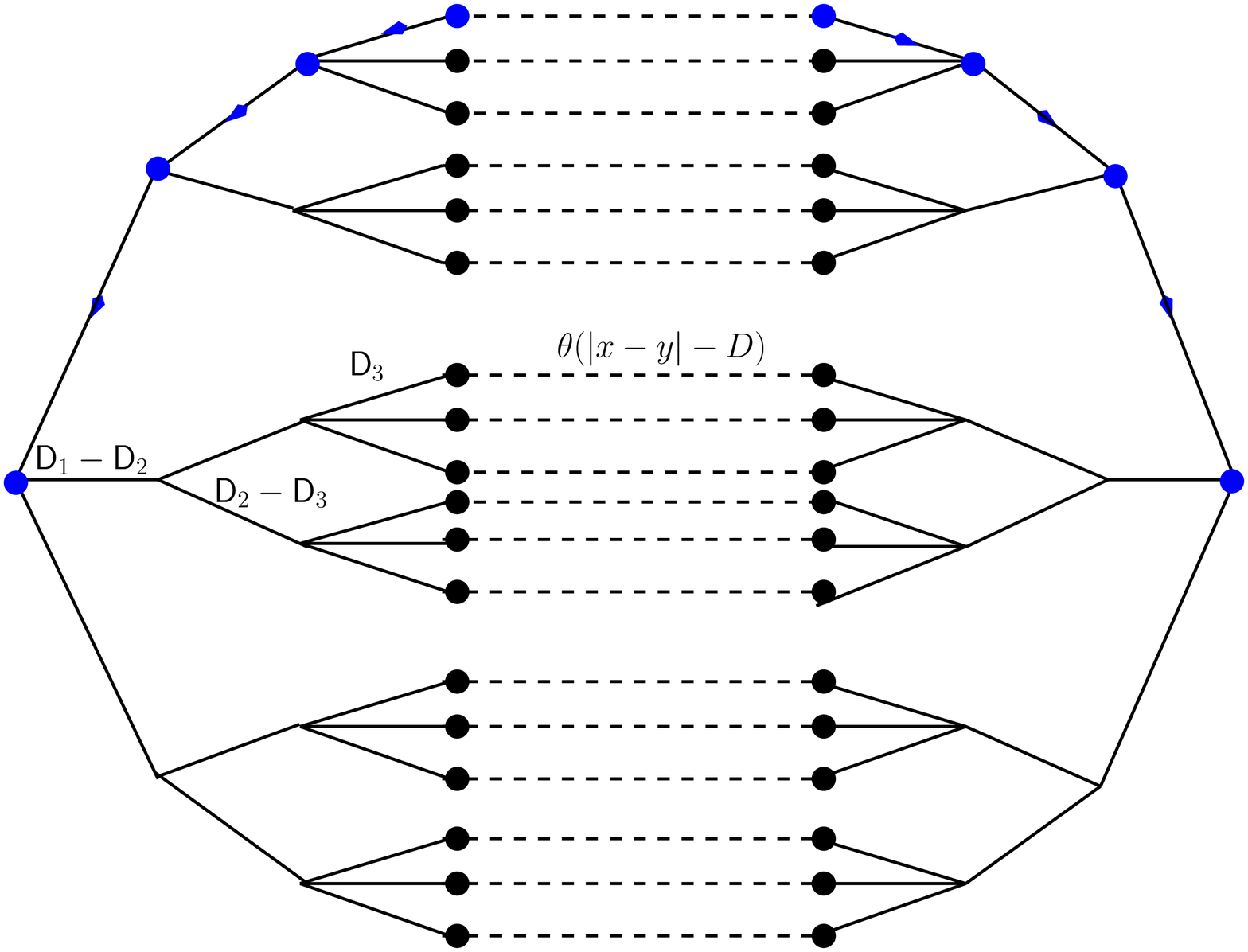}
\hskip50pt
\includegraphics[width=.4\textwidth]{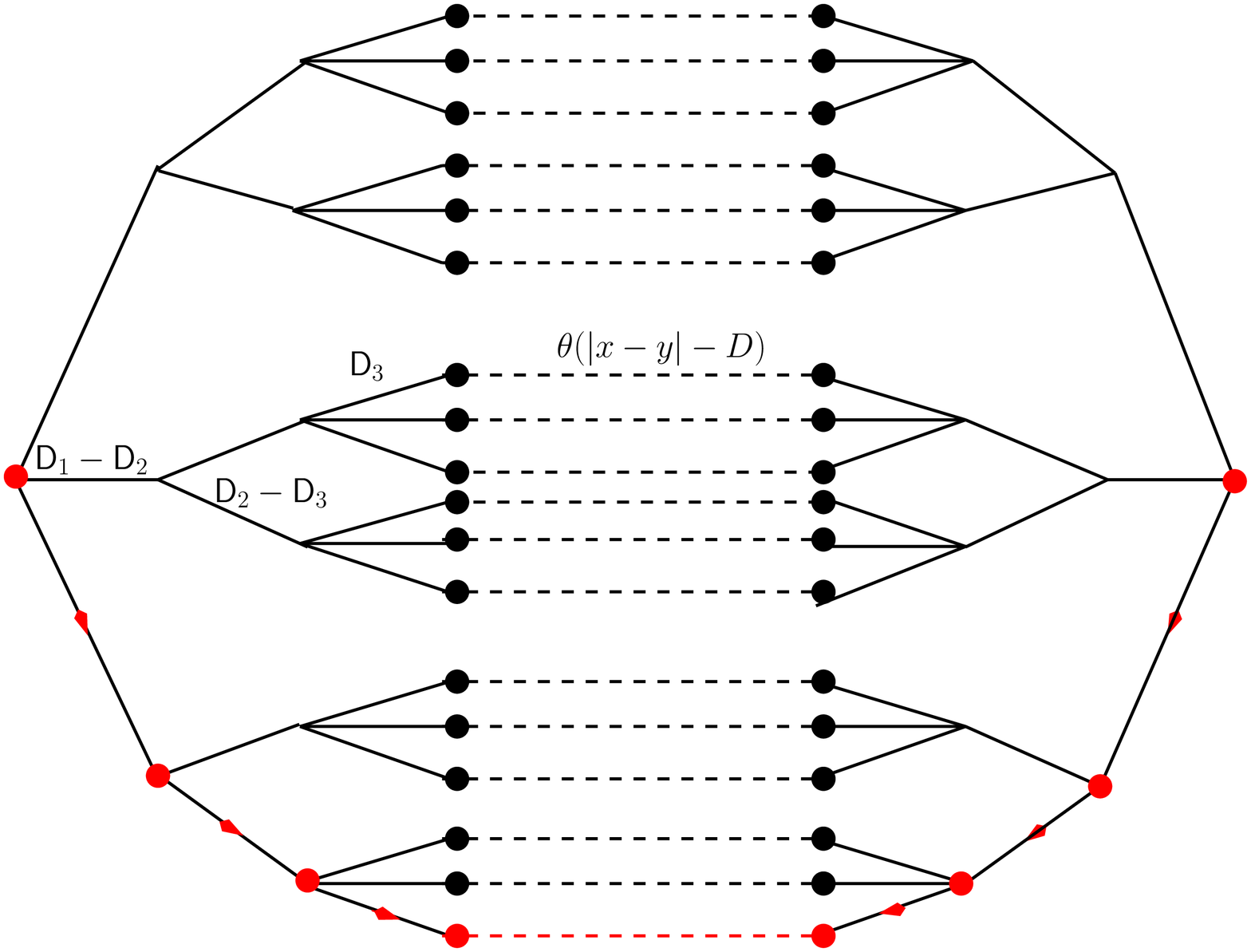}
\caption{Illustration of the relevant part of the interaction term, 
$\int \de \bar x \de \bar y \r(\bar x) \r(\bar y) \prod_{a=1}^m e^{-\b v(x_a - y_a)}$. Each of the two 
$k$RSB molecule densities is represented by a tree graph made by full lines. The hard sphere interaction between the atoms
of the two molecules is represented by dashed lines.
}
\label{fig:Gaussian}
\end{figure}

Generalizing this construction we easily obtain the structure of the $k$RSB Gaussian parametrization. We do not write it explicitly because this would
require introducing a heavy notation for block indices, however it is clear that one should couple each group of replicas in the innermost
blocks to reference points $X^k_i$, then group these reference points in blocks, each block being coupled to a reference $X^{k-1}_i$, and
so on until the most external reference points are coupled to a single $X^1$. This tree structure, and how it enters in the computation
of the interaction term, is illustrated in Fig.~\ref{fig:Gaussian}.

We now show how the computation of the interaction term is performed recursively. We note that the term $-1$ in the Mayer function is obviously
independent of the integration variables and we neglect it for the moment. 
The procedure is illustrated in the left panel of Fig.~\ref{fig:Gaussian}. Consider the innermost blue dots, which represent the coordinates of two atoms.
They interact through the ``bare'' interaction $e^{-\b v(x-y)}$. When we integrate over $x,y$, we generate a ``bare'' interaction between the
reference points $X^k,Y^k$, also marked by blue dots.
This interaction is
\beq\begin{split}
\GG(1,X^k - Y^k) &= \int \de x \de y \gamma^d_{\DE_k/2}(X^k - x)\gamma^d_{\DE_k/2}(Y^k- y) e^{-\b v(x-y)} = \int \de x \gamma^d_{\DE_k}(x) e^{-\b v(X^k - Y^k - x)} \\
&= ( \gamma^d_{\DE_k} \otimes e^{-\b v} ) (X^k - Y^k) \ ,
\end{split}\eeq
where we denoted by $\otimes$ the $d$-dimensional convolution.
Now, each of the $k$-level reference positions are coupled to $m_{k-1}/m_k$ atoms, therefore the total bare interaction between each pair
$X^k,Y^k$ is $\GG(1,X^k - Y^k)^{m_{k-1}/m_k}$. Now we can integrate over these variables to obtain the bare interaction between $(k-1)$-level
reference positions, and so on. It is easy to see that at each step of the iteration we have (we now use a similar notations as that of the previous sections):
\beq\label{eq:recGG}
\GG(m_i, r) =  \gamma^d_{\DE_i  - \DE_{i+1}} \otimes \GG(m_{i+1} ,r)^{m_i/m_{i+1}} \ .
\eeq 
The bare interaction at level $i$ is $\GG(m_i, r)^{m_{i-1}/m_i}$. Iterating this procedure, we finally obtain the bare interaction of the most external 
reference points which is $\GG(m_1,r)^{m/m_1}$. Therefore the interaction term per particle is
\beq\begin{split}
\frac1{2N} \int \de \bar x \de \bar y \r(\bar x) \r(\bar y) \left[ \prod_{a=1}^m e^{-\b v(x_a - y_a)} - 1 \right] &= 
\frac{\r^2}{2N} \int \de X^1 \de Y^1 \left[ \GG(m_1,X^1 - Y^1)^{m/m_1} - 1 \right] \\ &=
\frac{\r}2 \int \de r \left[ \GG(m_1,r)^{m/m_1} - 1 \right] \ .
\end{split}\eeq

Before we show that these expressions give exactly the same result obtained before in the limit $d\to\io$, 
let us consider the total ``dressed''
interaction between two $i$-level reference points. By dressed interaction we mean that we do not only integrate over the points ``down the tree'',
that are at level $j > i$, as we did before; we also integrate over all the other points. For the outermost reference points $X^1,Y^1$, the dressed
interaction is clearly given by
\beq\label{eq:PP1}
\PP(m_1, X^1 - Y^1) = \GG(m_1, X^1-Y^1)^{m/m_1} \ , 
\eeq
as we discussed before. Now, if we consider the points at level $2$, we have:
\beq
\PP(m_2, r) = \GG(m_2, r)^{m_1/m_2} \, \, \g^d_{\DE_1 - \DE_2} \otimes \GG(m_1, r)^{m/m_1-1} =
\GG(m_2, r)^{m_1/m_2} \, \, \g^d_{\DE_1 - \DE_2} \otimes \frac{\PP(m_1,r)}{\GG(m_1,r)}  
\ .
\eeq
The interpretation of this relation should be straightforward by looking at right panel of Fig.~\ref{fig:Gaussian}.
In fact, the first term is the bare interaction that comes from points down the tree, while the second term is the contribution that comes from points
up the tree, in which we divide $\PP$ by $\GG$ because we need to remove the contribution of the branch of the tree under consideration. 
Iterating, at level $i$ we have
\beq\label{eq:PPi}
\PP(m_i, r) = \GG(m_i, r)^{m_{i-1}/m_i} \, \, \g^d_{\DE_{i-1} - \DE_i} \otimes  \frac{\PP(m_{i-1},r)}{\GG(m_{i-1},r)}   \ .
\eeq
At level $k$ this procedure gives the dressed interaction between the innermost reference points $X^k,Y^k$. The last step allows
one to obtain the dressed correlations of two atoms $x,y$ with $r=x-y$. This is the two-body effective potential $\phi_{\rm eff}(r)$ 
of~\cite{PZ10,BJZ11}
and we obtain
\beq\label{eq:phieff1}
e^{-\phi_{\rm eff}(r)} = e^{-\b v(r)} \, \, \g^d_{\DE_k} \otimes \frac{\PP(m_k,r)}{\GG(m_k,r)} \ .
\eeq
One can easily check that in the 1RSB case this result gives back the effective potential used in~\cite{PZ10,BJZ11}.
This result is particularly interesting for two reasons: on the one hand, in the limit $d\to\io$ (or, in finite $d$, 
under the low-temperature approximation of Ref.~\cite{BJZ11}), the effective potential coincides with the pair correlation function
of the glass, $g_{\rm g}(r)$. On the other hand, in finite dimension one could plug this effective potential into some liquid theory integral
equations to compute an approximation of the replicated entropy, following the strategy of~\cite{PZ10}. 
We do not discuss this second issue here, leaving it for future work, and we keep our focus instead on the limit $d\to\io$.

\subsubsection{The limit $d\to\io$}

To take the $d\to\io$ limit, we assume that the potential has the form $e^{-\b v(r)} = e^{-\wh v[ d ( |r| - D )/D ]}$, and $e^{-\wh v(h)}$ is a finite
function\footnote{
Remember that in general we use a wide hat to denote quantities that are properly rescaled to be finite when $d\to\io$.
}
when $d\to\io$. 
The hard sphere potential, the only one that we consider in this paper, 
has this form with $e^{-\wh v(h)} = \th(h)$. Note, however, 
that the interaction potential enters only in the initial condition for the function $\GG(1,r)$.

We now show briefly that in the limit $d\to\io$ the equations we just derived give back Eq.~\eqref{eq:sDkRSB}.
First of all, we note that all the interaction functions $\GG(m_i,r)$ and $\PP(m_i,r)$ are rotationally invariant
and therefore depend only on $|r|$. Moreover, all these functions tend to 1 for $|r|\to\io$ and they decrease fast to zero
when $|r| \ll D$. Actually, as we show below, for $d\to\io$ the growth of these functions from 0 to 1 happens on a scale $\sim 1/d$
around $|r| =D$. We therefore define $|r| = D(1+h/d)$ and we consider $\GG(m_i,h)$ and $\PP(m_i,h)$ to be functions of $h$.
We will see that with this convention, $h$ is the same variable that enters in the equations of the previous sections.

Next, to obtain a non-trivial limit we scale the parameters introducing $\wh \D_i =  d^2 \DE_i / D^2$. Then,
following exactly the strategy of~\cite[Appendix C.2.d]{PZ10} through the use of bipolar coordinates, 
one can see that in the limit $d\to\io$, with the scaling
$r = D(1 + h/d)$ and $u = D(1+z/d)$, we have
\beq\label{eq:GG1def}
\begin{split}
\GG(1,h) &= \g^d_{ D^2 \wh\D_k / d^2} \otimes e^{-\wh v[ d ( |r| - D )/D ]} = \int_0^\io \de u \left(\frac{u}{r}\right)^{\frac{d-1}2} \frac{e^{-\frac{(r-u)^2}{2 \DE_k}}}{\sqrt{2\pi \DE_k}}
e^{- \frac{\DE_k}{8 d^2 r u }} e^{-\wh v[ d ( u - D )/D ]} \\
&\sim \int_{-\io}^\io \de z \, e^{(z-h)/2} \frac{e^{-\frac{(z-h)^2}{2 \wh\D_k}}}{\sqrt{2\pi \wh\D_k}}
e^{- \frac{\wh\D_k}{8 }} e^{-\wh v(z)}  =
\int_{-\io}^\io \de z \, \frac{e^{-\frac{(z-h -\wh\D_k/2)^2}{2 \wh\D_k}}}{\sqrt{2\pi \wh\D_k}}
 e^{-\wh v(z)} 
\\
&= ( \g_{\wh\D_k} \star e^{-\wh v}  ) (h + \wh\D_k/2) = g(1,h+\wh\D_k/2) \ .
\end{split}\eeq
With a similar reasoning, Eq.~\eqref{eq:recGG} becomes
\beq\begin{split}
\GG(m_i, h) &=  \gamma^d_{D_i  - D_{i+1}} \otimes \GG(m_{i+1} ,h)^{m_i/m_{i+1}} =
\int_{-\io}^\io \de z \, \g_{\wh\D_i - \wh\D_{i+1}}[h -z + ( \wh\D_i - \wh\D_{i+1})/2] \,
  \GG(m_{i+1} ,z)^{m_i/m_{i+1}} 
  \\ &=
\int_{-\io}^\io \de z \, \g_{\wh\D_i - \wh\D_{i+1}}(h +\wh\D_i/2  -z ) \,
  \GG(m_{i+1} ,z- \wh\D_{i+1}/2)^{m_i/m_{i+1}} \ .
\end{split}\eeq
This equation coincides with Eq.~\eqref{eq:gi} if one makes for all $i$ the identification 
\beq
\GG(m_i, h) = g(m_i, h +\wh\D_i/2) \ ,
\eeq
and recalling that for hard spheres $e^{-\wh v(h)} = \th(h)$.
Hence the interaction term is (using that the solid angle is $d \, V_d$ with $V_d$ the volume of a unit sphere)
\beq\begin{split}
&\frac{\r}2 \int \de r \left[ \GG(m_1,r)^{m/m_1} - 1 \right] = \frac{\r d\, V_d}{2} \int_0^\io \de r \, r^{d-1} \left[ g(m_1,h +\wh\D_1/2)^{m/m_1} - 1 \right] \\
 &= \frac{2^d \f}2  \int_{-\io}^\io \de h \, e^h \left[ g(m_1,h +\wh\D_1/2)^{m/m_1} - 1 \right] = 
 \frac{d \, \wh\f}2 e^{-\wh\D_1/2} \int_{-\io}^\io \de h \, e^h \left[ g(m_1,h)^{m/m_1} - 1 \right] \ ,
\end{split}\eeq
which provides the exact result of Eq.~\eqref{eq:sDkRSB}.

Next, we analyze the behavior of $\PP$ for $d\to\io$. The initial condition from Eqs.~\eqref{eq:PP1} and \eqref{eq:recP_b} is
\beq
\PP(m_1,h) = g(m_1, h  +\wh\D_1/2  )^{m/m_1} = e^{-h -\wh\D_1/2} P(m_1,h  +\wh\D_1/2) \ .
\eeq
With a similar reasoning as before, one can show that identifying 
\beq\label{eq:PPvsP}
\PP(m_i,h) = e^{-h - \wh\D_1/2} P(m_i,h  +\wh\D_i/2)
\eeq
for all $i$,
Eq.~\eqref{eq:PPi} becomes in fact identical to Eq.~\eqref{eq:recP} in the limit $d\to\io$.
We obtain from this identification a deep physical interpretation of the function $P(m_i,h)$, 
which turns out to be related to the dressed interaction of the reference positions at level $i$, $\PP(m_i,h)$, 
by Eq.~\eqref{eq:PPvsP}.

Finally, we can write Eq.~\eqref{eq:phieff1} for $d\to\io$. We get
\beq\label{eq:phieff_dinf}
e^{-\phi_{\rm eff}(h)} = e^{-\wh v(h)}  \int_{-\io}^\io \de z \, e^{z-h} \, \g_{\wh\D_k}(h-z) \, \frac{e^{-z-\wh\D_1/2} P(m_k,z)}{g(m_k,z)} \ .
\eeq
This is a very important result because it allows one to obtain structural information about the pair correlation from the knowledge
of the functions $P(m_i,h)$ and $g(m_i,h)$.

\clearpage

\part{Extraction of the results from the equations}

\section{Summary of the equations, and a numerically convenient formulation}

In the following sections, we will investigate the phase diagram that one obtains from the study of the $k$RSB equations, 
and we derive the scaling properties at large pressure. The way in which one has to extract physical information from the
replicated entropy has been explained in many reviews~\cite{MPV87,Mo95,MP09,PZ10}. 
Although we will give additional details along the way, we assume that the reader is familiar with this kind of computations.

Before extracting the physics, let us summarize here the $k$RSB equations~\eqref{eq:sDkRSB}
together with the variational equations~\eqref{eq:recP_b}-\eqref{eq:recP} and~\eqref{eq:varGfin}. These are
\beq
\begin{split}
\SS_{k{\rm RSB}} &=   \sum_{i=1}^k \left( \frac{m}{m_{i}} - \frac{m}{m_{i-1}}  \right) \log ( \wh G_i /m) - \wh\f \, e^{-\wh\D_1/2}\int_{-\infty}^\infty \de h\, e^{h} 
  \left\{ 1- e^{m  f(m_1,h)} \right\}  \ , \\
\wh\D_i &= \frac{\wh G_i}{m_i} + \sum_{j=i+1}^k \left(\frac1{m_j} - \frac1{m_{j-1}} \right) \wh G_j \ , \\
f(1,h) & = \log \g_{\wh\D_k} \star \th(h) = \log \Th\left(  \frac{h}{\sqrt{2\wh \D_k}}  \right) \ ,  \\
f(m_i,h) &= \frac1{m_i} \log \left[ \g_{\wh\D_i - \wh\D_{i+1}} \star e^{m_i f(m_{i+1},h)}  \right] \ , \hskip20pt i = 1 \cdots k-1 \ , \\
 P(m_1,h) &= e^h \, e^{m f(m_1,h) } \ , \\
P(m_i,h) &= \int \de z P(m_{i-1},z) e^{-m_{i-1} f(m_{i-1},z)} \g_{\wh\D_{i-1}-\wh\D_{i}}(h-z) e^{m_{i-1}  f(m_i,h)} 
\hskip20pt i = 2 , \cdots , k
\ , \\
 \frac{1}{\wh G_i}  &= \frac{ \wh\f}2 e^{-\wh\D_1 / 2} \int_{-\infty}^\infty \de h \, \left\{
  m_{i-1} P(m_i,h) f'(m_i,h)^2 -
\sum_{j=1}^{i-1} ( m_j - m_{j-1}) P(m_j,h) f'(m_j,h)^2
\right\} \ .
\end{split}
\eeq

\subsection{Scaled variables and the jamming limit}

Because (as we will see below) we are mostly going to work at small $m$, and the jamming limit corresponds to $m\to 0$,
it is convenient to write the equations in scaled variables that remain finite when $m\to 0$.
These are $y_i = m_i/m$ (keeping in mind that $y_0=1$ and that $y_k = 1/m$ diverges with $m$ and will play the role usually played by temperature), 
$\wh f(y_i,h) = m f(m_i,h)$, $\wh\g_i = \wh G_i /m$, from which it follows that $\wh\D_k = m \wh \g_k$ while all 
the other $\wh\D_i$ remain finite for $m\to 0$.
It will also be convenient for numerical reasons to introduce $\wh P(y_i,h) = e^{-\wh\D_1/2} e^{-h} P(m_i,h)$.
Note also that $\wh\D_i - \wh\D_{i+1} = (\wh\g_i - \wh\g_{i+1})/y_i$.
In terms of these variables, and introducing auxiliary variables 
$\wh\k_i$ (not to be confused with the exponent $\k$ discussed above)
we have:
\begin{align*}
\SS_{k{\rm RSB}} &=   \sum_{i=1}^k \left( \frac{1}{y_{i}} - \frac{1}{y_{i-1}}  \right) \log \wh \g_i  - \wh\f \, e^{-\wh\D_1/2}\int_{-\infty}^\infty \de h\, e^{h} 
  \left\{ 1- e^{  \wh f(y_1,h)} \right\}  \ , \\
\wh\D_i &= \frac{\wh \g_i}{y_i} + \sum_{j=i+1}^k \left(\frac1{y_j} - \frac1{y_{j-1}} \right) \wh \g_j \ , \\
\wh f(1/m,h) &= m \log \Th\left(  \frac{h}{\sqrt{2 m \wh \g_k}}  \right) \ ,  \\
\wh f(y_i,h) &= \frac1{y_i} \log \left[ \g_{(\wh\g_i - \wh\g_{i+1})/y_i} \star e^{y_i \wh f(y_{i+1},h)}  \right] \ , \hskip20pt i = 1 \cdots k-1 \ , \\
 \wh P(y_1,h) &= e^{-\wh\D_1/2} e^{\wh f(y_1,h) } \ , \\
\wh P(y_i,h) &= \int \de z \, e^{z-h} \, \wh P(y_{i-1},z) e^{-y_{i-1} \wh f(y_{i-1},z)} \g_{(\wh\g_{i-1} - \wh\g_{i})/y_{i-1}}(h-z) e^{y_{i-1} \wh f(y_i,h)} 
\hskip20pt i = 2 , \cdots , k
\ , \\
\wh\kappa_i &= \frac{ \wh\f}2  \int_{-\infty}^\infty \de h \, e^h \,  \wh P(y_i,h) \wh f'(y_i,h)^2 \ , \\
 \frac{1}{\wh \g_i}  &=  y_{i-1} \wh\kappa_i -
\sum_{j=1}^{i-1} ( y_j - y_{j-1}) \wh\kappa_j  \ .
\end{align*}
To solve numerically these equations it is convenient to have some control on the asymptotic behavior of
the functions when $h\to\pm\io$. We start by the function $\wh f(y_i,h)$. From the initial condition we
see that $\wh f(1/m,h\to\io) = 0$ and $\wh f(1/m,h\to -\io) \sim -h^2/(2 \wh \g_k)$.
Inserting these asymptotes in the evolution equation for $\wh f(y_i,h)$, one can show that
\beq\label{eq:fasymptotes}
\begin{split}
\wh f(y_i,h\to -\io) &\sim - h^2/(2\wh\g_i) \ , \\
\wh f(y_i,h\to\io) &= 0 \ . \\
\end{split}\eeq
From this, we obtain that $\wh P(y_1,h\to -\io) = 0$ while $\wh P(y_1,h\to \io) = e^{-\wh\D_1/2}$.
As a consequence, from the recurrence equation for
$\wh P(y_i,h)$ one can show that
\beq\label{eq:Pasymptotes}
\begin{split}
\wh P(y_i,h\to -\io) & = 0 \ , \\
\wh P(y_i,h\to \io) & = e^{-\wh\D_i/2} \ . 
\end{split}\eeq

To obtain a simpler asymptotic behavior it is convenient to make a change of variable:
\beq
\wh f(y_i,h) = -\frac{h^2 \th(-h)}{2\wh\g_i} + \wh j(y_i,h) \ ,
\eeq
in such a way that the leading asymptotic term of $\wh f(y_i,h)$ in Eq.~\eqref{eq:fasymptotes} is subtracted from
$\wh j(y_i,h)$.
Then we have
\beq\label{eq:Sscaledfinal}
\begin{split}
\SS_{k{\rm RSB}} &=   \sum_{i=1}^k \left( \frac{1}{y_{i}} - \frac{1}{y_{i-1}}  \right) \log \wh \g_i  - \wh\f \, e^{-\wh\D_1/2}\int_{-\infty}^\infty \de h\, e^{h} 
  \left\{ 1- e^{ -\frac{h^2\th(-h)}{2\wh\g_1} + \wh j(y_1,h) } \right\}  \ , \\
\wh\D_i &= \frac{\wh \g_i}{y_i} + \sum_{j=i+1}^k \left(\frac1{y_j} - \frac1{y_{j-1}} \right) \wh \g_j \ , \\
\wh j(1/m,h) & = m \log \Th\left(  \frac{h}{\sqrt{2 m \wh \g_k}}  \right) + \frac{h^2 \th(-h)}{2\wh\g_k}  \ ,  \\
\wh j(y_i,h) &= \frac1{y_i} \log \left[ \int_{-\io}^\io \de z\, K_{\wh\g_i, \wh\g_{i+1}, y_i}(h,z) \,e^{y_i \wh j(y_{i+1},z)}  \right] \ , \hskip20pt i = 1 \cdots k-1 \ , \\
 \wh P(y_1,h) &= \,e^{-\wh\D_1/2  -\frac{h^2\th(-h)}{2\wh\g_1}  + \wh j(y_1,h) } \ , \\
\wh P(y_i,h) &= \int \de z \, K_{\wh\g_{i-1}, \wh\g_i, y_{i-1}}(z,h) \,  \wh P(y_{i-1},z) \, e^{z-h} \, e^{ -y_{i-1} \wh j(y_{i-1},z)+y_{i-1} \wh j(y_i,h)} 
\hskip20pt i = 2 , \cdots , k
\ , \\
\wh\kappa_i & = \frac{ \wh\f}2  \int_{-\infty}^\infty \de h \, e^h \,  \wh P(y_i,h) \left(  -\frac{h \th(-h)}{\wh\g_i}  +   \wh j'(y_i,h) \right)^2 \ , \\
 \frac{1}{\wh \g_i}  &=  y_{i-1} \wh\kappa_i -
\sum_{j=1}^{i-1} ( y_j - y_{j-1}) \wh\kappa_j  \ .
\end{split}
\eeq
where
\beq\label{eq:Kscaledfinal}
\begin{split}
K_{\wh\g, \wh\g', y}(h,z) &= \frac{\exp\left[ -\frac{y}2 \left(  \frac{(z-h)^2}{\wh\g -\wh\g'} - \frac{h^2 \th(-h)}{\wh\g} + \frac{z^2 \th(-z)}{\wh\g'}
  \right) \right]}{\sqrt{2\pi(\wh\g -\wh\g')/y}} 
\end{split}\eeq
is {\it not} a symmetric function of $h$ and $z$, nor a function of $h-z$. However, the advantage of this formulation is that the kernel $K$ is
an almost Gaussian function which is well behaved, and all the other functions that appear in the integrals are smooth. This allows for a stable
numerical evaluation of the integrals.

Note that Eqs.~\eqref{eq:Sscaledfinal} admit a perfectly smooth $m\to 0$ limit. First of all one has to set $1/y_k = m$, $\wh\D_k=0$.
Then, using the large $\l$ development of $\Th(-\l/\sqrt{2})$, one can easily show that
\beq\label{limittheta}
\lim_{\mu\to 0} \Th\left(\frac{z}{\sqrt{\mu}} \right)^\mu = e^{-z^2 \th(-z) } \ ,
\eeq
and therefore $\wh j(y_k,h)=0$. All the other equations remain identical to the case $m>0$.

\subsection{The continuum limit}

It will be convenient for later purposes to write explicitly the continuum limit of the equations in terms of scaled variables. These
are
\beq\label{eq:Sscaledfinal_cont}
\begin{split}
\SS_{\io{\rm RSB}} & = 
-   \int_1^{1/m}\frac{\de y}{y^2}\log\left[ \g(y) \right] 
-   \wh \f \,
e^{-\D(1)/2}\int_{-\infty}^\infty \de h\, e^{h} 
  [1- e^{-\frac{h^2\th(-h)}{2\g(1)} + \wh j(1,h)}] \ , \\
  \D(y) &= \frac{\g(y)}y - \int_y^{1/m} \frac{\de z}{z^2} \g(z) \ , \hskip20pt \Leftrightarrow \hskip20pt
\g(y) =  y\D(y) + \int_y^{1/m} \de z  \D(z) \ , \\
\wh j(1/m,h) & = m \log \Th\left(  \frac{h}{\sqrt{2 m  \g(1/m)}}  \right) + \frac{h^2 \th(-h)}{2 \g(1/m) }  \ ,  \\
  \frac{\partial \wh j(y,h)}{\partial y} &= 
    \frac 12 \frac{\dot\g(y)}y \left[ - \frac{\th(-h)}{\g(y)} +  \frac{\partial^2 \wh j(y, h)}{\partial h^2}
  - 2y   \frac{h \th(-h)}{\g(y)}  \frac{\partial \wh j(y, h)}{\partial h}
  +y   \left(  \frac{\partial \wh j(y, h)}{\partial h}\right)^2    \right] \ , \\
 \wh P(1,h) &= \,e^{-\D(1)/2  -\frac{h^2\th(-h)}{2\g(1)}  + \wh j(1,h) } \ , \\
 \frac{\partial \wh P(y,h)}{\partial y} &=-\frac12 \frac{\dot\g(y)}y e^{-h} \left\{ 
\frac{\partial^2 [e^h \wh P(y, h)]}{\partial h^2} - 2y \frac{\partial}{\partial h} \left[ 
e^h \wh P(y,h)  \left(   - \frac{h \th(-h)}{\g(y)}  + \frac{\partial \wh j(y, h)}{\partial h}\right)
\right]
\right\} \ , \\
\kappa(y) & = \frac{ \wh\f}2  \int_{-\infty}^\infty \de h \, e^h \,  \wh P(y,h) \left(  -\frac{h \th(-h)}{\g(y)}  +   \wh j'(y,h) \right)^2 \ , \\
 \frac{1}{ \g(y) }  &=  y \kappa(y) -
\int_1^{y} \de z \kappa(z)
\ .
\end{split}
\eeq
Having formulated the $k$RSB equations in a convenient way, we now proceed to extract the physical results from them. 
However, because the numerical solution of these equations is not trivial, we first investigate a certain number of asymptotic limits in which
analytical results can be obtained.

\section{Perturbative 2RSB solution around the Gardner line}
\label{sec:perturbative}

The Gardner transition~\cite{Ga85} separates the region where the 1RSB solution is stable from the one where it is unstable.
In our problem, the 1RSB solution is stable in a certain region of the phase diagram, and it becomes unstable on a line 
that has been computed and characterized in the previous paper of this series~\cite{KPUZ13}.
The aim of this section, following the analysis of~\cite{Ri13}, is to perform a perturbative computation around the Gardner
line, in the region where the 1RSB solution is unstable, and discuss the existence of a 2RSB (or fullRSB) solution.
In Sec.~\ref{sec:VIIA} we perform the perturbative computation and show that a 2RSB solution only exists in a certain region
of the Gardner line; in Sec.~\ref{sec:VIIB} we discuss the behavior of the perturbative solution at large densities and pressures.

We start our analysis of the $k$RSB equations by examining what happens around the Gardner transition line that has been computed in~\cite{KPUZ13}.
We will follow closely the analysis of Refs.~\cite{MR03,Ri13}. Fig.~\ref{fig:diag_m} reports a schematic phase diagram in the $\wh\f$, $m$ plane. 
The reader should
keep in mind that $m \propto 1/p$ where $p$ is the reduced pressure~\cite{KPUZ13}. 
Let us summarize briefly the phase diagram. A 1RSB solution exists above the 1RSB dynamical line, i.e. for $\wh\f > \wh\f_{\rm d}^{\rm 1RSB}(m)$
or equivalently $m > m_{\rm d}^{\rm 1RSB}(\wh\f)$.
Above the Gardner line, i.e. for $m \geq m_{\rm G}(\wh\f)$ or $\wh\f > \wh\f_{\rm G}(m)$,
this 1RSB solution is stable~\cite{KPUZ13} and one can extract the physical results from it~\cite{PZ10}. 
Below this line, the 1RSB solution is unstable and we look for a $k$RSB solution with $k>1$. Since the instability is due to a vanishing mode~\cite{KPUZ13}, we
expect that the 1RSB solution will transform continuously in a $k$RSB solution with $k>1$, and we therefore start by looking at the new solution by doing perturbation
theory around the 1RSB solution in the vicinity of the Gardner line.
Based on the analogy with spin glass models~\cite{Ri13}, we expect the new solution to be a fullRSB one, but in perturbation theory it is enough to consider 
a 2RSB solution because the breaking is small and the two perturbative computations give identical results~\cite{CFLPR11}.

\subsection{Development around the 1RSB solution}
\label{sec:VIIA}

\begin{figure}[t]
\includegraphics[width=0.65\textwidth]{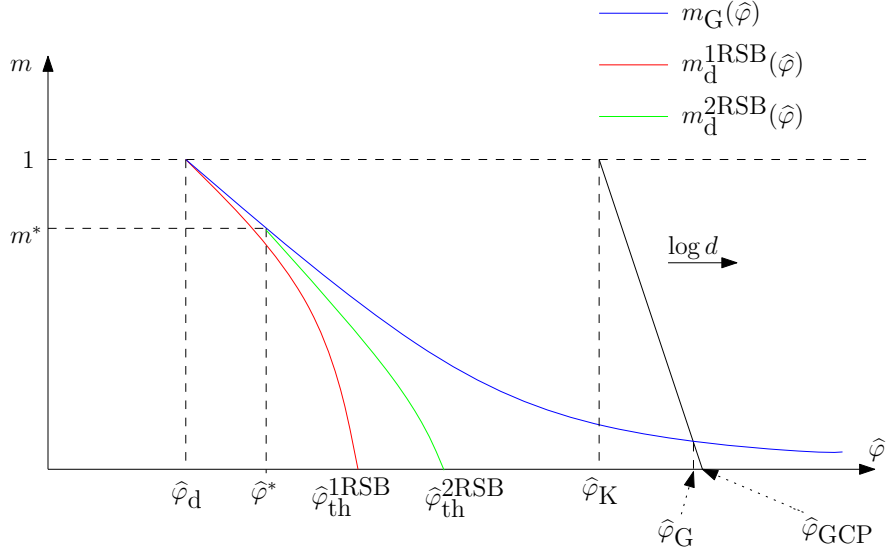}
\caption{Schematic phase diagram in the $(m,\wh\varphi)$ plane.}
\label{fig:diag_m}
\end{figure}

We have found in~\cite{KPUZ13} that the instability of the 1RSB solution is due to the vanishing of the ``replicon'' eigenvalue $\l_R(m)$. It is therefore
natural to consider a perturbation of the 1RSB matrix that is proportional to the subspace associated with the vanishing mode. We need to consider the cubic
expansion in this direction and check if the cubic terms can stabilize the negative quadratic part.
Hence we consider a perturbative 2RSB matrix of the form:
\beq\label{eq:perta}
\begin{split}
\alpha_{ab} &= \a_{ab}^{\rm 1RSB}+\delta \a_{ab} \ , \\
\delta \a_{ab} &= \delta \wh\alpha_1 (1-\delta_{ab})\left( \frac{m-m_1}{1-m_1}  I_{ab}^{m_1}   + I_{ab}^{m} - I_{ab}^{m_1}   \right) \equiv \delta \wh\alpha_1 r_{ab}
\end{split}\eeq
(remember that the matrices $I^{m_i}_{ab}$ are defined in Sec.~\ref{sec:interactionterm}).
The matrix $\delta \hat\a$ belongs to the replicon subspace~\cite{TDP02}. 
It has been shown in~\cite{TDP02} that the cubic terms of the expansion of the 2RSB entropy around the 1RSB solution are eight, 
but if the perturbation matrix $\delta \hat\a$ is in the replicon subspace, only the two terms proportional to the coefficients $w_1$ and $w_2$ 
that have been analyzed in~\cite{KPUZ13}
survive. In practice we have
\beq
\begin{split}
s[\hat\alpha^{\rm 1RSB}+\delta\hat\a]&=s[\hat\alpha^{\rm 1RSB}]+\frac{1}{2}\lambda_R(m)\sum_{a\neq b}\delta \a_{ab}^2+
\frac{1}{6}\left[w_1(m)\Tr \delta \hat\a^3+w_2(m)\sum_{a\neq b}\delta \a_{ab}^3\right] \\
&=s[\hat\alpha^{\rm 1RSB}]+\frac{1}{2}\wh \lambda_R(m,m_1)\delta\wh\a_1^2+\frac{1}{6}W[w_1(m),w_2(m),m,m_1]\delta\wh\a_1^3\:.
\end{split}\label{1rsb+espilon}
\eeq
where $\l_R(m)$ is the replicon eigenvalue that has been computed in~\cite{KPUZ13}.
The relation between $\lambda_R(m)$ and $\wh\lambda_R(m,m_1)$ can be exploited to obtain the replicon eigenvalue from the 2RSB entropy and it gives
\beq
\wh\l_R(m,m_1)=\frac{m(1-m)(m-m_1)}{1-m_1}\l_R(m)  = \left. \frac{\de^2 s[\hat\alpha^{\rm 1RSB}+\delta\hat\a]}{\de \d\wh\a_1^2} \right|_{\d\wh\a_1=0} \ .
\eeq
It expresses the fact that the replicon eigenvalue can be obtained from the second derivative with respect to $\delta\wh\a_1$ of the replicated entropy computed 
on a matrix of the form~\eqref{eq:perta}.
We want now to express $W$ explicitly in terms of $w_1$ and $w_2$ that we have already computed in~\cite{KPUZ13}. 
Note that $w_1(m)$ and $w_2(m)$ depend only on $m$ and not on $m_1$ as a consequence of the fact that they are computed on a 1RSB matrix
that does not depend on $m_1$.
To do this one can use the following relations between matrices $\hat I^{m_i}$:
\beq
(\hat I^{m_i})^2=m_i \hat I^{m_i} \ , \hskip20pt  \hat I^m \hat I^{m_1}= \hat I^{m_1} \hat I^m = m_1 \hat I^m \ ,
\eeq
to obtain that the matrix $\hat r$ defined in~\eqref{eq:perta} satisfies
\beq
\Tr \, \hat r^3 = \frac{m(m_1-m)(m-1)}{(m_1-1)^2}\left( m m_1+m_1-2m \right) \ , \hskip20pt  \sum_{a\neq b}r_{ab}^3=\frac{m(m_1-m)(m-1)}{(m_1-1)^2}(1+m-2m_1) \ ,
\eeq
and therefore
\beq
W[w_1,w_2,m, m_1]=\frac{m(m_1-m)(m-1)}{(m_1-1)^2}\left[ w_1(m)\left( m m_1+m_1-2m \right)+w_2(m)(1+m-2m_1)  \right] \ .
\eeq
To obtain the perturbative 2RSB solution in the $(m,\wh\varphi)$ plane we 
search for a non trivial stationary point solution for the expression (\ref{1rsb+espilon}). 
The trivial 1RSB solution $\delta\wh\a_1 = 0$ can always be found but it is unstable below the Gardner line where $\l_R(m)< 0$.
To find the non-trivial solution we first optimize over $\delta\wh\a_1$ 
and then we optimize over the breaking point $m_1$. The saddle point equation for $\delta\wh\a_1$ gives
\beq
\delta\wh\a_1=-\frac{2\wh\lambda_R}{W} \ .
\eeq
The entropy as a function of $m_1$ is obtained by plugging the above expression in (\ref{1rsb+espilon}) and it gives
\beq\
   s[\hat\alpha^{\rm 1RSB}+\delta\hat\a] = s[\hat\alpha^{\rm 1RSB}]  -\frac 13 \frac{\wh\l_R^3(m, m_1)}{W^2(m,m_1)}
\eeq
Now we should search for the extremum in $m_1$. We obtain that the breaking point is 
\beq
m_1=\frac{w_2(m)}{w_1(m)} \equiv \lambda(m) \ ,
\eeq
where $\lambda(m)$ is the Mode-Coupling theory exponent parameter that has been discussed in~\cite{KPUZ13}.
However, as it is usual in replica computations, the saddle point solution for the breaking point should satisfy
$m<m_1<1$. This implies that a 2RSB solution exists only if 
\beq
m \leq \frac{w_2(m)}{w_1(m)}\equiv \l(m) \ .
\label{frsbcondition}
\eeq
Note that, because we are perturbing around the 1RSB solution close to the Gardner line where the replicon mode vanishes,
all the quantities $\wh\lambda_R(m)$, $w_1(m)$ and $w_2(m)$ are computed on the Gardner line. 

We now follow the discussion of~\cite{Ri13}. We know that at $m=1$ the Gardner line corresponds to the dynamical transition point
and at that point $\l(m=1) = \l_{\rm MCT} = 0.70698 < 1=m$~\cite{KPUZ13}. Hence, by continuity, close to $m=1$ we have $\l(m) < m$ and
there is no non-trivial 2RSB solution. We conclude, following~\cite{Ri13}, that in this region only the 1RSB solution exists down to the Gardner
line, and there is no other solution below the Gardner line.
However, the condition (\ref{frsbcondition}) tells us that there might exist a point in the $(m,\wh\varphi)$ plane where 
\beq
m^*=\l(m^*) \ , \label{mstar}
\eeq
so that below this point a perturbative 2RSB solution can be found. 
The point $m^*$ can be computed using the expressions for $w_1(m)$ and $w_2(m)$ that we computed in~\cite{KPUZ13}. 
The numerical solution of equation (\ref{mstar}) gives
\beq\label{eq:mstarvalue}
m^*\simeq 0.414\implies \wh\varphi^* \simeq 5.84    \:.
\eeq 
We therefore obtain the schematic phase diagram represented in Fig.~\ref{fig:diag_m}, which is strongly similar to 
the one found in~\cite{Ri13} for the Ising $p$-spin glass model.
For $m>m^*$ or $\wh\f < \wh\f^*$, no solution exist below the Gardner line, which therefore delimits the region where the only non-trivial 1RSB solution exists.
Instead, for $m<m^*$ or $\wh\f > \wh\f^*$ a non-trivial $k$RSB solution with $k>1$ exists below the Gardner line. It is reasonable to expect that this solution will exist in a finite
region below the Gardner line. The region of existence of the 2RSB solution should be delimited by a dynamical line 
$\wh\f_{\rm d}^{\rm 2RSB}(m)$ or $m_{\rm d}^{\rm 2RSB}(\wh\f)$, shifted with respect
to the 1RSB dynamical line (see Fig.~\ref{fig:diag_m} for a schematic drawing of this line). We will show in Sec.~\ref{sec:2RSBnum}
how this line can be defined. Note however that the instability
of the replicon mode suggests that the 2RSB solution is also unstable towards 3RSB and so on, until the correct fullRSB solution is found.

\subsection{Asymptotic expression for $\l(m)$ at large densities on the Gardner line}
\label{sec:VIIB}

In this section we compute the asymptotic behavior of $\l(m)$ for $m\to 0$ or $\wh\f\to\io$ on the Gardner line. 
The importance of this computation is twofold. First, we want to check that the condition $m<\l(m)$ holds for all 
$m<m^*$ on the Gardner line, which implies that a non-trivial 2RSB solution
exists for all $\wh\f > \wh\f^*$. Second, we have shown that the breaking point of the 2RSB solution is $m_1 = \l(m)$ on the Gardner line.
Actually this is true also if we perform a perturbative fullRSB calculation around the instability line~\cite{CFLPR11}.
It follows that the asymptotic computation of $\l(m)$ tells us what is the behavior of the breaking point at large densities, 
which will be useful to investigate the general properties of the $k$RSB
solutions.

In the following we rely heavily on the notations and results of~\cite[Sec.V and VI]{KPUZ13}.
Let us first recall the expression for $\l(m)$, which can be written as
\beq
\l(m)=\frac{- \hat w^{({\rm I})}_2(\wh A_{\rm G}(m),m)}{\frac{2}{\wh \varphi_{\rm G}(m)}- \hat w^{({\rm I})}_1(\wh A_{\rm G}(m),m)}
\eeq
where $\wh\varphi_{\rm G}(m)$ is the Gardner line, $\wh A_{\rm G}(m)$ is the 1RSB cage radius on the Gardner line, and
\beq
\begin{split}
\hat w^{({\rm I})}_1(\wh A_{\rm G}(m),m)&=-\left<\Th_0^{m-1}(\l)\G_1(\l,m) \right>_{\wh A_{\rm G}(m)}\\
\hat w^{({\rm I})}_2(\wh A_{\rm G}(m),m) &=\frac 12 \left< \Th_0^{m-1}(\l)\Gamma_2(\l,m)\right>_{\wh A_{\rm G}(m)} 
\end{split}
\eeq
with
\beq
\begin{split}
\Gamma_2(\l,m)&=\left[ 2\left(\frac{\Th_1(\l)}{\Th_0(\l)}\right)^3 -3\frac{\Th_1(\l)\Th_2(\l)}{\Th_0^2(\l)}+\frac{\Th_3(\l)}{\Th_0(\l)}  \right]\left[  2\l^3+2(m-6)\left(\frac{\Th_1(\l)}{\Th_0(\l)}\right)^3  \right.  +\\
&\left.+ 3\frac{\Th_1(\l)}{\Th_0(\l)}\left[ 4\l\frac{\Th_1(\l)}{\Th_0(\l)}-(m-4)\frac{\Th_2(\l)}{\Th_0(\l)}\right] -6\l\left(\l\frac{\Th_1(\l)}{\Th_0(\l)}+\frac{\Th_2(\l)}{\Th_0(\l)}\right)+(m-2)\frac{\Th_3(\l)}{\Th_0(\l)}\right] \ , \\
\G_1(\l,m) &=\left[ 1+\frac{\Th^2_1(\l)}{\Th_0^2(\l)}-\frac{\Th_2(\l)}{\Th_0(\l)} \right]^2 \left[(m-3\l^2)+(m-6)\frac{\Th_1^2(\l)}{\Th_0^2(\l)}+6\l\frac{\Th_1(\l)}{\Th_0(\l)}-(m-3)\frac{\Th_2(\l)}{\Th_0(\l)}\right] \ ,
\end{split}
\eeq
where the functions $\Th_k(\l) = \frac1{\sqrt{2\pi}} \int_x^\io \de y \, y^k e^{-y^2/2}$ have been defined in~\cite[Eq.~(41)-(43)]{KPUZ13} and
\beq
\langle O(x) \rangle_{A}=\int_{-\infty}^\infty\frac{\de x}{\sqrt{2\pi}}O(x)e^{-\frac{1}{2}\left(x+\sqrt{2A}\right)^2} \ .
\eeq
Following the analysis and the notations of~\cite[Sec.V and VI]{KPUZ13},
at the instability line we have
$\wh \varphi_{\rm G}(m)=1/\mathcal F_m(\wh A_{\rm G}(m))$,
where $\wh A_{\rm G}(m)$ satisfies the equation
$2\mathcal F_m(\wh A_{\rm G}(m))=-\Lambda_m(\wh A_{\rm G}(m))$
and
\beq
\begin{split}
\Lambda_m(\wh A_{\rm G}(m))&=\left<\LL(\l,m)\right>_{\wh A_{\rm G}(m)}\\
\LL(\l,m) &= \Th_0^{m-1}(\l)
\left[\left(\frac{\Th_1(\l)}{\Th_0(\l)}\right)^2 - \l \frac{\Th_1(\l)}{\Th_0(\l)}\right]
\left[2 - 2 \l^2 + (m-4) \left(\frac{\Th_1(\l)}{\Th_0(\l)}\right)^2 + 
(6 - m) \l \frac{\Th_1(\l)}{\Th_0(\l)}\right]  \ .
\end{split}
\eeq
It follows that the expression for $\l$ can be put in the form
\beq
\l(m)=\frac{\hat w^{({\rm I})}_2(\wh A_{\rm G}(m),m)}{\Lambda_m(\wh A_{\rm G}(m))+ \hat w^{({\rm I})}_1(\wh A_{\rm G}(m),m)} \ .
\eeq
In~\cite[Sec.~V~D]{KPUZ13} it was shown that
in the limit $m\to 0$, $\sqrt{\wh A_{\rm G}(m)}\simeq 0.8m$. 
This means that we can hope to expand the numerator and the denominator in powers of $\sqrt{A_{\rm G}}$.
\beq
\begin{split}
\hat w^{({\rm I})}_2(\wh A_{\rm G}(m),m)&=w_2^{(0)}(m)+\sqrt{\wh A_{\rm G}(m)}w_2^{(1)}(m)+\wh A_{\rm G}(m)w_2^{(2)}(m)+\ldots\\
\hat w^{({\rm I})}_1(\wh A_{\rm G}(m),m)&=w_1^{(0)}(m)+\sqrt{\wh A_{\rm G}(m)}w_1^{(1)}(m)+\wh A_{\rm G}(m)w_2^{(2)}(m)+\ldots\\
\L_m(\wh A_{\rm G}(m))&=\L^{(0)}(m)+\sqrt{\wh A_{\rm G}(m)}\L^{(1)}(m)+\wh A_{\rm G}(m)\L^{(2)}(m)+\ldots
\end{split}
\eeq
Let us study the numerator first.
It happens that $w_2^{(0)}(m)=0$ with very good numerical accuracy, and the equality can be probably demonstrated by a series of integrations by parts
(see~\cite{KPUZ13} for a similar computation).
The first order term (multiplied by 2 for convenience) can be written as
\beq
2 w_2^{(1)}(m) = \int_{-\infty}^\infty\frac{\de \l}{\sqrt{2\pi}}e^{-\l^2/2}(-\l\sqrt{2})\Theta_0(\l)^{m-1}\Gamma_2(\l,m) \ .
\eeq
As it was discussed in~\cite{KPUZ13}, 
the behavior of the integral depends on how the function inside behaves as $\l\to\infty$ for small $m$. We have two possibilities: the integral decays as $\l^{\a}$ with $\a>1$ and in that case the integral is convergent; on the other case, we have a divergent contribution that has to be studied looking at the limit $m\to0$.
To see which of the two behaviors happens we develop asymptotically the integrand 
\beq
e^{-\l^2/2}(-\l\sqrt{2})\Theta_0^{m-1}(\l)\Gamma_2(\l,m)\sim e^{-m\l^2/2}(-\l\sqrt{2})(\l\sqrt{2\pi})^{1-m}\left(\frac{2m}{\l^6}-\frac{12(4m-1)}{\l^8}+\ldots\right)
\eeq
from which it follows that the integral is finite at $m=0$ and it is given by
\beq
2 w_2^{(1)}(0) = \int_{-\infty}^\infty\frac{\de \l}{\sqrt{2\pi}}e^{-\l^2/2}(-\l\sqrt{2})\Theta_0^{-1}(\l)\Gamma_2(\l,0)\simeq-0.134
\eeq
Let us now study the denominator. Also in this case the zeroth order term is zero with very good accuracy.
The first order term of the denominator is
\beq\begin{split}
2\L^{(1)}(m) + 2 w_1^{(1)}(m) &= 
\int_{-\infty}^\infty\frac{\de\l}{\sqrt{2\pi}}(-\l \sqrt{2})e^{-\l^2/2}\Theta_0^{m-1}(\l)\left[\LL(\l,m)+\G_1(\l,m)\right] \\
&\sim\int_{-\infty}^\infty\frac{\de\l}{\sqrt{2\pi}}e^{-m\l^2/2}(\l\sqrt{2\pi})^{1-m}(-\l\sqrt{2})\left[\frac{m}{\l^2}+\frac{2-8m}{\l^4}+\ldots\right] \ .
\end{split}\eeq
Also in this case the integral is finite at $m\to0$ and it is given by
\beq
2\L^{(1)}(0) + 2 w_1^{(1)}(0) = 
\int_{-\infty}^\infty\frac{\de\l}{\sqrt{2\pi}}e^{-\l^2/2}(-\l\sqrt 2)\Theta_0^{-1}(\l)\left[\LL(\l,0)+\G_1(\l,0)\right]\simeq-1.067 \ .
\eeq
We therefore obtain the final result
\beq
\lim_{m\to 0}\l(m)=0.124 \ ,
\eeq
which shows that $\l(m)$ has a finite limit on the Gardner line and therefore the condition $m<\l(m)$ holds for all $m<m^*$.

\section{The jamming limit of the 2RSB solution}

The opposite limit, with respect to the perturbative computation of Sec.~\ref{sec:perturbative}, in which the problem simplifies a lot
is the jamming limit $m\to 0$. In this limit, pressure diverges and one approaches the jamming point where
particles are in contact. This has been investigated in full details at the 1RSB level~\cite{PZ10,BJZ11}:
at the 1RSB level, in the limit $m\to 0$ the parameter $\wh\g_1 = 2\wh\a_1$ remains finite, in such a way that the mean square displacement
in the glass, $\wh\D_1 = m \wh\g_1$ vanishes proportionally to $m$, $\wh\D_1 \sim m \sim 1/p$.

In this section we discuss what happens at the 2RSB level. This is interesting because it allows us to determine the endpoint of the 2RSB dynamical
line, which corresponds to the 2RSB threshold $\wh\f_{\rm th}^{\rm 2RSB}$, see Fig.~\ref{fig:diag_m}. In general, we can expect (based on the experience
accumulated on the spin glass models~\cite{MPV87}) that the 2RSB computation is an extremely good quantitative approximation for the fullRSB result 
as far as thermodynamic quantities are concerned, therefore the 2RSB threshold should be a very good approximation of the fullRSB one.
Moreover, we will encounter here most of the numerical difficulties that will also be relevant for the study of the fullRSB solution.

In Sec.~\ref{sec:2RSB_m0} we obtain the expression of the 2RSB entropy and the associated variational equations in the limit $m\to 0$.
In Sec.~\ref{sec:highphi} we show that the variational equations can be solved analytically 
in a systematic high density expansion.
In Sec.~\ref{sec:2RSBnum} we discuss the results of a numerical solution of the variational equations; we show that the numerical results
are consistent with the high density expansion, and we discuss how the 2RSB threshold is computed.
Finally, in Sec.~\ref{sec:2RSBpd} we summarize the phase diagram obtained from the 2RSB solution.

\subsection{The 2RSB equations at $m=0$}
\label{sec:2RSB_m0}

In order to discuss the behavior of the 2RSB entropy at small $m$, it is convenient to make a change of variables as follows. 
We eliminate $\wh \g_2$ and $m_1$ in favor of
\beq\begin{split}
\h &= 1 - \frac{\wh \g_2}{\wh \g_1} \ , \\
\n &= \frac{m}{m_1} = \frac{1}{y_1} \ .
\end{split}\eeq
In terms of the variables $\wh\g_1$, $\h$, $\n$, and using Eqs.~\eqref{eq:Da2RSB} we can reconstruct the other parameters as follows:
\beq\label{conversion2RSB}
\begin{split}
m_1 &=  m/\n \, , \\
\wh \g_2 &= \wh\g_1 (1-\h) \, , \\
\wh\D_1 &=  m \wh\g_1 (1-\h)  + \wh\g_1 \h\n \ , \\
\wh\D_2   &=   m \wh\g_1 (1-\h) \ .  \\
\end{split}\eeq
Furthermore, from the condition that $\wh\D_1 \geq \wh\D_2 \geq 0$ 
we see that $\h \in [0,1]$, while from $1 \geq m_1 \geq m$ we have that $\n \in [m,1]$.
Writing the entropy \eqref{eq:sD2RSB} in terms of these variables, and writing explicitly the convolutions with some simple changes of variables, 
we obtain
\beq\label{eq:S2RSBsimple}
\begin{split}
s_{\rm 2RSB} &=1-\log \r +\frac{d}2m \log m +\frac{d}{2}(m-1)\log(\pi eD^2/d^2) \\
&+\frac{d}{2} \left\{
(m - 1) \log \wh \g_1 +  (m-\n)  \log (1-\h)
- \wh \varphi
 \sqrt{\nu \h \wh\g_1 } \, e^{- \wh\g_1  [m (1-\h) + \h \n]/2} 
\int_{-\io}^\io \de x \, e^{x \sqrt{\nu \h \wh\g_1 } }  \left[ 1 - (I_2^m(x))^{\n} \right] \right\} \\
I_2^m(x) & =  \int_{-\io}^\io \DD z \,\Theta^{m/\n}\left[ \sqrt{\frac{\n}{m} \frac{\h}{1-\h}} \frac{x - z}{\sqrt{2}}  \right] 
\end{split}
\eeq
As we will see later, this expression is also convenient to perform a numerical computation of the optimal values
of the parameters $\wh\g_1$, $\eta$, $\nu$.

We now want to check that Eq.~\eqref{eq:S2RSBsimple} has a finite limit $m\to 0$ if 
all the other parameters $\wh\g_1, \h, \n$ are fixed (i.e. they do not scale with $m$).
Note that in this limit both $\h, \n \in [0,1]$ and moreover, according to Eq.~\eqref{conversion2RSB},
$\wh\D_2 \to 0$ while $\wh\D_1$ remains finite. 
The limit $m \to 0$ of Eq.~\eqref{eq:S2RSBsimple} can be taken easily. The only non-trivial part is the function $I_2^{m}(x)$.
Using Eq.~\eqref{limittheta}, we have
\beq\begin{split}
I_2(x) &= \lim_{m\to 0} I_2^m(x) = \int_{-\io}^\io  \DD z \exp\left[ - \frac\h{2(1-\h)} \left( x - z \right)^2  \th\left(  z-x \right) \right]
 \\
& = \Theta\left( \frac{x}{\sqrt{2}} \right) +e^{-\frac\h2 x^2}\sqrt{1-\h}\,\Theta\left( - \sqrt{\frac{1-\h}2} x\right) \ ,
\end{split}\eeq
and therefore
\beq\label{eq:s2RSBm0}
\begin{split}
\lim_{m\to 0} s_{\rm 2RSB} &=1-\log \r - \frac{d}{2}\log(\pi eD^2/d^2) \\ &+\frac{d}2 \left\{-  \log \wh \g_1 -\nu \log(1-\h)   -  \wh \varphi  \sqrt{  \h\n\wh\g_1} 
e^{ -\wh\g_1 \h\n/2 } \int \de x \, e^{ x \sqrt{ \h\n\wh\g_1} } \,  [1 - I_2(x)^\nu]  \right\} \\
I_2(x)&= \Theta\left( \frac{x}{\sqrt{2}} \right) +e^{-\frac\h2 x^2}\sqrt{1-\h}\,\Theta\left( - \sqrt{\frac{1-\h}2} x\right) \ . \\
\end{split}\eeq
We see that we obtain a finite limit that corresponds to the complexity at $m=0$~\cite{PZ10}, and
we also conclude that the three parameters $\wh\g_1$, $\h$, $\n$ have finite values at $m=0$. 
From this, we reach the important physical conclusion that the mean square displacement
{\it inside a glass}, $\wh\D_2$, vanishes at jamming, as it should, but at the same time, the mean square displacement $\wh\D_1$ 
{\it between different sub-glasses inside a
meta-glass remains finite}. Hence, sub-glasses inside a meta-glass are microscopically distinct.
Finally, we note that $m_1$ vanishes proportionally to $m$ because $\nu$ is finite.

\subsection{The high density limit of the 2RSB solution at $m=0$}
\label{sec:highphi}

The expression \eqref{eq:s2RSBm0} is still quite difficult to handle numerically. Therefore, 
before discussing the numerical optimization
it is convenient to obtain some asymptotic results for 
large density.
We first observe that in the 1RSB case $\wh\g_1 \sim \wh\f^{-2}$ and we therefore expect the same scaling also in the 2RSB case. Furthermore,
on the Gardner transition line we showed in section~\ref{sec:perturbative} that $m_1 = \l(m) \to 0.124$ and that $m \sim 1.98  \wh\f^{-2}$, hence
$\nu = m/m_1 \sim 16.0 \wh\f^{-2}$.
We therefore seek for a small $\wh\g_1$ and small $\nu$ expansion of Eq.~\eqref{eq:s2RSBm0}.
Because both $\wh\g_1$ and $\nu$ are of the same order of magnitude, we use $1/\wh\f$ as the small expansion parameter.
Note that we write
$\wh\g_1, \n \sim \OO(2)$ to indicate that these quantities are of order 2 in $1/\wh\f$, and similarly for other quantities.

According to the definition in Eq.~\eqref{eq:SSKRSBdef}, 
the part of the entropy that has to be optimized is
\beq
\SS_{\rm 2RSB} =  - \log \wh \g_1 - \nu \log(1-\h) -\wh\f  \II \ ,
\eeq
and we have to expand the integral
\beq
\II = \sqrt{ \wh\g_1 \h\n}  \int_{-\io}^\io \de x \, e^{ x \sqrt{ \wh\g_1 \h\n} -\wh\g_1 \h\n/2  } \,  [1 - I_2(x)^\nu]  \ .
\eeq
When $\nu$ is small, it is very useful to separate two contributions to this integral as follows:
\beq\label{eq:IIan}
\begin{split}
\II &= \II^{(a)} + \II^{(n)} \\
\II^{(n)} & = 
\sqrt{ \wh\g_1 \h\n}  \int_{-\io}^\io \de x \, e^{ x \sqrt{ \wh\g_1 \h\n} -\wh\g_1 \h\n/2  } \,  [ \th(-x)  e^{-\frac\h2 x^2 \nu} \sqrt{1-\h}^\nu + \th(x)  - I_2(x)^\nu ] \\
 & = 
\sqrt{ \wh\g_1 \h\n} e^{ -\wh\g_1 \h\n/2  } 
\int_{0}^\io \de x \left[
e^{ x \sqrt{\wh\g_1 \h\n}  } (1 -  I_2(x)^\nu) +  e^{ -x \sqrt{ \wh\g_1 \h\n}  }
(  e^{-\frac\h2 x^2 \nu} \sqrt{1-\h}^\nu - I_2(-x)^\nu ) \right] \\
\II^{(a)} &  =\sqrt{\wh\g_1 \h\n}  \int_{-\io}^0 \de x \, e^{ x \sqrt{ \wh\g_1 \h\n} -\wh\g_1 \h\n/2  } \, 
 [1 - e^{-\frac\h2 x^2 \nu}  \sqrt{1-\h}^\nu ] \\
&=
 \sqrt{\wh\g_1 \h}  \int_{-\io}^0 \de y \, e^{ y \sqrt{\wh\g_1 \h} -\wh\g_1 \h\n /2  } \,  [1 - e^{-\frac\h2 y^2 }  \sqrt{1-\h}^\nu ] \\
& = e^{-\wh\g_1 \h \n/2} [1 - e^{\wh\g_1/2} \sqrt{ \pi \wh\g_1/2} (1-\h)^{\n/2} (1- \erf(\sqrt{\wh\g_1/2}))]
\end{split}\eeq
The term $\II^{(a)}$ can be easily handled and expanded in a power series of $\nu$ and $\wh\g_1$. 
Moreover, the integrand
of $\II^{(n)}$ is well behaved at large $|x|$ (it decays as a Gaussian)
for all values of $\nu$, therefore we can expand the integrand.

We can expand $\II^{(n)}$ as follows
\beq\begin{split}
\II^{(n)} & = 
 e^{ -\wh\g_1 \h\n/2  } \int_0^\io \de x
\sum_{l=0}^\io \frac{x^l}{l!} ( \wh\g_1 \h\n)^{(l+1)/2}
\sum_{k=1}^\io \frac{\n^k}{k!}
\left\{
- \log I_2(x)^k + (-1)^l \left[ 
\left( \frac12\log(1-\h) - \frac\h2 x^2 \right)^k  - \log I_2(-x)^k \right]
\right\} \\
& = 
 e^{ -\wh\g_1 \h\n/2  }
\sum_{l=0}^\io\sum_{k=1}^\io \frac{1}{l! k!} ( \wh\g_1 \h\n)^{(l+1)/2} \n^k
 \II_{l,k}(\h) \ , \\
\II_{l,k}(\h) & = 
\int_0^\io \de x \, x^l 
\left\{
- \log I_2(x)^k + (-1)^l \left[ 
\left( \frac12\log(1-\h) - \frac\h2 x^2 \right)^k  - \log I_2(-x)^k \right]
\right\}
\end{split}\eeq
The functions $\II_{l,k}(\h)$ are defined by well convergent integrals, hence they are
well behaved functions of $\h$ and they can be
differentiated as many times as one wishes by exchanging the derivative with the integration
over $x$. In this way, expanding all the variables in powers of $1/\wh\f$ as follows,
\beq\begin{split}
\wh\g_1 = \sum_{k=2}^\io \wh\g_{1,k} \wh\f^{-k} \ , \\
\nu = \sum_{k=2}^\io \nu_k \wh\f^{-k} \ , \\
\h = \sum_{k=0}^\io \h_k \wh\f^{-k} \ , \\
\end{split}\eeq
one can perform a systematic expansion of the entropy in powers of $1/\wh\f$ and determine
the coefficients $\wh\g_{1,k}$, $\nu_k$, $\h_k$ by optimizing order by order in inverse density.
This computation can be very easily performed with the help of some algebraic manipulation software
(we used Mathematica). Below we just give an example of the lowest orders in the computation.

At the lowest order we have, recalling that $\OO(n)$ denotes an order $n$ in $1/\wh\f$:
\beq\begin{split}
\II^{(n)} & =  \nu^{3/2} \sqrt{\wh\g_1 \h} \, \II_{0,1}(\h) + \OO(6) \ , \\
\II_{0,1}(\h) & =  \int_{0}^\io \de x \left[
- \log I_2(x) + \frac12\log(1-\h) - \frac\h2 x^2 \ - \log I_2(-x) \right] \ .
\end{split}\eeq
and
\beq
\II^{(a)} = - \sqrt{\pi \wh\g_1/2} + \wh\g_1 - \sqrt{\pi (\wh\g_1/2)^3} - \frac12 \sqrt{\frac{\pi \wh\g_1}2} \nu \log(1-\h) + \frac13 \wh\g_1^2 + \frac12 \wh\g_1 \nu (\log(1-\h) - \h) + \OO(5) \ . 
\eeq
Collecting all together and rearranging the terms in increasing order in the expansion we have
\beq\label{eq:Sphiexp}
\begin{split}
\SS_{\rm 2RSB} &= - \log \wh \g_1 + \wh\f \sqrt{\pi \wh\g_1/2} -\wh\f   \wh\g_1 +\left(  - \nu \log(1-\h)  + \wh\f 
   \sqrt{\pi (\wh\g_1/2)^3} + \frac12 \wh\f  \sqrt{\frac{\pi \wh\g_1}2} \nu \log(1-\h)  \right) \\
&   -\wh\f  \left(
   \frac13 \wh\g_1^2 + \frac12\wh\g_1 \nu (\log(1-\h) - \h) 
+ \nu^{3/2} \sqrt{ \wh\g_1 \h} \, \II_{0,1}(\h) 
\right) + \OO(4)
\end{split}\eeq

\begin{figure}
\includegraphics[width=.3\textwidth]{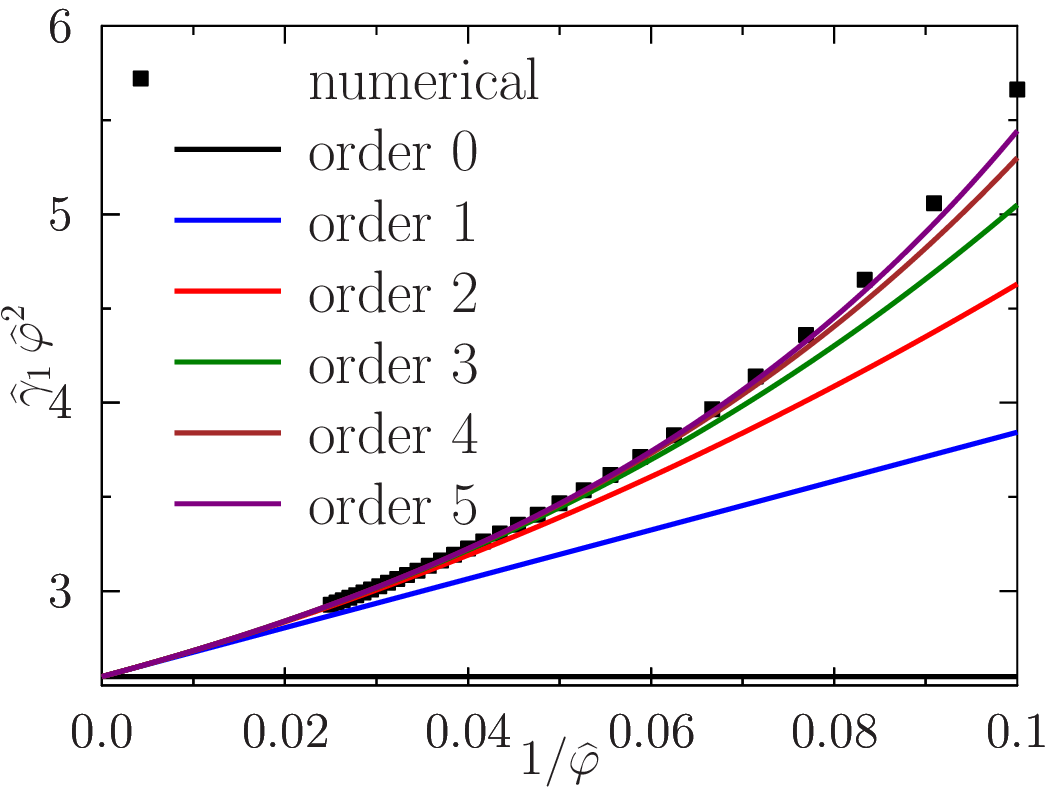}
\includegraphics[width=.3\textwidth]{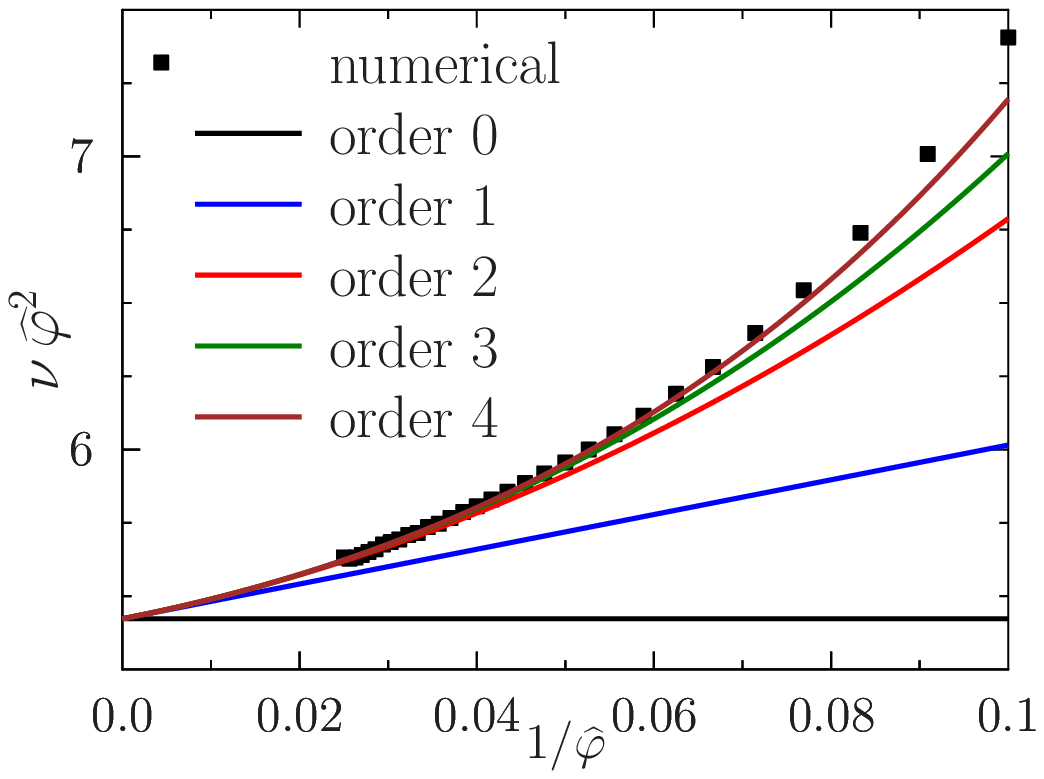}
\includegraphics[width=.3\textwidth]{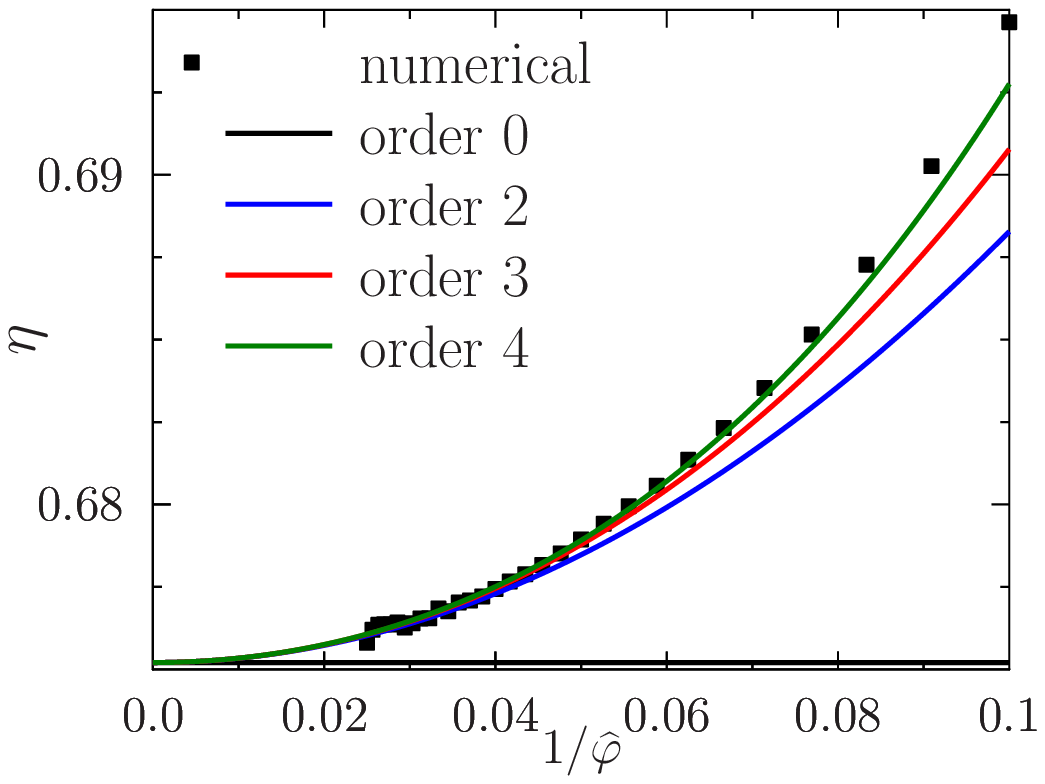}
\caption{
Results for $\wh\g_1$, $\nu$ and $\eta$ in the high density expansion at $m=0$.
Full lines are the analytical results at different orders in the small $1/\wh\f$ expansion. Points
are the result of a direct numerical optimization of the 2RSB entropy at $m=0$.
}
\label{fig:expm0}
\end{figure}

We have now to optimize Eq.~\eqref{eq:Sphiexp} order by order. At the leading order we obtain an equation for $\wh\g_1$ which coincides with the 1RSB one:
\beq
-\frac2{\wh\g_1} + \frac12 \wh\f\sqrt{\frac{2\pi}{ \wh\g_1}} = 0 
\ \ \ \
\Rightarrow
\ \ \ \
\wh\g_1 = \frac8\pi \wh\f^{-2} + \OO(3) 
\ \ \ \
\Rightarrow
\ \ \ \
\wh\g_{1,2} = \frac8\pi
\ .
\eeq
We therefore have to look for a solution of the form $\wh\g_1 = \frac8\pi \wh\f^{-2} + \wh\g_{1,3}  \wh\f^{-3} + \OO(4)$. Plugging this in
Eq.~\eqref{eq:Sphiexp} and expanding we get
\beq
\SS_{\rm 2RSB} = \text{const.} + \left(-  \wh\g_{1,3}  +\frac{\pi^2}{256} \wh\g_{1,3}^2 \right) \wh\f^{-2} + \OO(3)
\eeq
which allows to determine $\wh\g_{1,3} = 128/\pi^2$. Finally we look for $\wh\g_1 = \frac8\pi \wh\f^{-2} + (128/\pi^2)  \wh\f^{-3} + \wh\g_{1,4}  \wh\f^{-4} + \OO(5)$
and $\nu = \nu_2  \wh\f^{-2} + \OO(3)$, plug this in Eq.~\eqref{eq:Sphiexp} and expand to get
\beq
\SS_{\rm 2RSB} = \text{const.} -\frac{2}{3\pi} \left(
-6  \n_2 (\h + \log(1-\h)) + 3  \sqrt{2 \pi \h} \n_2^{3/2} h(\h) 
\right)  \wh\f^{-3} + \OO(4)
\eeq
This function must be optimized numerically and we obtain 
$\nu_2 = 5.4226$ and $\h_0 = 0.6752$.
We therefore obtain the asymptotic 2RSB solution for $m=0$ and $\wh\f\to\io$:
\beq\begin{split}
\wh\g_1 &= \frac8\pi \wh\f^{-2} + (128/\pi^2)  \wh\f^{-3} +  \OO(4) \ , \\
\nu &= 5.4226 \wh\f^{-2} + \OO(3) \ , \\
\eta &= 0.6752 + \OO(1) \ , \\
\end{split}\eeq
and
\beq
s[\hat\a_{2RSB}] = s[\hat\a_{1RSB}] - \frac{d}2 \, 1.03416 \, \wh\f^{-3} + \OO(4) \ .
\eeq
This last result shows that at the 2RSB level the value of $\wh\f_{\rm GCP}$, which corresponds to the point where the complexity (equal to the replicated entropy) at $m=0$ vanishes,
is slightly reduced with respect to the 1RSB level. However, this happens only at subdominant orders in large $d$, because the dominant orders are defined by a term $\log d$ that comes
from the ideal gas term~\cite{PZ10}.

Higher orders in the calculation can be easily obtained by iterating the above procedure.
This requires adding more terms in the expansion in Eq.~\eqref{eq:Sphiexp} which can be easily
automatized with Mathematica. In Fig.~\ref{fig:expm0} we report the results of the
calculation done at order 11 in density, which allows to obtain $\wh\g_1$ to order $\wh\f^{-7}$,
$\nu$ to order $\wh\f^{-6}$ and $\h$ to order $\wh\f^{-4}$. The results are in perfect agreement
with a numerical optimization of the 2RSB entropy \eqref{eq:s2RSBm0} at $m=0$, that we describe in Sec.~\ref{sec:2RSBnum}.

We expect that this strategy to construct a high density expansion (which corresponds to the small cage expansion of~\cite{PZ10} in the
1RSB case) could be generalized to $m>0$ and to $k$RSB solutions with $k>2$ with a little bit of additional work. However, having
tested the accuracy of our numerical optimization code, we do not pursue this strategy further and we turn to the discussion of the numerical
results.

\begin{figure}
\includegraphics[width=.45\textwidth]{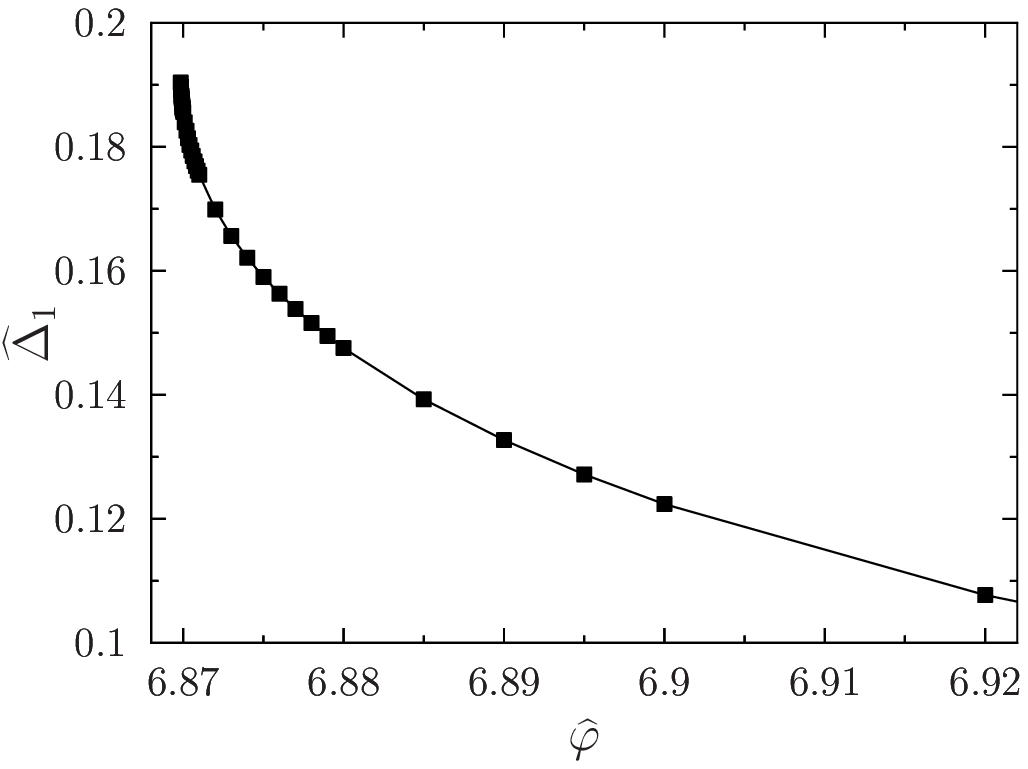}
\includegraphics[width=.45\textwidth]{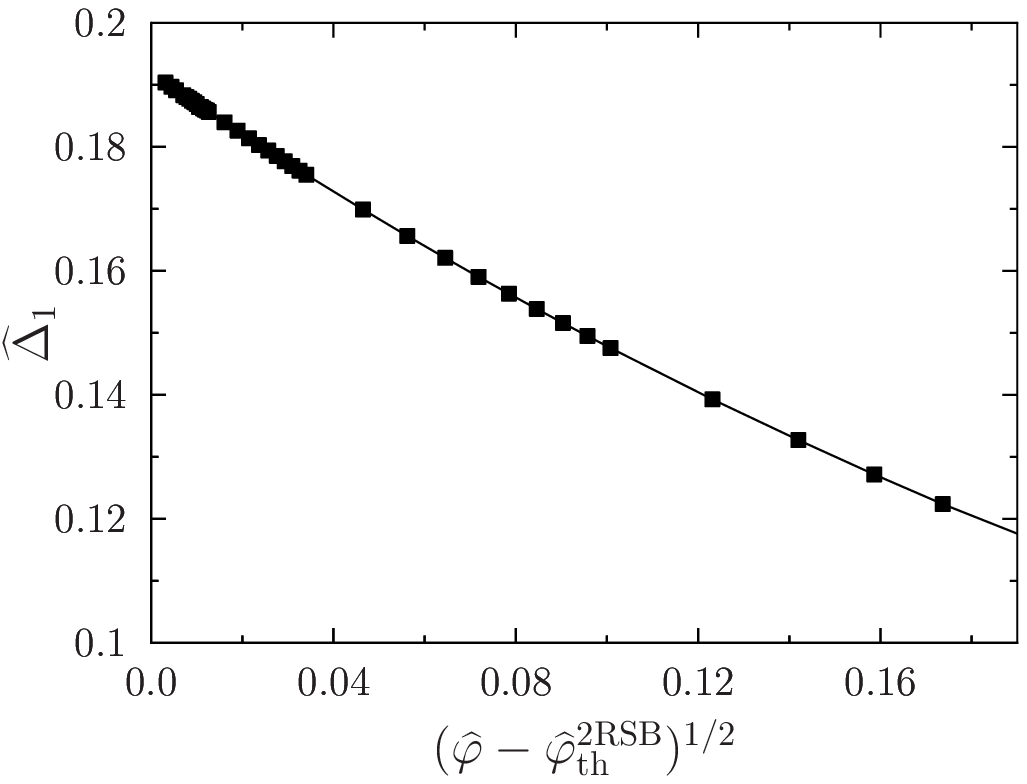}
\caption{
The inter-state overlap of the 2RSB solution at $m=0$, $\wh\D_1 = \wh\g_1 \h\n$ as a function of $\wh\f$. A square-root singularity 
is seen, and the singular point can be located to $\wh\f_{\rm th}^{\rm 2RSB} = 6.86984$.
}
\label{fig:th2RSB}
\end{figure}

\subsection{Numerical solution of the 2RSB equations at $m=0$: the 2RSB threshold}
\label{sec:2RSBnum}

We report here results from the full numerical optimization of the 2RSB entropy at $m=0$, given in Eq.~\eqref{eq:s2RSBm0}.
The code we used makes explicit use of the decomposition~\eqref{eq:IIan}, in such a way that $\II^{(a)}$ is computed easily
and $\II^{(n)}$ is a numerically stable integral (some care should be taken to write the error functions in a numerically stable way). 
Taking derivatives with respect to $\h$ and $\wh\g_1$, we obtain recurrence
equations for these quantities, that we do not report because they are the specialization of Eqs.~\eqref{eq:Sscaledfinal} to
the case $k=2$ and $m=0$. For each fixed value of density $\wh\f$ and breaking point $\nu = 1/y_1$, these two equations can be solved
by iteration to obtain $\h$, $\wh\g_1$, and the entropy $s_{\rm 2RSB}$.

We make now a few remarks on the 2RSB equation at $m=0$:
\begin{itemize}
\item 
When $\n=1$ (corresponding to $m_1=m$), the 2RSB entropy reduces to the 1RSB one, function of $\wh \g_2 = \wh\g_1 (1-\h)$.
\item
Similarly, when $\nu\to 0$ (corresponding to $m_1=1$), the 2RSB entropy reduces to the 1RSB one, function of $\wh\g_1$.
\item
Finally, when $\h\to 0$ (corresponding to $\wh\g_1 = \wh\g_2$), one again recovers the 1RSB entropy.
\end{itemize}
Let us call $f(\n,\h ; \wh\f) = \min_{\wh\g_1} s_{2RSB}$.
This function has to be minimized with respect to $\{\n,\h\} \in [0,1]^2$. 
Based on the considerations above, $f(\n,\h ; \wh\f)$
is a constant equal to the 1RSB value of the free energy
when $\n=0$, $\n=1$ or $\eta=0$, i.e. on three sides of the box $[0,1]^2$.
Moreover its derivative in $\eta=0$ is always strictly negative (except for $\n=0$ and $\n=1$) 
as a consequence of the instability of the replicon mode (it is easy to see that the replicon is related to the derivative with respect to $\h$ of $s_{\rm 2RSB}$. 
We conclude that a minimum must exist at non trivial values of $\n$ and $\h$ (remember that as usual in replica computations, the entropy should be minimized
and not maximized~\cite{MPV87}). However, it is also important to remark that
for some values of the parameters $\{\n,\h\}$ the solution for $\wh\g_1$ might not exist
formally corresponding to $\wh\g_1 = \io$, which corresponds to losing the non-trivial 2RSB solution to a 
trivial solution where all replicas are uncorrelated.

The best way to find the minimum is to optimize over $\wh\g_1$ and $\h$ at fixed $\nu$ and plot the resulting entropy $s_{\rm 2RSB}$
as a function of $\nu$ to find the minimum, recalling that
based on the above considerations we must have $s_{\rm 2RSB} = s_{\rm 1RSB}$ at $\n = 0,1$.
We performed this procedure at high density (up to $\wh\f=40$) and compared the results to the high density expansion of Sec.~\ref{sec:highphi}, finding
excellent agreement and confirming therefore the validity of our numerical code, see Fig.~\ref{fig:expm0}. We note however that solving the equations in the
very high density regime is hard and the data are quite noisy, especially for $\h$.

We then focus on the low density regime. We observe that, upon decreasing $\wh\f$, a secondary maximum appears when the 2RSB entropy is plotted
as a function of $\nu$. At a critical density $\wh\f_{\rm th}^{\rm 2RSB}$, the physical minimum coalesces with this secondary maximum and disappears. Below this point,
the 2RSB solution does not exist anymore. This does not contradict the previous statement, that a minimum should exist for $\{\nu,\h\} \in [0,1]^2$.
In fact, when the maximum and minimum disappear, we also observe that in a region of values of $\n$ the solution for $\wh\g_1$ and $\h$ does not exist, which
means that the 2RSB is not defined.

If we follow the values of the parameters $\wh\g_1$, $\n$ and $\h$ corresponding to the physical 2RSB solution, we see that they display a square root
singularity on approaching $\wh\f_{\rm th}^{\rm 2RSB}$ from above, 
which confirms that when the maximum and minimum coalesce, a longitudinal mode
of the 2RSB solution vanishes signaling the singularity that marks the disappearance of the solution. 
This is illustrated in Fig.~\ref{fig:th2RSB} using a physical quantity, the inter-state overlap $\wh\D_1$, but the same behavior is observed in all the three
parameters $\wh\g_1$, $\n$ and $\h$.
We conclude therefore that $\wh\f_{\rm th}^{\rm 2RSB}= 6.86984$
can be taken as a definition of the threshold state within a 2RSB calculation. We stress once again that, however, we expect that the 2RSB solution is unstable 
and we expect that one should perform a fullRSB calculation to obtain the correct value of the threshold, following~\cite{Ri13}.
The same analysis could be repeated at $m>0$ but we do not report the calculation here.

\subsection{The 2RSB phase diagram}
\label{sec:2RSBpd}

\begin{figure}
\includegraphics[width=0.45\textwidth]{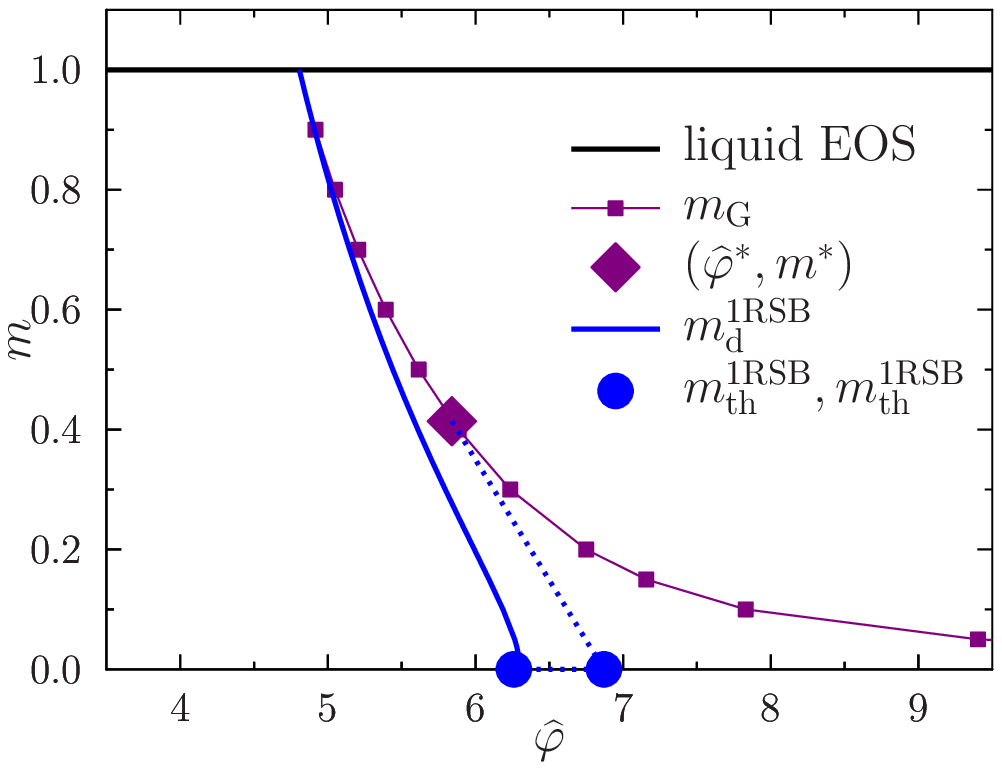}
\includegraphics[width=0.45\textwidth]{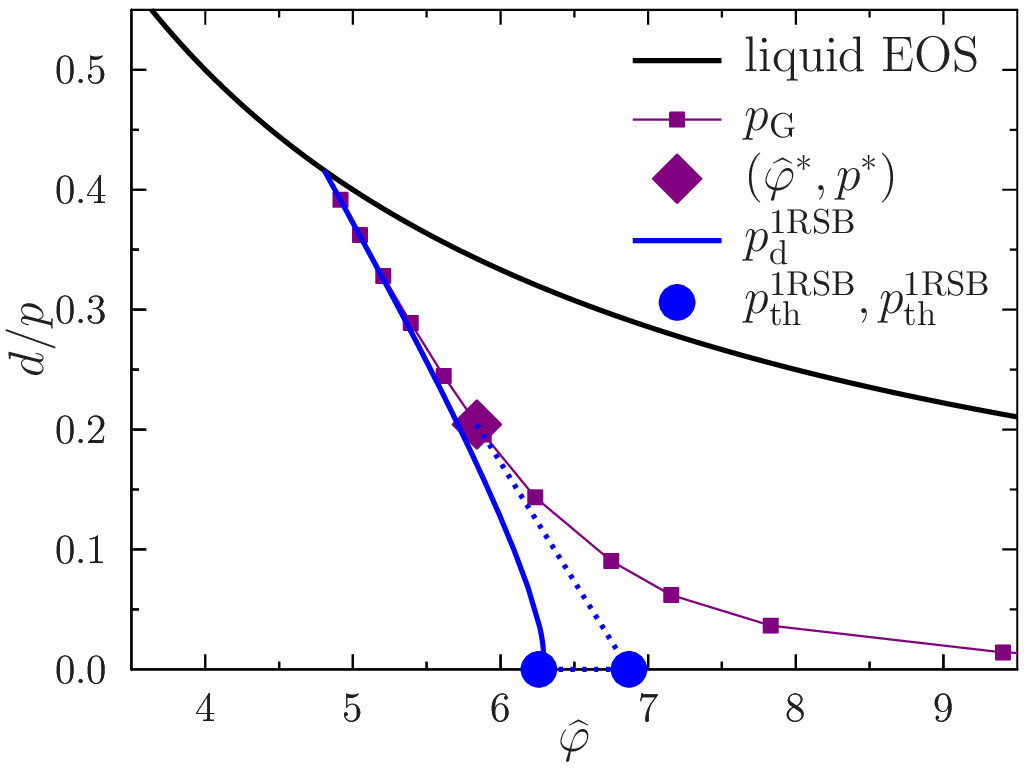}
\caption{Phase diagram in the $(m,\wh\varphi)$ plane (left) and in the $(d/p,\wh\varphi)$ plane (right).
In both cases the black full line corresponds to the liquid state, the purple line is the Gardner line, the large purple diamond
is the point $(m^*,\wh\f^*)$ in Eq.~\eqref{eq:mstarvalue}, the blue line is the 1RSB dynamical line, and the two blue dots are 
the 1RSB and 2RSB thresholds. The blue dashed line is just a straight line joining the point $(m^*,\wh\f^*)$ with the 2RSB threshold;
it is drawn as a guide to the eye and should be a decent approximation to the 2RSB dynamical line at finite pressure or finite $m$.
}
\label{fig:diag_p}
\end{figure}

\subsubsection{Phase diagram in the $(m,\wh\f)$ plane}

In Fig.~\ref{fig:diag_p} we summarize the phase diagram that we can infer from the previous discussions. 
A schematic version of this phase diagram was already presented in Ref.~\cite[Fig.~1]{KPUZ13}, 
and here we substantiate this proposal with actual computations.
At the 1RSB level, the dynamical line ends at $\wh\f_{\rm th}^{\rm 1RSB} = 6.25967$, as computed in Ref.~\cite{PZ10}.
However, we find here that this 1RSB dynamical line falls into the unstable region and therefore has no physical meaning.
The 2RSB calculation indeed gives a higher value for the threshold point, $\wh\f_{\rm th}^{\rm 2RSB}= 6.86984$.
We expect that a dynamical line $\wh\f_{\rm d}^{\rm 2RSB}(m)$ connects the 2RSB threshold to the point $(m^*,\wh\f^*)$ defined in Sec.~\ref{sec:VIIA},
as
illustrated in the schematic Fig.~\ref{fig:diag_m}. We did not compute this line, but its location can be reasonably approximated
by joining the two points by a straight line, as we did in Fig.~\ref{fig:diag_p}.
By analogy with spin-glass systems, we expect that further RSB will not strongly affect the location of the dynamical line, so the 2RSB computation should give
a good approximation of the exact result. This dynamical line and the Gardner line, that join at the point $(m^*,\wh\f^*)$,
delimit the region of existence of the fullRSB solution.

The 1RSB solution remains correct around the liquid phase (corresponding to $m=1$)
so that when glassy states form, they have a 1RSB structure (as described in~\cite{PZ10}). They only undergo the Gardner transition at higher
densities. In the region where the 1RSB solution is stable, all the results of Ref.~\cite{PZ10} remain valid.
In particular, note that the Kauzmann point~\cite{PZ10}, depicted schematically in Fig.~\ref{fig:diag_m}, shifts to infinite density on the scale
of Fig.~\ref{fig:diag_p} and for that reason it is not depicted in the figure. This point nonetheless falls within the region where the 1RSB solution is stable
and therefore none of its properties is changed with respect to the discussion of Ref.~\cite{PZ10}.
The glass close packing (GCP) point introduced in~\cite{PZ10}, corresponding to the densest amorphous packing that can be obtained
by compressing liquid configurations, is also located at infinite density ($\wh\f \propto \log d$) on the $m=0$ (infinite pressure) line of 
Fig.~\ref{fig:diag_p}. Because the Gardner transition occurs at infinite pressure when $\wh\f \to\io$, the equilibrium {\it ideal} glass
only undergoes the Gardner transition exactly at infinite pressure, when GCP is reached. The GCP point somehow lies exactly
on the Gardner line, which may explain why previous results obtained in~\cite{PZ10} for GCP (like the fact that the GCP point is isostatic)
were quite accurate despite neglecting the Gardner transition.
We would like to stress, however, that experiments and numerical simulations are typically conducted in the vicinity of the dynamical line, and therefore never approach the Kauzmann nor the GCP points.

\subsubsection{Phase diagram in the $(1/p,\wh\f)$ plane: isocomplexity assumption}

It would be nice to convert the $(m,\wh\f)$ phase diagram into a physical $(1/p,\wh\f)$ phase diagram where $p$ is the reduced pressure
of the glassy states visited at a given $m$ and $\wh\f$. Doing this exactly requires a so-called ``state following'' calculation
where we adiabatically follow the evolution of a given state in density to compute the pressure. Although this is certainly possible, we do not report
this computation here and we resort to a much simpler ``isocomplexity'' assumption~\cite{MR03,PZ10} where we assume that states
can be followed by fixing the value of the complexity. If this is the case, we can reason as follows, see~\cite{PZ10} for a more detailed discussion.
First we recall that $s_{\rm 2RSB} = m s^* + \Si(s^*)$, where $s^*$ is the internal entropy of the state and $\Si(s^*)$ the associated complexity. 
Then, we note that the value of $m$ corresponding to a given level of compexity $\Si_{\rm g}$ is given by the point where
\beq
\frac1m [ s_{\rm 2RSB} - \Si_{\rm g} ]  = s^* + \frac1m [ \Si(s^*) - \Si_{\rm g} ] 
\eeq
is maximum with respect to $m$. In fact the maximum condition
\beq
-m^2 \frac{\de}{\de m} \{ s^* + \frac1m [ \Si(s^*) - \Si_j ] \} = \Si(s^*) - \Si_{\rm g} = 0 
\eeq
is equivalent to the isocomplexity condition.
Let us call the solution $m_{\rm g}(\wh\f)$. We also note that the corresponding internal entropy of the state is 
\beq
s_{\rm g} = \left. \frac1m [ s_{\rm 2RSB} - \Si_{\rm g} ] \right|_{m = m_{\rm g}(\wh\f)} \ .
\eeq
It follows that the pressure of the glass is
\beq\label{eq:pg}
p_{\rm g} = -\wh\f \frac{\de s_{\rm g}}{\de \wh\f} = -\frac{\wh\f}{m_{\rm g}(\wh\f)} \left. \frac{\partial s_{\rm 2RSB}}{\partial \wh\f}\right|_{m = m_{\rm g}(\wh\f)} 
= \frac{1}{m_{\rm g}} \left[ 1 + \frac{d}2 \wh\f \FF(2 \hat\a^* ) \right]
\ ,
\eeq
where $\FF(2 \hat\a^* )$ is the interaction part of the entropy computed in the matrix $\hat\a^*$ corresponding to the 2RSB solution
at $m_{\rm g}$. This formula generalizes the one in~\cite[Eq.~(17)]{KPUZ13}, it reduces to that one for the equilibrium glass which
corresponds to the choice $\Si_{\rm g}=0$, and it allows us to convert the $(m,\wh\f)$ phase diagram into a pressure-density one under
the isocomplexity approximation. Note that we must scale pressure by dimension plotting $d/p$ to obtain a finite result for $d\to\io$ as it is evident
from Eq.~\eqref{eq:pg}. 
The result is shown in Fig.~\ref{fig:diag_p} and is qualitatively similar to the $(m,\wh\f)$ one.


\section{Critical scaling of the fullRSB solution at jamming ($m=0$)}
\label{sec:scaling}

In the previous section we have delimited approximately the region where a non-trivial $k$RSB solution with $k>1$ is found.
We now assume that this phase is a fullRSB phase, which we confirm below through a numerical solution of Eq.~\eqref{eq:Sscaledfinal}, 
and that the 2RSB calculation provides a quite good approximation to the fullRSB one, which is usually correct in spin glasses. 
Then the fullRSB region is, in Fig.~\ref{fig:diag_p}, the one below the Gardner line, at densities above the 2RSB dynamical 
line that connects the 2RSB threshold at $m = 0$ (or infinite pressure) and the point
$(m^*,\wh\f^*)$.
We are now in the position to explore the scaling of the fullRSB solution in the jamming limit $m\to 0$.
This limit is formally very similar to the zero-temperature limit in the Sherrington-Kirkpatrick model which has been thoroughly
investigated in the past~\cite{PT80,SD84,MPV87,Pa06,CD12}, and we follow here a very similar strategy.

Our results can be considered as a generalization of the arguments of Pankov~\cite{Pa06} to the more complex case under investigation,
as it will be clear below.
Our starting point are Eq.~\eqref{eq:Sscaledfinal} and its continuum version Eq.~\eqref{eq:Sscaledfinal_cont}, and we begin 
the discussion by making some conjectures on the behavior of these equations for $m=0$ and
large $y$. These conjectures are partly based on the results of the numerical solution of the equations, that we discuss later,
and partly on physical intuition. The aim of this section will be to show that they are indeed correct.

In Sec.~\ref{sec:IXA} and \ref{sec:IXB}
we will show that in the limit $m\to 0$ (in which the variable $y$ extends from 1 to $\io$) 
a scaling regime of Eq.~\eqref{eq:Sscaledfinal} and Eq.~\eqref{eq:Sscaledfinal_cont} appears,
and is characterized by non-trivial scaling exponents.
In Sec.~\ref{sec:IXC}, we consider a simplified set of equations (a toy model)
for which the scaling regime can be fully characterized,
that is instructive to discuss the scaling regime of the complete equations.
In Sec.~\ref{sec:IXD}, we extend the results of the toy model to the complete fullRSB equations.
Finally, in Sec.~\ref{sec:IXE} we determine analytically all the critical exponents that characterize
the scaling regime.

\subsection{Scaling of $\D(y)$, and asymptotic scaling of $\wh P(y,h)$ for $h\to-\io$}
\label{sec:IXA}

We look for an asymptotic solution at large $y$ characterized by $\D(y) \sim \D_\io y^{-\k}$.
From the relation between $\D(y)$ and $\g(y)$ in Eq.~\eqref{eq:Sscaledfinal_cont} it follows that
$\g(y) \sim \g_\io y^{-c}$ with $c=\k-1$ and $\g_\io = \frac{\k}{\k-1} \D_\io$.
We will later show that the exponent $c$ is close to $1/2$.

Before studying the full scaling of $\wh P(y,h)$, let us look to its asymptotic behavior at $h\to-\io$, which provides some very
useful insight. First, we note that $\wh P(y_i,h) \sim A_i e^{B_i h - h^2 D_i}$ when $h \to -\io$. In fact, this is true for $y_1$,
and the iteration~\eqref{eq:Sscaledfinal} for $\wh P(y_i,h)$ preserves this asymptotic behavior. From the analysis
of Eq.~\eqref{eq:Sscaledfinal} in the limit $h\to-\io$, we obtain the discrete recurrence equations
\beq\begin{split}
D_i &= \frac{D_{i-1} y_{i-1} \g_{i-1}^2}{\g_i [2 D_{i-1} \g_{i-1} (\g_{i-1} - \g_i) + y_{i-1} \g_i]} \ , \\
B_i &= \frac{y_{i-1} \g_{i-1} + B_{i-1} y_{i-1} \g_{i-1}}{2 D_{i-1} \g_{i-1} (\g_{i-1} - \g_i) + y_{i-1} \g_i} -1  \ , \\
A_i &= A_{i-1} \sqrt{ D_i / D_{i-1} } \exp\left[
\frac{(1+B_{i-1})^2 \g_{i-1} ( \g_{i-1} - \g_i)}{2[2 D_{i-1} \g_{i-1} (\g_{i-1} - \g_i) + y_{i-1} \g_i]}
\right] \ ,
\end{split}\eeq
which in the continuum limit become
\beq\label{eq:ABD}
\begin{split}
\dot D &= \frac{2 D \dot \g}{y} ( D - y/\g) \ , \\
\dot B &= \frac{(- y \dot\g +2 D \g\dot\g) }{y \g}(1+B) \ , \\
\dot A &= A \left( - \frac{(1+B)^2 \dot\g}{2y} + \frac{D\dot \g}{y} - \frac{\dot\g}{\g} \right) \ .
\end{split}\eeq
Note in fact that Eqs.~\eqref{eq:ABD} can also be obtained directly from the continuum equation for $\wh P(y,h)$.
Under the assumption that $\g(y) \sim \g_\io y^{-c}$ with $0<c<1$, Eqs.~\eqref{eq:ABD} admit a solution 
with $D(y) \sim D_\io y^{2c}$, $B \sim B_\io y^{c}$ and $A(y) \sim A_\io y^{c}$ for $y\to\io$.
Hence we conclude that for $h\to -\io$ and large $y$ we have
\beq\label{eq:p0as}
 \wh P(y_i,h) \sim A_\io y^{c} e^{B_\io h y^{c} - D_\io h^2 y^{2c}} = y^{c} p_0( h y^{c})
\eeq
with $p_0(z \to -\io) = A_\io \exp(B_\io z -D_\io z^2)$.

\subsection{Complete scaling of $\wh P(y,h)$}
\label{sec:IXB}

Let us conjecture that the scaling of Eq.~\eqref{eq:p0as} holds for all $h<0$. This means that for $h<0$ and large $y$, 
$\wh P(y,h) \sim y^c$ diverges,
while we know from Eq.~\eqref{eq:Pasymptotes} that for large $h>0$, $\wh P(y,h) \sim \exp(-\D(y)/2)$, therefore it
remains finite for large $y$.
Combining this information we expect that, on increasing $h$ from $-\io$, $\wh P(y,h)$ increases up to a value $\approx y^{c}$ 
on a scale $|h| \approx y^{-c}$, it
reaches a peak and then it decreases fast around $h \approx 0$ to approach its asymptotic limit $\exp(-\D(y)/2) \approx 1$ at $h\to\io$.

\begin{figure}[t]
\includegraphics[width=.45\textwidth]{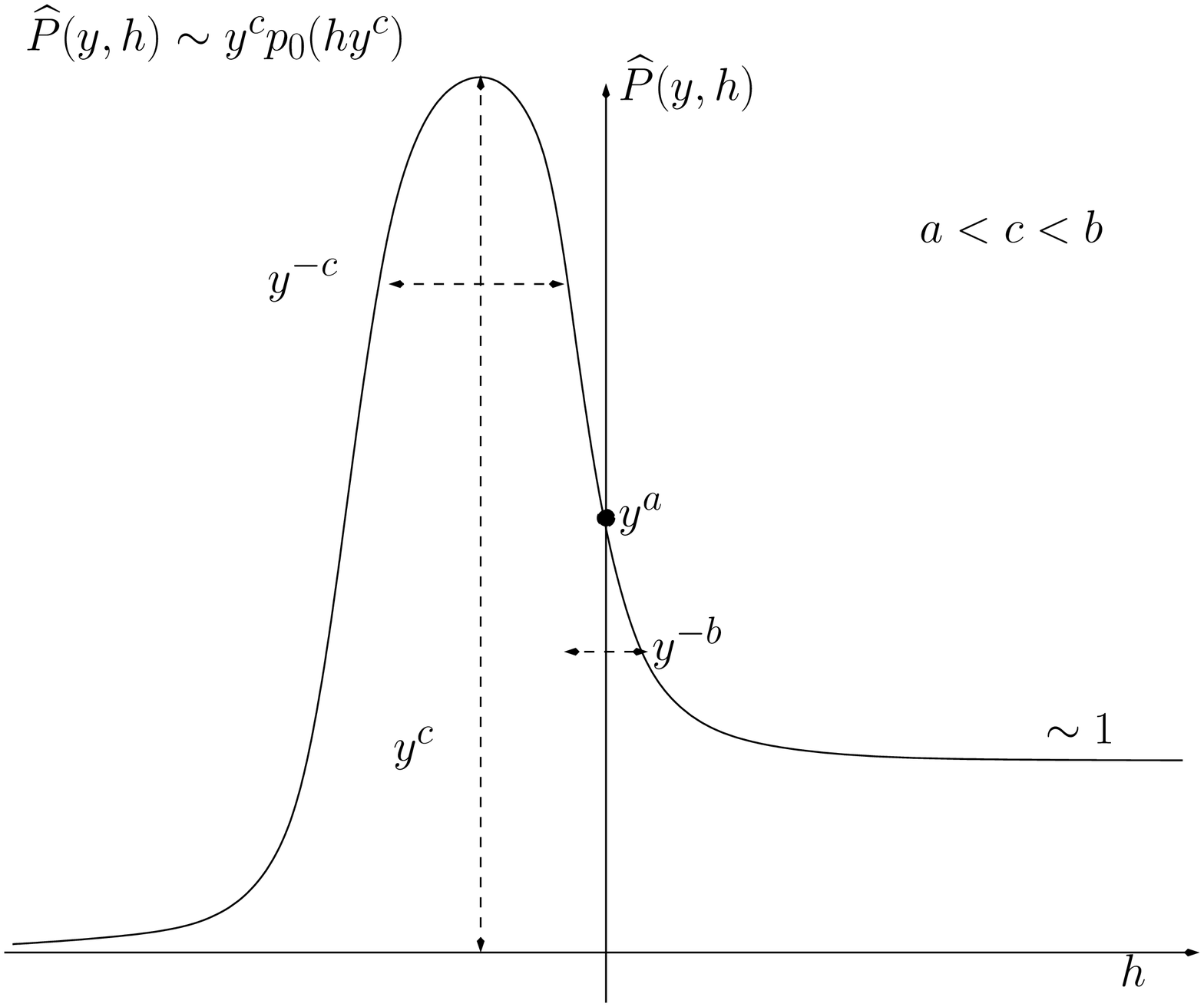}
\includegraphics[width=.45\textwidth]{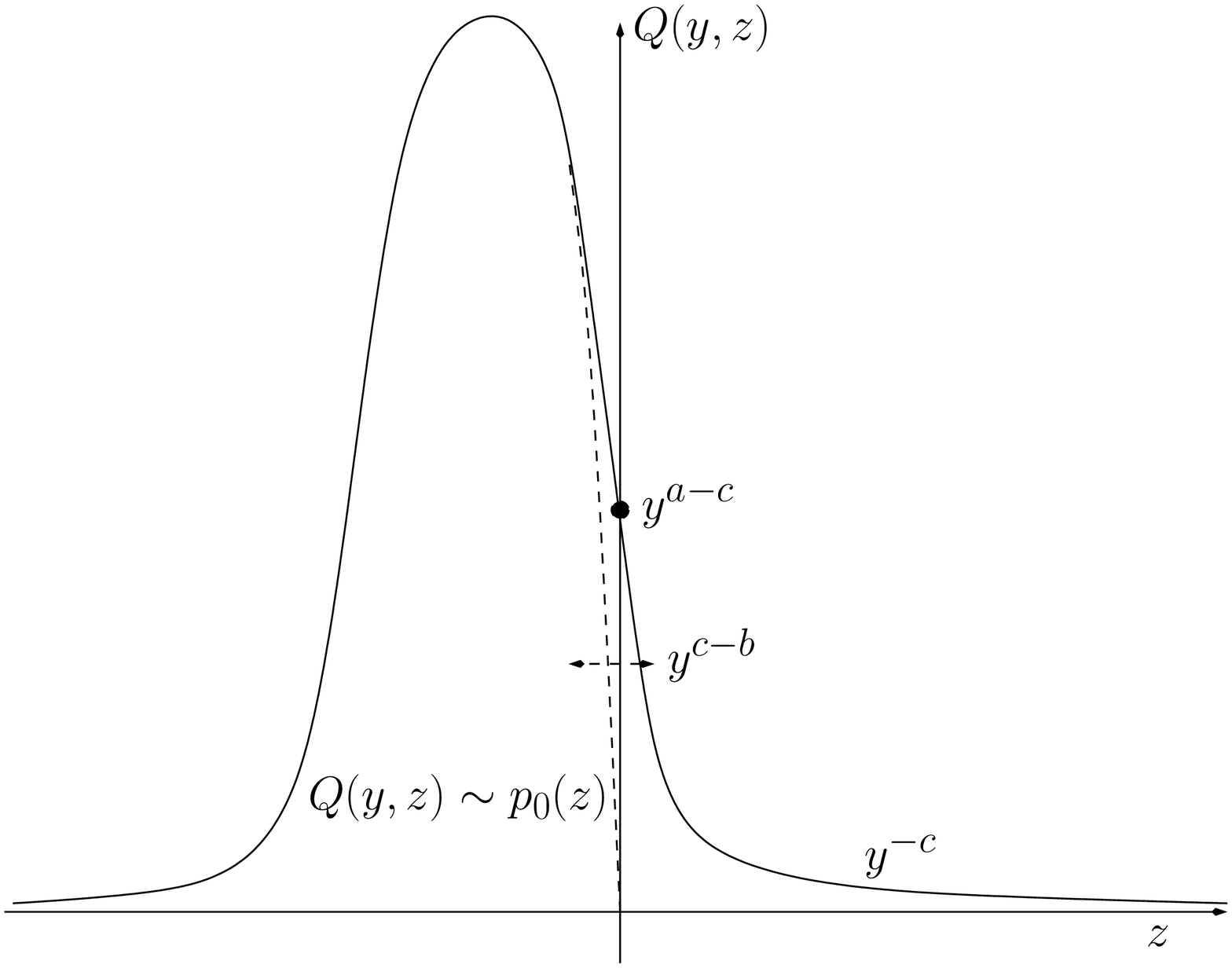}
\caption{
Scaling of $\wh P(y,h)$ and $Q(y,z)$.
}
\label{fig:Pscal}
\end{figure}

It is natural therefore to conjecture that
the decrease from the peak down to values of order 1 happens around $h\sim 0$
on another scale $|h| \approx y^{-b}$ with $b>c$, which matches between the behavior at $h<0$ and $h>0$.
We pose that in this regime $\wh P(y,h \sim 0) \sim y^a$ with $a < c$.  
In summary, we have
\beq\label{eq:Pscal}
\wh P(y,h)  \sim  \begin{cases}
y^{c} p_0( h y^{c} ) & \text{for } h \sim -y^{-c} \\
y^a p_1(h y^b) & \text{for } |h| \sim y^{-b} \\
p_2(h) & \text{for } h \gg y^{-b}
\end{cases}
\eeq
with the condition $a<c<b$, and this scaling is illustrated in Fig.~\ref{fig:Pscal}.

Assuming this scaling, we can match the different regimes. Note first that obviously the scaling~\eqref{eq:Pscal} requires that $p_0(z=0) =0$. We can assume
that $p_0(z) \sim |z|^\th$ for small $z$. Then, to match with $p_1(z)$, we must assume that $p_1(z\to-\io) \sim |z|^\th$ too. Matching requires that
$y^{c} | h y^{c} |^\th \sim y^a |h y^b|^\th$, therefore $c (1+\th) = a + b \th$, which implies $\th = \frac{c- a}{b -c}$.
Similarly, in order to match with the regime of positive $h$, we must have $p_1(z \to\io) \sim z^{-\alpha}$, and $y^a (h y^b)^{-\alpha} \sim O(1)$, from which
we obtain that $a - b\alpha =0$, hence $\alpha = a/b$, and $p_2(h) \sim h^{-\alpha}$ for $h\to 0$.
In summary, we obtain the following scaling relations between exponents\footnote{The reader should not confuse the exponent $\a$ introduced here
with the previously used matrix $\hat\a$.}:
\beq\label{eq:exprel}
\begin{split}
\a &= \frac{a}{b} \hskip70pt \th = \frac{c-a}{b-c} 
\hskip70pt 
\k = c + 1
\ . 
\end{split}\eeq
We will see later that the exponents $\a,\th,\k$ are directly related to the scaling of physical observables (the cage radius and the pair correlation function).

Eq.~\eqref{eq:Pscal} suggests to
define the scaled variable $z = h y^{c}$ and the
functions
\beq\begin{split}
H(y_i, z) = y_i \wh j(y_i, z y_i^{-c} ) \ ,
\hskip20pt &\Leftrightarrow \hskip20pt
\wh j(y_i, h) = \frac{1}{y_i} H(y_i, h y_i^{c}) \ , \\
Q(y_i,z) = y_i^{-c} \wh P(y_i , zy_i^{-c}) \ ,
\hskip20pt &\Leftrightarrow \hskip20pt
\wh P(y_i, h) = y_i^{c} Q(y_i, h y_i^{c}) \ .
\end{split}\eeq
In terms of $Q(y,z)$, the scaling~\eqref{eq:Pscal} becomes
\beq\label{eq:Qscal}
Q(y,z)  \sim  \begin{cases}
p_0( z ) & \text{for } z < 0 \ , \\
y^{a-c} p_1(z y^{b-c}) & \text{for } |z| \sim y^{-b+c} \ , \\
0 & \text{for } z > 0 \ .
\end{cases}
\eeq
A plot of $Q(y,z)$ is a scaled plot of $y^{-c} \wh P(y,h)$ versus $z = h y^{c}$;
this plot approaches a master function $p_0(z)$ at negative $z$, while around $z =0$ there is a region where $|z| \approx y^{-b+c}$,
in which $Q(y,z) \approx y^{a-c}$ is small, which
matches the scaling function $p_0(z)$ for negative $z$ with the
behavior $Q \sim y^{-c} \wh P(y,h) \to 0$ at positive $z$.

With this change of variable, the leading terms at large $y$ in 
the continuum equation for $Q(y,z)$ are
\beq\label{eq:contQ}
\begin{split}
\dot Q(y,z) &= 
-\frac{\dot\g(y)}2 y^{2c-1} Q''(y,z) 
- \left( \frac{c}{y} + \frac{\dot \g(y)}{\g(y)} \th(-z) \right) [ z Q(y,z)  ]'   
+ \dot\g(y) y^{2c - 1}  \left[  Q(y,z) H'(y,z) \right]' \ .
\end{split}\eeq

\subsection{A toy model}
\label{sec:IXC}

Before analyzing the complete fullRSB equations, we discuss here a toy model that gives a lot of insight about how the scaling of $Q(y,z)$ can
be studied analytically.
The toy model is obtained as a strong simplification of Eq.~\eqref{eq:contQ}, obtained by
{\it setting} $\g(y) = y^{-c}$ for all $y \in [1,\io)$, and $H=0$. 
We have $\dot \g(y) = -  c y^{-1-c}$ and $\dot\g(y)/\g(y) = -c/y$,
therefore we obtain
\beq\label{eq:toy}
\dot Q(y,z) = \frac{c}{2 y^{2-c}}  Q''(y,z) 
- \frac{c}{y} \th(z) [ z Q(y,z)  ]' \ ,
\eeq
and we keep a generic initial condition $Q(1,z) = Q_{\rm i}(z)$.
Eq.~\eqref{eq:toy} is a Fokker-Planck equation with diffusion and drift, and it corresponds to the Langevin equation
\beq
\frac{\de z}{\de y} = \frac{cz}{y} \th(z) + \frac{\sqrt{c}}{y^{(2-c)/2}} \eta(y)
\eeq
where $\eta(y)$ is a white noise with $\langle \eta(y) \eta(y') \rangle  = \d(y-y')$. The Langevin equation shows that on the $z>0$ side,
the random walkers drift towards $z=\io$, and for this reason $Q(y,z)\to 0$ for large $y$. Instead, for $z<0$ the equation corresponds
to a simple random walk with a diffusion coefficient that is reduced when $y$ grows.

\subsubsection{Scaling in the matching region}

The numerical study of Eq.~\eqref{eq:toy} is straightforward and shows that $Q(y,z)$ satisfies the scaling~\eqref{eq:Qscal}.
To obtain analytical insight on the scaling, we plug the {\it ansatz} $Q(y,z) = y^{a-c} p_1(z y^{b-c})$ in
Eq.~\eqref{eq:toy}; then we see that a non-trivial equation is obtained only if $b=(1+c)/2$, otherwise the diffusion term has a different scaling
from the other terms.
Calling $t = z y^{b-c} = h y^{b}$, with $b=(1+c)/2$,
one obtains a simple differential equation for the master function $p_1(t)$:
\beq\label{eq:toyp1}
\frac{c}2 p_1''(t) = \left(a-c \th(-t) \right) p_1(t) + \left( \frac{1+c}2 - c\th(-t) \right) t p_1'(t)   \ .
\eeq
This equation admits solutions with the correct asymptotic behavior of $p_1(t)$. In fact, one can show
that for $t\to \io$ there is a solution $p_1(t) \sim t^{-\a}$ with $\a = a/b$, and for $t\to -\io$ there is a solution $p_1(t) \sim |t|^\th$, with
$\th = (c-a)/(b-c) = 2(c-a)/(1-c)$. 
However, a solution that has {\it both} correct asymptotic behaviors does not exist for generic values of $a$. Imposing the existence of a solution with the correct
asymptotes therefore defines the value of $a$ as a function of $c$, which can be easily determined with very high precision 
by solving Eq.~\eqref{eq:toyp1} numerically
and using a bisection method to find the correct $a$.
As an example, we obtain $a(c=0.5)=  0.38923\ldots$ and $a(c=0.4)=0.29110\ldots$. A decent fit to have an idea of the trend is
$a(c) \approx 0.53 c + 0.49 c^2$, but of course one needs a much higher precision on $a$ to have the correct behavior of $p_1(t)$.
Note that $a$ is not a simple rational function of $c$,
because the condition that determines it is quite involved. 
The solution for $p_1(t)$ corresponding to this value of $a$ gives
the scaling function in the matching regime.

\begin{figure}[t]
\includegraphics[width=.45\textwidth]{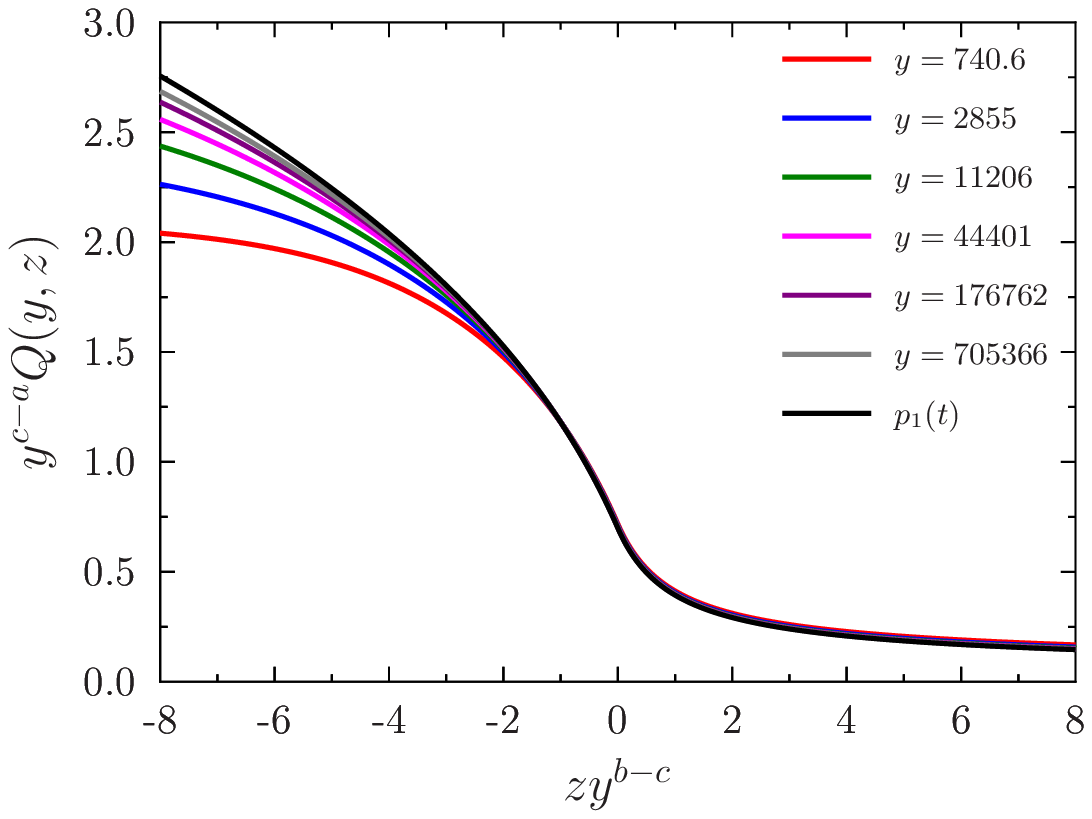}
\includegraphics[width=.45\textwidth]{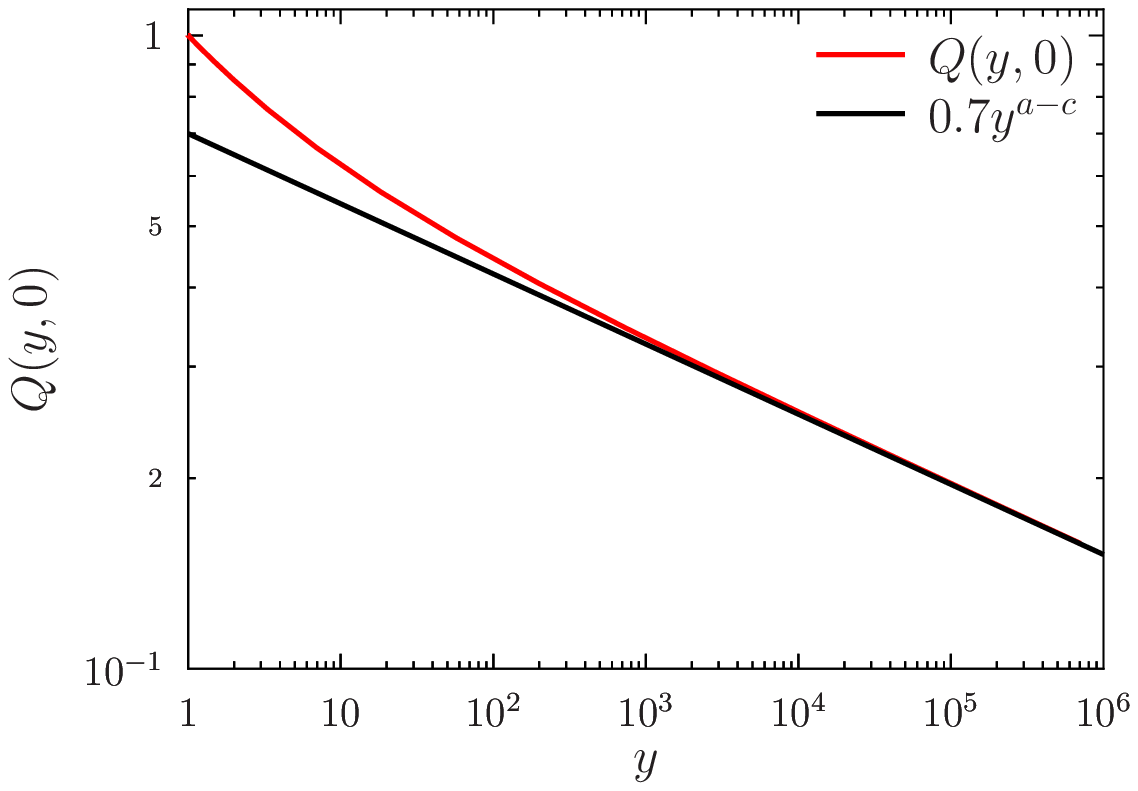}
\caption{
{\it (Left)}
Scaling plot of $y^{c-a} Q(y,z)$ with $c=1/2$, $b=3/4$ and $a = 0.38923\ldots$
versus $t = z y^{b-c}$ for different values of $y$ in the toy model Eq.~\eqref{eq:toy}, compared with the solution for $p_1(t)$ with the same $a$.
The agreement is excellent.
{\it (Right)}~Plot of $Q(y,z=0)$ versus $y$ which shows that $Q(y,z=0) \sim y^{a-c}$ as expected.
}
\label{fig:toy_scal}
\end{figure}

In Fig.~\ref{fig:toy_scal}
we compare the results of this scaling analysis with direct numerical resolution of Eq.~\eqref{eq:toy} for $c=1/2$. 
We chose initial condition $Q_{\rm i}(z) = 1$ and solved numerically Eq.~\eqref{eq:toy}. For large $y$, we observe
the appearance of a scaling regime around $z=0$.
Note that other initial conditions can be considered and the choice does not affect the scaling around $z=0$. 
In the scaling regime, we find excellent agreement with the predictions of the scaling analysis presented above,
as discussed in the caption of Fig.~\ref{fig:toy_scal}.

\subsubsection{Computation of $p_0(z)$}

We know that $Q(y\to\io,z) = p_0(z)$ for $z<0$, while $Q(y\to\io ,z)$ vanishes for $z>0$.
To compute the function $p_0(z)$ it is convenient to change variables to $\t = 1 - y^{c-1}$, and write Eq.~\eqref{eq:toy} for $z<0$ as
\beq\label{eq:toytau}
\frac{\partial Q(\t,z)}{\partial \t} = \frac{c}{2(1-c)}  Q''(\t,z) = \frac{D}2 Q''(\t,z)  \ ,
\eeq
which is the standard diffusion equation. The corresponding Green's function is
\beq
G(\t,z-z') = \frac1{\sqrt{2\pi D \t}} e^{-\frac{(z-z')^2}{2 D\t}} \ ,
\eeq
with $G(\t=0,z-z') = \d(z-z')$.
We look for a solution of Eq.~\eqref{eq:toytau} in the region $z<0$.
We start by constructing a solution that satisfies the boundary condition $Q(\t=0,z) = Q_{\rm i}(z)$ and
$Q(\t,z=0) = 0$. It has the form
\beq
Q_{\rm reg}(\t,z) = \int_{-\io}^0 \de z' \left[ G(\t,z-z') - G(\t,z+z') \right]  Q_{\rm i}(z') \ .
\eeq
The regions $z<0$ and $z>0$ admit different solutions, and we know that $Q(y,z=0) \sim y^{a-c}$ hence
$Q(\t,z=0) \sim (1-\t)^{(c-a)/(1-c)} = (1-\t)^{\th/2}$ using the relations $b=(1+c)/2$ and $\th = (c-a)/(b-c) = 2 (c-a)/(1-c)$.
Therefore, we should add a second contribution to $Q_{\rm reg}(\t,z)$ in order to match the boundary condition at $z=0$.
This contribution can be chosen, for $z<0$, as
\beq\begin{split}
Q_{\rm sing}(\t,z) &= \int_0^\t \de t \, G(\t-t,z) B(1-t) =
\int_0^\t \de t \frac{B(1-t)}{\sqrt{2\pi D (\t - t)}} e^{-\frac{z^2}{2D(\t-t)}}  \\ 
&  = \int_{1-\t}^1 \de s \frac{B(s)}{\sqrt{2\pi D [s - (1-\t)]}} e^{-\frac{z^2}{2 D[s - (1-\t)]}} 
 = \int_{1-\t}^1 \de s  \frac{s^{(\th-1)/2} \BB(s)}{\sqrt{2\pi D [s - (1-\t)]}}   e^{-\frac{z^2}{2D[s - (1-\t)]}} 
 \ .
\end{split}\eeq
In fact, it is a solution of Eq.~\eqref{eq:toytau}, such that $Q_{\rm sing}(\t=0,z)=0$, hence it does not affect the initial condition in $\t=0$.

We want to show that choosing $B(s) = s^{(\th-1)/2} \BB(s)$, where $\BB(s) = \BB_0 +  \BB_1 s + \cdots$ is an analytic function of $s$,
such that
\beq\label{eq:appQsing}
Q_{\rm sing}(\t=1,z=0) = \int_{0}^1 \frac{\de s}{\sqrt{2\pi D} s^{1 - \th/2}} \BB(s) = 0 \ ,
\eeq
provides a solution with the correct scaling $Q(\t,z=0) \sim (1-\t)^{\th/2}$. 
The function $\BB(s)$ should be determined from the matching with the solution at $z>0$, 
but this is not relevant for the rest of this discussion\footnote{
One can choose for illustrative purposes $\BB(s) = - \BB_0 [ 1 - s (2+\th)/\th]$, which has the required properties.
}.
In fact, for $z=0$ we have at leading order for small $\ee=1-\t$, using Eq.~\eqref{eq:appQsing} 
\beq\begin{split}
Q_{\rm sing}(\t,z=0) &= \int_{\ee}^1 \de s  \frac{s^{(\th-1)/2} \BB(s)}{\sqrt{2\pi D [s - \ee]}}  
= \int_{0}^{1-\ee} \de s  \frac{(s+\ee)^{(\th-1)/2} \BB(s+\ee)}{\sqrt{2\pi D s}}
\sim \int_{0}^{1} \de s  \frac{(s+\ee)^{(\th-1)/2} \BB(s)}{\sqrt{2\pi D s}} \\
&=  \int_{0}^{1} \de s  \frac{\BB(s)}{\sqrt{2\pi  D s}} \left[ (s+\ee)^{(\th-1)/2} - s^{(\th-1)/2} \right] 
\sim \int_{0}^{1} \de s  \frac{\BB_0 + \BB_1 s + \cdots}{\sqrt{2\pi D} s^{1 - \th/2}} \left[ (1+\ee/s)^{(\th-1)/2} - 1 \right] \\
& \sim \BB_0 \ee^{\th/2} \int_{0}^{1/\ee} \frac{\de u}{\sqrt{2\pi D} u^{1 - \th/2}} \left[ (1+1/u)^{(\th-1)/2} - 1 \right] \propto (1-\t)^{\th/2}  \ ,
\end{split}\eeq
and repeating the same argument at vanishingly small $z$ we obtain
\beq\begin{split}
Q_{\rm sing}(\t,z) & \sim \int_{0}^{1} \de s  \frac{\BB_0 + \BB_1 s + \cdots}{\sqrt{2\pi D} s^{1 - \th/2}} \left[ (1+\ee/s)^{(\th-1)/2} - 1 \right] e^{-\frac{z^2}{2D s}}
\\
& \sim \BB_0 \ee^{\th/2} \int_{0}^{1/\ee} \frac{\de u}{\sqrt{2\pi D} u^{1 - \th/2}} \left[ (1+1/u)^{(\th-1)/2} - 1 \right] e^{-\frac{z^2}{2 Du \ee}}
= (1-\t)^{\th/2} q[ z/(1-\t)^{1/2} ] \ ,
\end{split}\eeq
which translated back to the variable $y$ implies, recalling that $b-c = (1-c)/2$,
\beq
Q_{\rm sing} = y^{a-c} q[ z y^{b-c} ] \ ,
\eeq
as expected from Eq.~\eqref{eq:Qscal}.
Finally,  using again Eq.~\eqref{eq:appQsing}, we have that for small $z$
\beq\begin{split}
Q_{\rm sing}(\t=1,z) &= \int_{0}^1 \de s  \frac{s^{(\th-1)/2} \BB(s)}{\sqrt{2\pi D s}}   e^{-\frac{z^2}{2Ds}} = 
\int_{0}^1 \frac{\de s}{\sqrt{2\pi D} s^{1 - \th/2}}  \left[ e^{-\frac{z^2}{2Ds}}  - 1 \right] (\BB_0 + \BB_1 s + \cdots) \\
& \sim \BB_0   z^\th \int_{0}^{1/z^2} \frac{\de u}{\sqrt{2\pi D} u^{1 - \th/2}}  \left[ e^{-\frac{1}{2Du}}  - 1 \right] \ .
\end{split}\eeq

We conclude that, even if the function $\BB(s)$ remains undetermined by this analysis, the resulting function 
$Q(\t,z) = Q_{\rm reg}(\t,z) + Q_{\rm sing}(\t,z)$ has all the correct scaling properties. 
The resulting fuction $p_0(z)$ is given by
\beq
p_0(z) = Q(\t=1,z) = 
\int_{-\io}^0 \de z' \left[ G(1,z-z') - G(1,z+z') \right]  Q_{\rm i}(z') + 
\int_{0}^1 \de s  \frac{s^{(\th-1)/2} \BB(s)}{\sqrt{2\pi D s}}   e^{-\frac{z^2}{2Ds}} \ ,
\eeq
which shows that {\it the function $p_0(z)$ depends on non-universal details of the evolution of $Q(y,z)$}, such as the initial
condition $Q_{\rm i}(z)$ and the function $\BB(s)$. 

\subsubsection{Summary}

From the analysis of the toy model Eq.~\eqref{eq:toy}, we obtain several important informations:
\begin{itemize}
\item
In the matching regime, the scaling is controlled by a universal function $p_1(t)$, solution of Eq.~\eqref{eq:toyp1}, with exponents
$b=(1+c)/2$ and $a(c)$ determined by the condition that $p_1(t)$ has the correct asymptotic behavior. However, the exponent $c$ remains undertermined
by this analysis, with the only condition that $a<c<b$ which implies $c\in [0,1]$.
\item 
There is no hope of computing $p_0(z)$ from scaling arguments, because this function depends on non-universal quantities
that contain information on the whole evolution of $Q(y,z)$. Hence we have to obtain it through the full numerical solution of the equation for $Q(y,z)$.
\end{itemize}
These informations are very useful to understand the complete fullRSB equations.

\subsection{Scaling of the functions $\wh j(y,h)$ and $\wh P(y,h)$ in the matching region}
\label{sec:IXD}

Using the analogy with the toy model investigated in the previous section, we can start to discuss
the asymptotic scaling of the solution of the complete fullRSB equations.
We consider first the equation for $\wh j(y,h)$. For $m=0$, its initial condition for $y=\io$ becomes $\wh j(y,h)=0$, see
Eq.~\eqref{eq:Sscaledfinal}.
One can show from Eq.~\eqref{eq:Sscaledfinal} that
\beq
\wh j(y_i,h\to -\io) = \sum_{j=i}^{k-1} \frac{1}{2 y_j} \log(\wh\g_{j+1}/\wh \g_j) \ ,
\eeq
and in the continuum limit
\beq
\wh j(y,h\to-\io) = \int_y^\io \frac{\de u}{2u} \frac{\dot \g(u)}{\g(u)} \ ;
\eeq
this relation can also be derived directly from Eq.~\eqref{eq:Sscaledfinal_cont}.
Similarly, one can easily show that $\wh j(y,h\to\io) = 0$ for all~$y$.

At large $y$, under the assumption that $\g(y) \sim \g_\io y^{-c}$, we then have
$\wh j(y,h\to -\io) \sim -c/(2 y)$. We therefore conjecture the scaling form
\beq\label{eq:Jscal}
\wh j(y,h) = -\frac{c}{2y} J( h y^b /\sqrt{\g_\io} ) \ ,
\hskip20pt 
H(y,z) = -\frac{c}2 J(z y^{b-c} /\sqrt{\g_\io} ) \ ,
\eeq
and we insert it in Eq.~\eqref{eq:Sscaledfinal_cont} for $\wh j(y,h)$.
As in the case of the toy model, the only choice that leads to a non-trivial equation
is $b=(1+c)/2$, which leads to the following equation for $J(t)$ with $t = h y^{b}/\sqrt{\g_\io} =z y^{b-c}/\sqrt{\g_\io} $:
\beq\label{eq:J}
 \frac{c}2 J''(t) =  t J'(t) \left(-\frac{1+c}2 + c \th(-t) \right) + J(t) - \th(-t) + \frac{c^2}{4} J'(t)^2  \ .
 \eeq
This equation must be solved with boundary conditions $J(-\io)=1$ and $J(\io)=0$, 
which can be done easily by a shooting method and leads to a unique solution for each value of $c$.

We consider next the scaling of $\wh P(y,h)$ 
in the intermediate scaling regime that matches around $h=0$.
According to Eq.~\eqref{eq:Pscal} and \eqref{eq:Qscal}, we have 
$\wh P(y,h) = y^a p_1(h y^b/\sqrt{\g_\io})$ and $Q(y,z) = y^{a-c} p_1(z y^{b-c}/\sqrt{\g_\io})$, and
we added here the term $\sqrt{\g_\io}$ because in this way the dependence on $\g_\io$ disappears from
the scaling equations.
We choose again $b=(1+c)/2$, because, as in the case of the toy model, 
it is the only choice that leads to a non-trivial equation for $p_1(t)$,
that depends in a non-trivial way on the function $J(t)$ that varies on the same scale. 
If we call once again
$t = z y^{b-c}/\sqrt{\g_\io} = h y^{b}/\sqrt{\g_\io}$,
we plug the scaling form of $Q(y,z)$ in Eq.~\eqref{eq:contQ}, and we use Eq.~\eqref{eq:Jscal},
then we obtain, at the leading order
\beq\label{eq:p1}
\frac{c}2 p_1''(t) = \left(a-c \th(-t) \right) p_1(t) + \left( \frac{1+c}2 - c\th(-t) \right) t p_1'(t)  
 - \frac{c^2}2 \left[ p_1'(t) J'(t) + p_1(t) J''(t) \right]  \ ,
\eeq
which correctly coincides with Eq.~\eqref{eq:toyp1} of the toy model if one chooses $J=0$.
Note that for $|t|\to\io$, $J' \sim J'' \to 0$, therefore one can repeat
the same analysis as for  Eq.~\eqref{eq:toyp1} to show that Eq.~\eqref{eq:p1}
admits solutions with the correct asymptotic behavior of $p_1(t)$, namely $p_1(t \to\io) \sim t^{-\a}$ with $\a = a/b$, 
and $p_1(t\to-\io) \sim |t|^\th$, with
$\th = (c-a)/(b-c)$. 
Like for Eq.~\eqref{eq:toyp1}, a solution that has both correct asymptotic behaviors exists only for a specific value of $a$, that can be determined
through a bisection method. 

In summary, for given $c$ and $b = (1+c)/2$, we have to solve the two Eqs.~\eqref{eq:J} and \eqref{eq:p1}
with their appropriate boundary condition. As in the toy model, this fixes the value of $a$ as a function of $c$.
We find that the presence of $J \neq 0$ is only a small perturbation of Eq.~\eqref{eq:p1} with respect to Eq.~\eqref{eq:toyp1},
and the resulting values of $a$ are just slightly smaller than the ones of the toy model. 

\subsection{Determination of the critical exponents}
\label{sec:IXE}

Up to now the exponent $c$ has remained undetermined. In this section, we derive a condition that allows us to determine
$c$ and from it $a$ and $b$, hence obtaining analytical predictions for all the critical exponents.

\subsubsection{Some exact relations}

We start by deriving a few useful exact relations, following~\cite{Ri13}.
To do so, we introduce
\beq
\wt f(y,h)=\g(y) \wh f(y,h) = -\frac{h^2 \th(-h)}2 + \g(y) \wh j(y,h) 
\eeq
and we consider the equation for $\k(y)$ in Eq.~\eqref{eq:Sscaledfinal_cont}, which is:
\beq\label{eq:kdiy}
\k(y)=\frac{\widehat\varphi}{2\gamma^2(y)}\int_{-\infty}^\infty\de h\, e^h \wh P(y,h) \wt f'(y,h)^2 \ .
\eeq
We take the derivative with respect to $y$, and using the equation of motion~\eqref{eq:Sscaledfinal_cont}
for $\wh P(y,h)$ and $\wh f(y,h)$ we obtain
\beq
\dot \k(y)=-\frac{\widehat \varphi \dot \g(y)}{2 y \g^2(y)}\int_{-\infty}^{\infty}\de h \, e^h \wh P(y,h)
\wt f''(y,h)^2 \ .
\eeq
From the relation between $\k(y)$ and $\g(y)$ we have
$\dot \k(y)=-\frac{\dot \g(y)}{y \g^2(y)}$,
so we obtain the exact expression
\beq\label{exact_rel}
1=\frac{\wh\varphi}{2}\int_{-\infty}^\infty\de h\, e^h \wh P(y,h) \wt f''(y,h)^2 \ .
\eeq
Note that to derive Eq.~\eqref{exact_rel} we have assumed that $\dot\k(y) \neq 0$ in a finite interval of $y$.
Hence, for any finite $k$RSB solution, Eq.~\eqref{exact_rel} does not hold because $\dot\k(y)=0$ except in
a finite number of isolated points.
For a fullRSB solution, instead, 
Eq.~\eqref{exact_rel} holds for all $y$. Indeed, if $\dot\k(y)=0$ then $\wh P(y,h)$ and $\wt f''(y,h)$ do not
depend on $y$, hence if Eq.~\eqref{exact_rel} holds in the region where $\dot\k(y) \neq 0$, it must also hold
(trivially) where $\dot\k(y)=0$.

If we derive once more Eq.~\eqref{exact_rel} with respect to $y$, we obtain (using once more the equations of motion)
\beq
0=\int_{-\infty}^\infty \de h\, e^h \wh P(y,h)\left[2\left( \wt f''(y,h)^2+ \wt f''(y,h)^3\right)-\frac{\g(y)}y \wt f'''(y,h)^2 \right] \ ,
\eeq
from which it follows that
\beq\label{exact_rel2}
y = \frac{\g(y)}2 \frac{\int_{-\infty}^\infty \de h\, e^h \wh P(y,h)\, \wt f'''(y,h)^2}{ \int_{-\infty}^\infty \de h\, e^h \wh P(y,h)\left( \wt f''(y,h)^2+ \wt f''(y,h)^3\right)} \ .
\eeq
Note that Eq.~\eqref{exact_rel2} can only hold in the region where $\dot\k(y) \neq 0$; this is obvious because the
right hand side is constant when $\dot\k(y)=0$.

\subsubsection{Scaling regime}

These equations have important consequences in the scaling regime. 
First of all, Eq.~\eqref{exact_rel} is a kind of normalization condition for $\wh P(y,h)$. In the limit $y\to\io$, at the leading
order we have $\wt f''(y,h) = -\th(-h)$, and using Eq.~\eqref{eq:Pscal} one can show that Eq.~\eqref{exact_rel} becomes
\beq\label{eq:normp0}
1=  \frac{\wh\f}2 \int_{-\infty}^{0} \de z \, p_0(z) \ ,
\eeq
which will be crucial, in the following, to obtain isostaticity. Also, in the same regime $\wt f'(y,h) = - h \th(-h)$ and
Eq.~\eqref{eq:kdiy} gives
\beq
\k_\io = \k(y \to \io) =\frac{\widehat\varphi}{2\gamma_\io^2 }\int_{-\infty}^0 \de z \, p_0(z) z^2 \ .
\eeq

Eq.~\eqref{exact_rel2}, instead, receives non-vanishing contributions only from the matching regime, both in the numerator
and in the denominator, as one can check that the other contributions vanish.
In the matching regime for large $y$, using $\g(y) \sim \g_\io y^{-c}$ and Eq.~\eqref{eq:Jscal}, one can show easily that
\beq\label{wtf2}
\begin{split}
&\wt f''(y,h) = - \th(-t) - \frac{c}{2} J''(t) \ ,  \\
&\wt f'''(y,h) = \frac{y^{(1+c)/2}}{\sqrt{\g_\io}} \frac{\de}{\de t} \left[ - \th(-t) - \frac{c}{2} J''(t) \right] \ .
\end{split}\eeq
Plugging this in Eq.~\eqref{exact_rel2}, we get
\beq\label{eq:cond_c}
\frac12 = \frac{
\int_{-\io}^\io \de t \, p_1(t) \left[ \th(-t) + \frac{c}2 J''(t) \right]^2 \left[ \th(t) - \frac{c}2 J''(t) \right]
}{
\int_{-\io}^\io \de t \, p_1(t) \left[ \frac{\de}{\de t} \left( \th(-t) + \frac{c}2 J''(t) \right) \right]^2
} \ .
\eeq
Now, remember that for a given $c$, we can determine $a$, $b$, $J(t)$ and $p_1(t)$ through Eqs.~\eqref{eq:J} and \eqref{eq:p1}.
Hence, Eq.~\eqref{eq:cond_c} becomes a condition on the exponent $c$. Solving this condition numerically, 
and also using Eq.~\eqref{eq:exprel},
we obtain the values
of the exponents:
\beq\label{eq:exp_num}
\begin{split}
&a=0.29213\ldots \hskip30pt b=0.70787\ldots
\hskip30pt 
c=0.41574\ldots
\ , \\
&\a = 0.41269\ldots \hskip30pt \th = 0.42311\ldots
\hskip30pt 
\k = 1.41574\ldots
\ . 
\end{split}\eeq
The precision of the determination of these exponents depends on the cutoffs that are used to discretize Eqs.~\eqref{eq:J} and \eqref{eq:p1}:
we estimate (conservatively) that the error in Eq.~\eqref{eq:exp_num} is smaller than $\pm 1$ on the last reported digit.
These are our analytic predictions for the scaling exponents, and they complete the analysis of the scaling regime of
Eq.~\eqref{eq:Sscaledfinal_cont}. Note that within our error we find that $a=1-b$, which, together with $b=(1+c)/2$, implies the relation
$\a = 1/(2+\th)$ that has been derived in~\cite{Wy12} using scaling arguments. We will come back on this point later.
In the next sections we test the correctness of our scaling analysis, and
we relate these exponents to observable quantities.

\section{Numerical test of the critical scaling of the fullRSB solution}

Having obtained analytical results for the exponents
that characterize the asymptotic scaling of the fullRSB solution at jamming, 
we now solve the fullRSB numerically to test them and check the pre-asymptotic corrections.

We consider Eq.~\eqref{eq:Sscaledfinal} at $m=0$, which means that $y_k = \io$ and $\wh j(y_k,h) =0$, 
and we solve the recurrence equations numerically. We use here a
different code than in the 2RSB computation, which is not optimized to work at large densities, hence it does not make use
of the decomposition~\eqref{eq:IIan}. This code just solves the equations by iterating them,
carefully taking into account
the behavior of the various functions for $h\to\pm\io$. The code can work for any number $k$ of RSB. Note that even if $m=0$,  
numerically solving the equations necessitates working at finite $k$, hence
we effectively introduce a cutoff $y_{k-1} = y_{\rm max}$ which is akin to a finite $m$ (we will come back to this point later).
To study the fullRSB solution at $m=0$ we therefore have to set $y_{\rm max}$ and $k$ to be as large as possible. 
Cutoffs are also used to discretize the integrals, but we checked that the results we report 
are independent of their choice so we do not further discuss this issue below.

In the following, all numerical results are obtained for $\wh\f=10$,
which is a value large enough that we are sure to be above the threshold, but low enough that
our code works well.

\subsection{Dependence on the cutoffs $k$ and $y_{\rm max}$}

\begin{figure}[t]
\includegraphics[width=.45\textwidth]{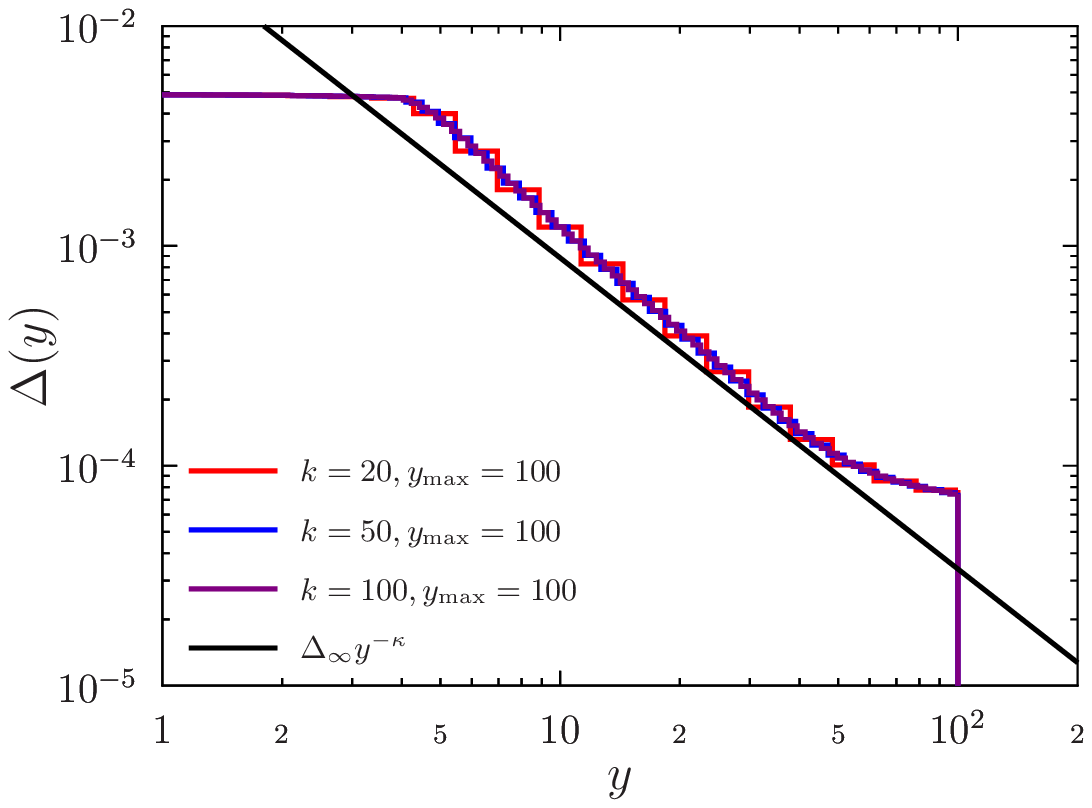}
\includegraphics[width=.45\textwidth]{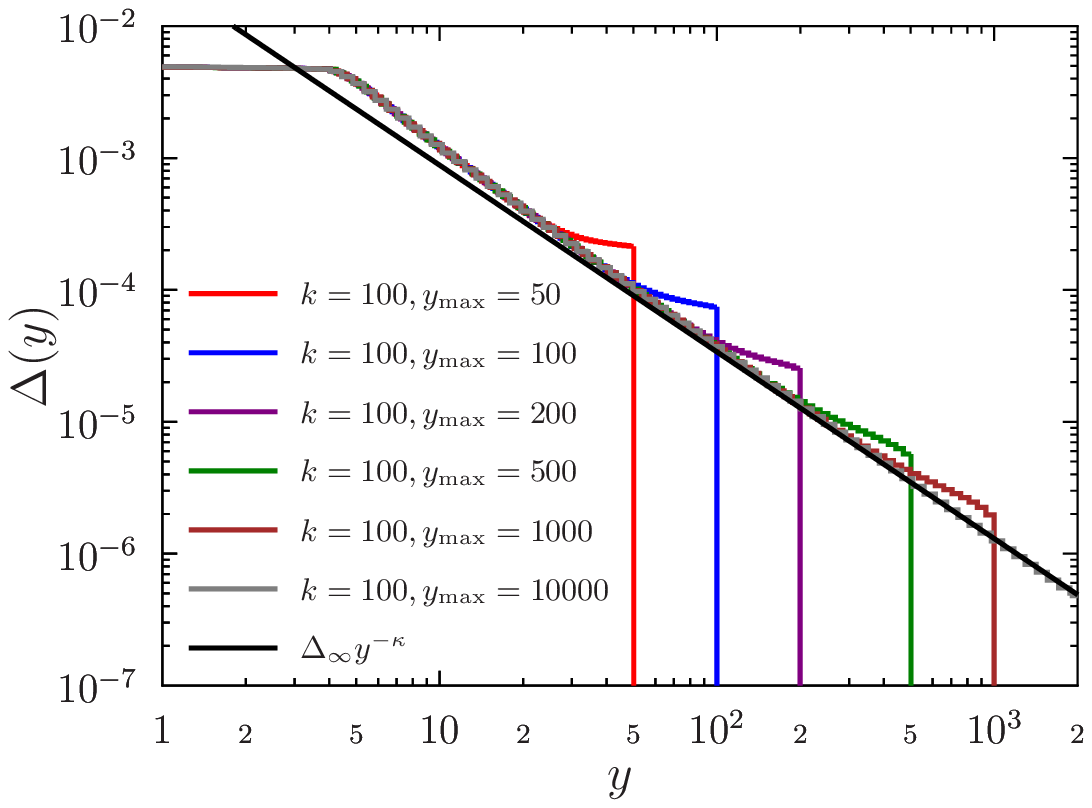}
\caption{
The function $\D(y)$ at $m=0$ and $\wh\f=10$. In left panel we report results for fixed $y_{\rm max}=100$ and $k=20, 50, 100$, which show
that $k=100$ is large enough to be considered infinite. In right panel we report results for fixed $k=100$ and $y_{\rm max}=50,100,200$,
which show that for large $y_{\rm max}$ the cutoff at large $y$ disappears. The power law regime 
$\D(y) \sim \D_\io y^{-\k}$ with $\D_\io \approx 0.023$ and $\k$ given in Eq.~\eqref{eq:exp_num} 
is approached at large $y$.
}
\label{fig:Deltafull}
\end{figure}

We start our discussion with a brief study of the dependence of the function $\D(y)$ on the cutoffs $k$ and $y_{\rm max}$.
For our numerical studies, we chose logarithmically spaced $y_i$, between $y_1=1$ and $y_{k-1} = y_{\rm max}$. 
In Fig.~\ref{fig:Deltafull} results for several
$k$ at fixed $y_{\rm max}$, and for fixed $k$ at several $y_{\rm max}$, show that we can reach the limit where 
both $k$ and $y_{\rm max}$ can be considered as infinite.
Clearly, $\D(y)$ is a non-trivial continuous function of $y$, 
which confirms that we are in a fullRSB phase, as we conjectured at the beginning of Sec.~\ref{sec:scaling}.
Our results are also perfectly compatible with the expected behavior at large $y$, $\D(y) \sim \D_\io y^{-\k}$ with $\k$ given in
Eq.~\eqref{eq:exp_num}. A deviation is observed for values of $y$ close to the cutoff at $y_{\rm max}$, but the value of $y$
at which we observe the deviation grows proportionally to $y_{\rm max}$.
This observation will be important later. Note that in the region where this deviation is observed, we expect that $\D(y)$ should
tend to be approximately constant. The reason why this is not the case is that the convergence of our code to the fixed point
of Eq.~\eqref{eq:Sscaledfinal} is very slow in that region. 
We could perform more iterations to observe full convergence but we did not do so, because this calculation is computationally hard and because the behavior of this cutoff-dependent region is irrelevant for our analysis.

\subsection{Test of the critical exponents}

\begin{figure}[h]
\includegraphics[width=.45\textwidth]{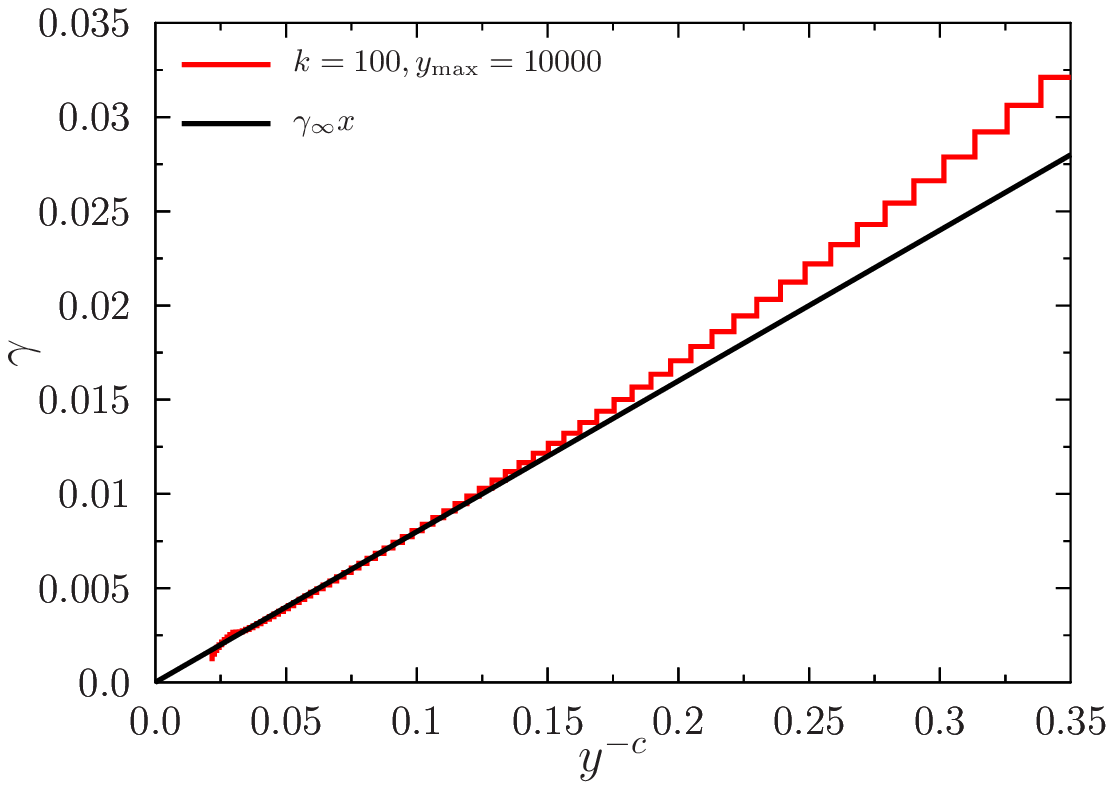}
\includegraphics[width=.45\textwidth]{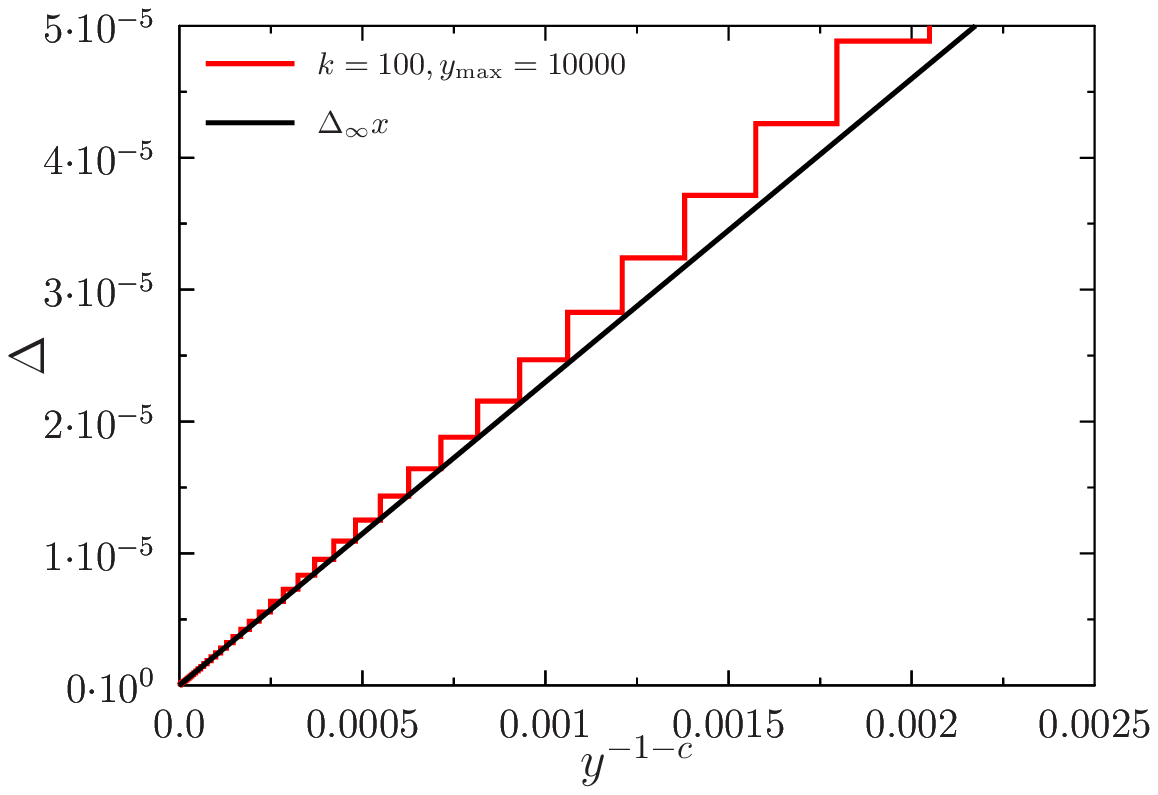}
\caption{
Plot of $\g$ versus $y^{-c}$ ({\it left}), and of $\D$ versus $y^{-1-c}$ ({\it right}), with $c$ given in Eq.~\eqref{eq:exp_num}. 
The linear fits give $\g_\io \approx 0.080$ and $\D_\io = \frac{c}{c+1} \g_\io \approx 0.023$.
}
\label{fig:gammascal}
\end{figure}

We now perform a more detailed test of the critical scaling derived in Sec.~\ref{sec:scaling} using the data with the largest
available cutoff, $y_{\rm max}=10000$, and $k=100$.
In Fig.~\ref{fig:gammascal} we plot $\g$ as a function of $y^{-c}$ and $\D$ as a function of $y^{-c-1}$; 
the plots are linear at large $y$ and provide an estimate
of $\g_\io \approx 0.080$ and $\D_\io = \frac{c}{c+1} \g_\io \approx 0.023$, which is also perfectly compatible with Fig.~\ref{fig:Deltafull}.

\begin{figure}[t]
\includegraphics[width=.5\textwidth]{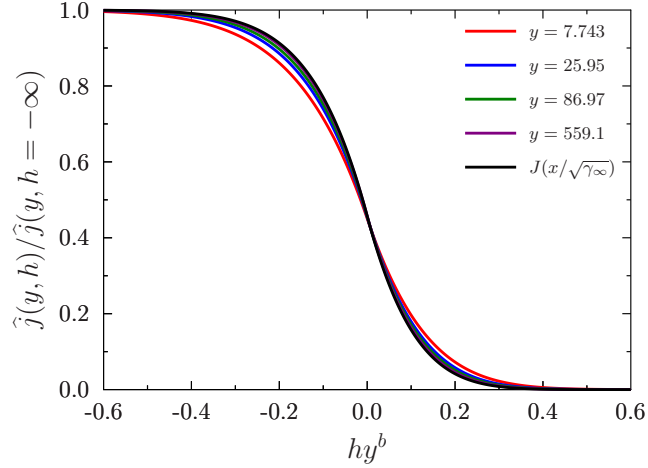}
\caption{
Scaling plot of the function $\wh j(y,h)$ according to Eq.~\eqref{eq:Jscal}. 
We plot $\wh j(y,h)/\wh j(y,h\to-\io)$ versus $h y^b$, with $b$ given in Eq.~\eqref{eq:exp_num}. We also report the (properly scaled) 
solution $J(z)$ of Eq.~\eqref{eq:J} corresponding to the choice of exponents in Eq.~\eqref{eq:exp_num}.
On increasing $y$ the curves approach the master function.
}
\label{fig:Jscal}
\end{figure}

\begin{figure}[t]
\includegraphics[width=.45\textwidth]{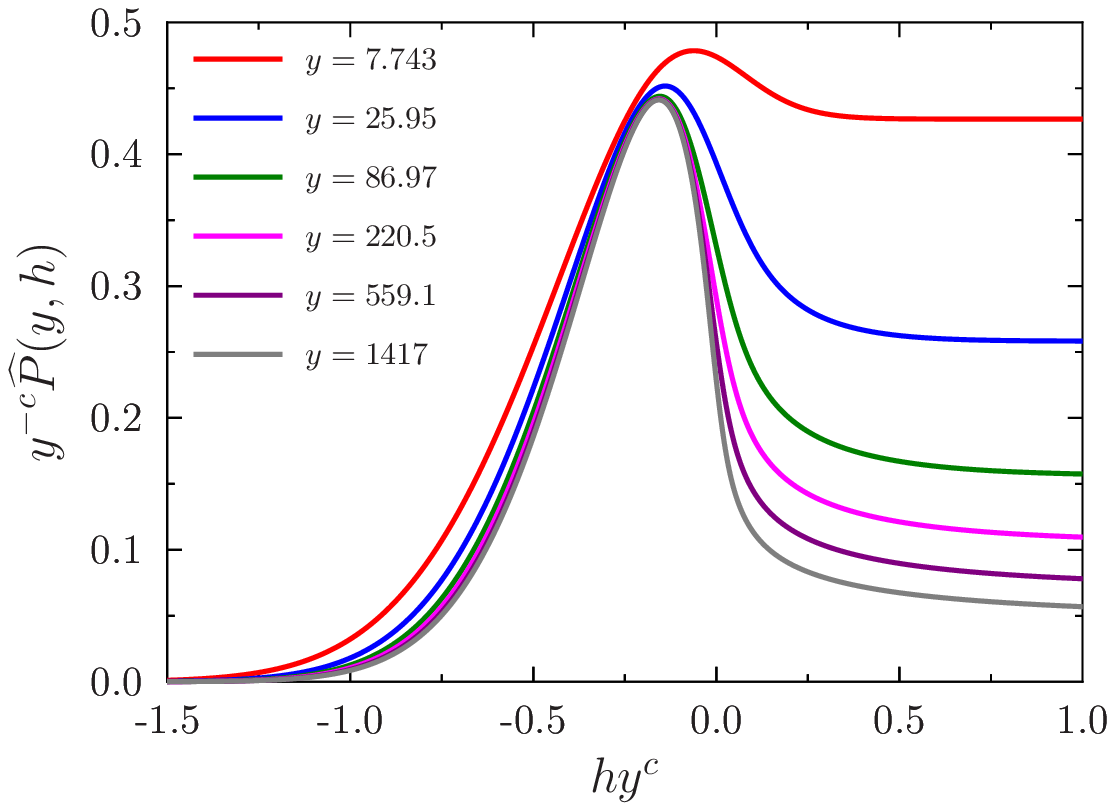}
\includegraphics[width=.45\textwidth]{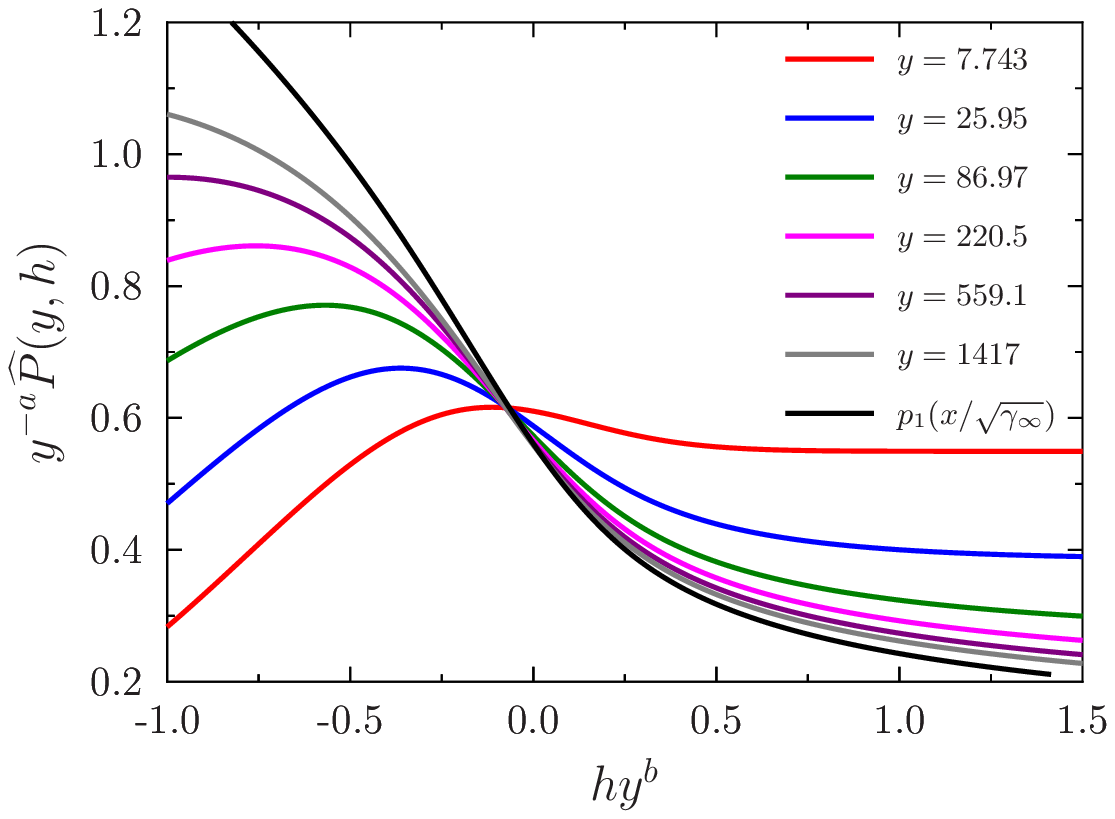}
\caption{
Test of Eq.~\eqref{eq:Pscal} and \eqref{eq:Qscal}, 
with exponents given by Eq.~\eqref{eq:exp_num}. 
{\it (Left)} Convergence to the function $Q(y,z)$ to $p_0(z)$. 
{\it (Right)} Scaling in the region around $h=0$ where 
$\wh P(y,h) \sim y^a p_1(h y^b)$, compared with the solution $p_1(z)$ of Eq.~\eqref{eq:p1}.
}
\label{fig:Pscalnum}
\end{figure}

Having an estimate of $\g_\io$, we can test the scaling of Eq.~\eqref{eq:Jscal}, see Fig.~\ref{fig:Jscal}, and
the scaling in Eq.~\eqref{eq:Pscal}, see Fig.~\ref{fig:Pscalnum}. In both cases we find 
excellent agreement between the analytical
results of Sec.~\ref{sec:scaling} and the numerical solution of the equations.

\section{Critical scaling of physical observables}

We have confirmed that the asymptotic solution of Eqs.~\eqref{eq:Sscaledfinal} and \eqref{eq:Sscaledfinal_cont} found
in Sec.~\ref{sec:scaling} is realized by the numerical solution of the fullRSB equations.
We now start to investigate the physical consequences, in particular to identify observables that display critical scaling
controlled by the exponents in Eq.~\eqref{eq:exp_num}.

In Sec.~\ref{sec:Deltap} we show that the exponent $\k$ is related to the scaling of the cage radius with pressure.
In Sec.~\ref{sec:paircorr} we discuss the scaling of the pair correlation function of the glass on approaching jamming
and we show that it is determined by the exponents $\a$ and $\th$. 
In Sec.~\ref{sec:forcedis} we show that the distribution of forces in the packing can be obtained from the pair correlation
function and that it is characterized by the exponent $\th$.
Finally, in Sec.~\ref{sec:iso} we show that the fullRSB solution predicts that jammed packings are isostatic, i.e. particles
touch on average $2d$ other particles.

\subsection{Scaling of the cage radius with pressure}
\label{sec:Deltap} 

First, we discuss the physical meaning of the exponent $\k$.
We have already explained that finite pressures are equivalent to finite $m$ with $m \sim 1/p$. Moreover, we see from 
Eq.~\eqref{eq:Sscaledfinal} that a finite (small) $m$ is equivalent to a cutoff for $\D(y)$ at $y_{\rm max} \sim 1/m \sim p$.
In fact, the only difference between the equations at finite $m$ and the ones at $m=0$ with a cufoff at $y_{\rm max}$ is the initial
condition for $\wh j(y,h)$, which is non-vanishing (but small) for finite small $m$. However, this difference is completely irrelevant
because the scaling regime is universal and independent of the initial conditions.
From this argument we conclude that the intra-state cage radius (or mean square displacement, or Debye-Waller factor) scales as
$\wh\D_{\rm EA} \sim \D(y\propto1/m) \propto m^{\k} \propto p^{-\k}$, which provides a way to measure the exponent $\k$. 
Note that instead, at any level of $k$RSB with finite $k$, we have $\wh\D_{\rm EA} = \wh\D_{k} \propto m \propto 1/p$ 
as in the 1RSB solution. As in the Sherrington-Kirkpatrick model, the presence of a fullRSB solution is the signature of a 
marginally stable phase where the scaling of the intra-state overlap is changed.

\subsection{Pair correlation function}
\label{sec:paircorr}

We now want to study the scaling of the effective potential
given by Eq.~\eqref{eq:phieff_dinf}, that in $d\to\io$ coincides with the glass correlation function\footnote{The reader
should not confuse this function, which is a standard object of liquid theory~\cite{hansen} with the auxiliary function
$g(m,h)$ that has been introduced above.} $g(r)$
as a function of $h = d (r-D)/D$ (recall that $m_k=1$ and $y_k=1/m$):
\beq\label{eq:phieffHS}
\begin{split}
e^{-\phi_{\rm eff}(h)} &= \th(h) \int_{-\io}^\io \de z \, e^{z-h} \, \g_{\wh\D_k}(h-z) \, \frac{e^{-z-\wh\D_1/2} P(m_k,z)}{g(m_k,z)}
= \th(h) \int_{-\io}^\io \de z \, e^{z-h} \, \g_{\wh\D_k}(h-z) \, \frac{\wh P(y_k,z)}{\Th\left(z/\sqrt{2 \wh\D_k}\right)} \\
&= \th(h) \int_{-\io}^\io \de z \,  \, \frac{\g_{\wh\D_k}(h + \wh\D_k-z)}{\Th\left(z/\sqrt{2 \wh\D_k}\right)}  \wh P(y_k,z) e^{\wh\D_k/2}
 \ .
\end{split}\eeq

\subsubsection{Finite $k$RSB}

Before looking at the fullRSB solution, it is instructive to examine what happens at any finite level of $k$RSB. In that case, when $m\to 0$,
we have $\wh \D_k \sim m$ and $\wh P(m_k,h)$ tends to a finite function for all $h$. In this situation, if $h>0$, the kernel $\g_{\wh\D_k}(h + \wh\D_k-z)$
forces $| z - h | \sim m^{1/2} \to 0$, hence $z > 0$. The function $\Th\left(z/\sqrt{2 \wh\D_k}\right) \to 1$, and we conclude that
\beq\label{eq:largeh}
e^{-\phi_{\rm eff}(h)} =  \wh P(y_k,h) e^{\wh\D_k/2} \ .
\eeq
In particular, in the 1RSB case, $\wh P(y_1,h) e^{\wh\D_1/2} = 1$ and the effective potential is 1 corresponding to no correlations at all for any $h>0$.
Note however that for $k>1$RSB, the function $\wh P(y_k,h)$ has qualitatively the shape of Fig.~\ref{fig:Pscal}, although the peak is not divergent.
We therefore obtain that the glass correlation has a peak for small $h$.

On top of that, there is a highly non-trivial behavior when $h \sim m$~\cite{PZ10}. In fact, in that regime $z$ is not always positive. The positive part of the integral over $z$
gives a small contribution, but for negative $z$
the function $\Th$ can be computed in large and negative arguments where it goes to zero. This small denominator changes completely the behavior of the function
inducing a large peak. In this regime, defining $\l = h/m$, making use of the expansion of the error function
$\Th(s\to\io) \sim \frac{e^{-s^2}}{2\sqrt{\pi} |s|}$~\cite{PZ10}, and omitting subleading terms, we get:
\beq\label{eq:bella}
\begin{split}
e^{-\phi_{\rm eff}(h)} &=  \int_{-\io}^0 \de z \,  
\frac{e^{-\frac{(m \l + \wh\D_k - z)^2}{2\wh\D_k}}}{\sqrt{2\pi \wh\D_k}} |z| \, 
e^{\frac{z^2}{2\wh\D_k}} \sqrt{\frac{2\pi}{ \wh\D_k}}
 \wh P(y_k,z) 
 =  \int_{-\io}^0 \de z \, |z|  \, \frac{1}{\wh\D_k}
e^{z \left( \frac{ m \l}{\wh\D_k} +1 \right)} 
 \wh P(y_k,z) 
 \ .
\end{split}\eeq
Now, using that $\wh\D_k  = m \wh\g_k$, and that $\wh\g_k$ remains finite for $m\to 0$, we have
\beq\label{eq:effpotk}
\begin{split}
e^{-\phi_{\rm eff}(h)}  &= \frac1m \int_{-\io}^0 \de z \, |z|  \, \frac{1}{\wh\g_k}
e^{z \left( \frac{ \l}{\wh\g_k} +1 \right)} 
 \wh P(y_k,z)  = \frac1{m \wh\g_k} \FF_k\left(\frac{\l}{\wh\g_k}\right)
 =\frac1{m \wh\g_k} \FF_k\left(\frac{h}{m \wh\g_k }\right)
 \ ,
\end{split}\eeq
which shows that the contact value of the pair correlation 
$g(r)$ diverges as $1/m$ and the peak is characterized by a scaling function\footnote{
We called the scaling function $\FF_k$ to keep the same notation
used in previous papers~\cite{PZ10}. This function should not be confused with the function $\FF(\hat\D)$ 
that has been introduced in the expression of the replicated entropy at the beginning of the paper.
}
$\FF_k$ on a scale $h \sim m$.
This function is finite for $\l=0$ (remember that $P(y_k,z)$ is a finite function, and it decays as a Gaussian for large $z$),
while for $\l\to\io$ it decays as $1/\l^2$, because $P(y_k,z)$ is finite in $z=0$. These results were already derived in~\cite{PZ10}
at the 1RSB level.

\subsubsection{FullRSB}

Let us now see how this scenario is profoundly modified in the fullRSB case. 
As discussed in Sec.~\ref{sec:Deltap}, at finite $m$ the scaling discussed in Sec.~\ref{sec:scaling} holds, but there is a
cutoff at $y =  x_{\rm max}/m$ with some constant factor $x_{\rm max}$, after which $\D(y)$ is constant, and so is $\wh P(y,h)$. 
Therefore, $\wh\D_k  \sim \D_\io (m/x_{\rm max})^{1+c}$. At the same time, $\wh P(y_k,h)$ is not finite anymore: it satisfies
the scaling~\eqref{eq:Pscal} and in particular for $h<0$ we have
\beq
\wh P(y_k,h) \sim (x_{\rm max}/m)^{c} p_0( h (x_{\rm max}/m)^{c}) \ .
\eeq
The reasoning that leads to Eq.~\eqref{eq:bella} still holds, but now we have to take into account these modifications.
We get, calling $t = - z (x_{\rm max}/m)^{c}$,
\beq\label{eq:effpotio}
\begin{split}
e^{-\phi_{\rm eff}(h)} &
 =  \int_{-\io}^0 \de z \, |z|  \, \frac{1}{\D_\io (m/x_{\rm max})^{1+c}}
e^{z \left( \frac{ m \l}{\D_\io (m/x_{\rm max})^{1+c}} +1 \right)} 
  (x_{\rm max}/m)^{c} p_0( z (x_{\rm max}/m)^{c}) \\
&  = \frac{x_{\rm max}}{m \D_\io} \int_{0}^\io \de t \, t  \, p_0(- t) 
e^{- t  \frac{\l x_{\rm max}}{\D_\io } }  = \frac{x_{\rm max}}{m \D_\io} \int_{0}^\io \de t \, t  \, p_0(- t) 
e^{- t  \frac{h x_{\rm max}}{m \D_\io } } 
 \ .
\end{split}\eeq
The divergent part of the pressure is given by the contact value of $g(r)$, hence
\beq\begin{split}
p &= 1 + 2^{d-1}\f g(D) \sim \frac{d}2 \wh\f e^{-\phi_{\rm eff}(0)} = \frac{d}2 \wh\f \frac{x_{\rm max}}{m \D_\io} \int_{0}^\io \de t \, t  \, p_0(- t)
= \frac{d x_{\rm max}}{m \D_\io} \frac{\int_{0}^\io \de t \, t  \, p_0(- t)}{\int_{0}^\io \de t \, p_0(- t)} \equiv \frac{d x_{\rm max}}{m \D_\io} \overline{t}
\ ,
\end{split}\eeq
where we made use of Eq.~\eqref{eq:normp0} and introduced $\overline{t}$.
Note that this result confirms that the pressure is indeed proportional to $1/m$ as we have already discussed above.
Eq.~\eqref{eq:effpotio} can therefore be written as
\beq
\frac{g(r)}{g(D)} = \frac{e^{-\phi_{\rm eff}(h)}}{e^{-\phi_{\rm eff}(0)}} = \frac{ \int_{0}^\io \de t \, t  \, p_0(- t) 
e^{- t  \frac{h x_{\rm max}}{m \D_\io } } }{ \int_{0}^\io \de t \, t  \, p_0(- t) }
=\frac{ \int_{0}^\io \de t \, t  \, p_0(- t) 
e^{- \frac{t}{\overline{t}}  \frac{h p }{ d } } }{ \int_{0}^\io \de t \, t  \, p_0(- t) } \ .
\eeq
We now define the scaling variable $\l = (h p)/d = p (r-D)/D$ and the function
\beq\label{eq:Pp0}
P_\io(f) = \frac{p_0(-f \overline{t})}{\int_0^\io  \de f \, f \, p_0(-f \overline{t})} \ ,
\eeq
and we obtain the final result
\beq\label{eq:deltapeak}
\frac{g(r)}{g(D)} = \int_0^\io  \de f \, f \, P_\io(f) \, e^{-\l f} \equiv \FF_\io(\l) = \FF_\io\left(p \frac{r-D}D \right) \ .
\eeq
Note that the function $P_\io(f)$ is normalized in such a way that 
\beq\label{eq:Pfnorm}
\int_0^\io  \de f \, P_\io(f) = \int_0^\io  \de f \, f \, P_\io(f) = 1 \ , 
\eeq
therefore $\FF_\io(\l=0) = 1$ as it should. Furthermore,
because $p_0(z) \sim |z|^\th$ for small $z$, one has $P_\io(f) \sim f^\th$ for small $f$ and
$\FF_\io(\l) \sim \l^{-2-\th}$ for large $\l$, which provides a way
to measure $\th$ from the scaling of the contact peak of the pair correlation function.

Finally, Eq.~\eqref{eq:largeh} remains true also in the fullRSB case, but we have seen that the function $p_2(h)$ that describes the behavior of $\wh P(y_k,h)$
at finite $h$ diverges as $h^{-\a}$ when $h\to 0$. We therefore {\it predict} that the pair correlation function, at jamming, 
should diverge as $h^{-\a}$ for $h\to 0^+$, which provides a way to measure the exponent $\a$.

\subsection{Force distribution}
\label{sec:forcedis}

It has been shown in~\cite{DTS05} that the function $P(f)$ that enters in the scaling function $\FF(\l)$ 
in Eq.~\eqref{eq:deltapeak} is exactly the probability distribution of the forces between particles.
Here forces are scaled by the average force, in such a way that the average force is 1,
which is consistent with the normalization~\eqref{eq:Pfnorm}.
Obviously, we cannot compute directly $P(f)$ for hard spheres, because forces between hard particles are due
to collisions: they have dynamical origin and cannot be obtained through a static computation without making 
assumptions about the connection between structure and forces~\cite{DTS05,BW06}.

Interestingly, however, in the case of soft harmonic spheres the forces are simply linear functions of the overlaps, 
and therefore their distribution $P(f)$ is related to $g(r)$ by a straightforward relation~\cite{CCPZ12}.
In Appendix~\ref{app:A} we present a simple extension
of the theory to soft harmonic spheres, following~\cite{BJZ11}, that allows us to compute $P(f)$ directly.
Because the distribution of forces at jamming is only determined by the contact network~\cite{Wy12,LDW13}, 
$P(f)$ must be the same if one approaches jamming from below using hard spheres or from above using soft spheres.
Indeed, as in the 1RSB case~\cite{CCPZ12}, we find that the soft sphere computation gives the result in Eq.~\eqref{eq:Pp0},
thus identifying $P_\io(f)$ with the (scaled) force distribution at jamming in the fullRSB solution.
This result provides an indepenent proof of the relation \eqref{eq:deltapeak} between $P(f)$ and $\FF(\l)$ derived in~\cite{DTS05}.

Note that a similar manipulation starting from Eq.~\eqref{eq:effpotk} shows that at any finite level of $k$RSB
\beq\label{eq:Pkf}
P_k(f) = \frac{ e^{-\overline{t} f} \wh P(y_k,-f \overline{t}) }{ 
\int_0^\io  \de f \, f \, e^{-\overline{t} f} \wh P(y_k,-f \overline{t})
} \ ,
\eeq
with $\overline{t}$ defined by the same normalization condition as in Eq.~\eqref{eq:Pfnorm}.
The resulting force distribution is finite for $f\to 0$ and decays as a Gaussian at large forces, as found at the 1RSB level in~\cite{PZ10}.
At the fullRSB level, thanks to the appearance of the scaling regime, 
Eq.~\eqref{eq:Pkf} becomes Eq.~\eqref{eq:Pp0}.
This is very interesting, because for finite forces we obtain a function that is still qualitatively similar to the $k$RSB result (it is close to a Gaussian), but for
small forces we have a large deviation and in particular $P_\io(f) \sim f^\th$ with $\th$ given in Eq.~\eqref{eq:exp_num}. 
As noted in~\cite{Wy12}, the opening of this ``pseudogap'' in the
force distribution is exactly the same phenomenon as the opening of a pseudogap in the frozen field distributions of the 
SK model~\cite{SD84,MP07}, and it fully reflects the presence of fullRSB.

\subsection{Isostaticity}
\label{sec:iso}

Finally, we can compute the integral of the delta peak to obtain the coordination number.
The number of neighbors at distance $h$ is
\beq\label{eq:ZZdef}
Z(h) = \r V_d \, d \int_0^{1+h/d} \de s s^{d-1} e^{-\phi_{\rm eff}} = 
d \wh\f \int_{-\io}^h \de u e^u e^{-\phi_{\rm eff}(u)} 
=d \wh\f \int_{0}^h \de u e^u e^{-\phi_{\rm eff}(u)} 
\eeq
The rise of $Z(h)$ from 0 to the plateau happens on the scale $\l = h/m$. We obtain therefore for the contact number,
using Eq.~\eqref{eq:effpotio}:
\beq
z =  d \wh\f \int_0^\io \de \l  \frac{x_{\rm max}}{ \D_\io} \int_{0}^\io \de t \, t  \, p_0(- t) 
e^{- t  \frac{\l x_{\rm max}}{\D_\io } } = d \wh\f \int_0^\io \de t \, p_0(-t) \ .
\eeq
Now, we know from Eq.~\eqref{eq:normp0} that $\int_0^\io \de t \, p_0(-t) = 2/\wh\f$, we therefore {\it predict} 
that jammed packings are isostatic with $z = 2 d$, indepentently of their density. 
Note that isostaticity appears as a direct consequence of Eq.~\eqref{exact_rel}.

\section{Marginal stability}

In this section we want to prove that the fullRSB solution is marginally stable. This means that the matrix
of small fluctuations around the optimum selected by the variational equations has at least one zero eigenvalue.
The complete proof of this statement in the case of the SK model has been given in~\cite{TAK80,DK83,Go83,KD86}.
Here we do not discuss the complete calculation of all the eigenvalues but we focus our attention to the one 
that is responsible for the marginal stability of the fullRSB state.
We consider the replicon eigenvalue that is responsible for the stability of the fullRSB solution with 
respect to small fluctuations of the mean square displacement matrix that 
are localized in the innermost blocks. 

We start the computation from the replicated entropy defined by Eq.~\eqref{eq:gauss_r}. We want to study the eigenvalues of the matrix defined by
\beq\label{eq:Mkabcd}
M^{(k)}_{ab;cd}=\frac 2d\frac{\partial^2 s}{\partial \D_{a<b}\partial \D_{c<d}}=\frac{\partial^2}{\partial \D_{a<b}\partial \D_{c<d}}\left[\log \det (\hat \a^{m,m})-\wh \varphi \mathcal F(\hat \D) \ ,\right]
\eeq
where all the four indexes $a$, $b$, $c$, $d$ are in the same innermost block in a $k$RSB ansatz and the derivatives are computed on the $k$RSB solution. We follow the same
convention as in~\cite{KPUZ13} by assuming that the variation of $\D_{ab}$ also induces an identical variation of $\D_{ba}$, thereby preserving the symmetry of the matrix. We
denote this ``symmetric'' derivative by $\partial/\partial \D_{a<b}$.
Replica symmetry implies that all the innermost blocks are equivalent, and that the form of the stability matrix in each of these blocks is
\beq\label{eq:M1k}
M^{(k)}_{ab;cd}=M_1\left(\frac{\delta_{ac}\delta_{bd}+\delta_{ad}\delta_{bc}}{2}\right)+M_2\left(\frac{\delta_{ac}+\delta_{bd}+\delta_{ad}+\delta_{bc}}{4}\right)+M_3 \ .
\eeq
The replicon eigenvalue responsible for the stability of the innermost states is given by \cite{Ga85,TDP02}
\beq
\l_R^{(k)}=M_1 \ .
\eeq
If we consider a $k$RSB matrix and make a variation of the innermost element $\wh\D_k$, we have (here $B$ denotes one of the innermost blocks)
\beq\begin{split}
\frac2d \frac{\partial^2 s}{\partial \wh\D_{k}^2} &= \frac{m}{m_{k-1}} \sum_{a\neq b , c \neq d \in B} 
\frac2d \frac{\partial^2 s}{\partial \D_{ab} \partial \D_{cd}} 
+ \frac{m}{m_{k-1}} \left(\frac{m}{m_{k-1}} -1 \right) [ m_{k-1} (m_{k-1}-1)]^2  
\left. \frac2d\frac{\partial^2 s}{\partial \D_{ab} \partial \D_{cd}} \right|_{a\neq b \in B, c\neq d \in B'\neq B} \\
& = \frac14 m (m_{k-1}-1) M_1 + O( (m_{k-1} - 1)^2 )
\ ,
\end{split}\eeq
where in the first step we used that $\frac{\partial^2 s}{\partial \D_{ab} \partial \D_{cd}} $ is independent of the choice of indexes if they belong to different blocks,
and in the second step we used Eq.~\eqref{eq:M1k} and neglected higher orders in $m_{k-1}-1$. This is due to the fact that eventually we want to take the continuum limit
$k\to \io$ in which $m_{k-1}-1 \to 0$, and we only want to keep the leading order. The factor $1/4$ in the second line comes from the symmetrization of the derivative.
We therefore obtain the result for the replicon eigenvalue in the continuum limit
\beq
\l_R = \frac{4}{m (m_{k-1}-1) } \frac2d \frac{\partial^2 s}{\partial \wh\D_{k}^2} =
 \frac{4}{m (m_{k-1}-1) }  \frac{\partial^2 }{\partial \wh\D_{k}^2} \left[\log \det (\hat \a^{m,m})-\wh \varphi \mathcal F(\hat \D) \right] \ .
\eeq

For the first term (the entropic term) we use the $k$RSB expression given in Eq.~\eqref{eq:entropickRSB}. We have
\beq
\frac{\partial }{\partial \wh\D_{k}} \log \det (\hat \a^{m,m}) = \frac{1}{\wh\D_k} \left( \frac{m}{m_k} - \frac{m}{m_{k-1}} \right)
+  O( (m_{k-1} - 1)^2 )
 \eeq
 and
 \beq
\frac{\partial^2 }{\partial^2 \wh\D_{k}} \log \det (\hat \a^{m,m}) = -\frac{1}{\wh\D_k^2} \left( \frac{m}{m_k} - \frac{m}{m_{k-1}} \right)
+  O( (m_{k-1} - 1)^2 )  \ ,
 \eeq
so that the contribution to the replicon of the entropic term is in the continuum limit (remember that $m_k=1$ and $m_{k-1}\to 1$):
\beq
\l_R^{({\rm E})}=-\frac{4}{\D(1)^2} \ .
\eeq

We have now to compute the interaction part of the replicon eigenvalue. 
This can be done following exactly the same lines of Sec.~\ref{sec:vark}.
The result is
\beq
\begin{split}
\l_R^{({\rm I})}&=2\int_{-\infty}^\infty \de h e^h\frac{\de }{\de h}\left[\g_{-\wh\D_1}\star\left[P(1,h)\left[\frac{(g''(1,h))^2}{g^2(1,h)}-2\frac{(g'(1,h))^2g''(1,h)}{g^3(1,h)}+\frac{(g'(1,h))^4}{g^4(1,h)}\right]\right]\right] \\
&=2\int_{-\infty}^\infty \de h e^h\g_{-\wh\D_1}\star\left[\frac{\de }{\de h}\left[P(1,h)(f''(1,h))^2\right]\right]=-2\int_{-\infty}^\infty \de h e^h\wh P\left(\frac{1}{m},h\right)(f''(1,h))^2\:.
\end{split}
\eeq

Putting all the pieces together, we obtain
\beq
\l_R=-\frac{4}{\D(1)^2}+2\wh\varphi \int \de h e^h\wh P\left(\frac{1}{m},h\right)(f''(1,h))^2 \ .
\eeq
By using the fact that $\D(1)=m\g(1/m)$, and $f(1,h)=\wh f(1/m,h)/m$ we can rewrite this expression in the following form
\beq
\l_R=-\frac{4}{m^2\g(1/m)^2}\left[1-\frac{\wh\varphi}{2} \g(1/m)^2 \int_{-\infty}^\infty \de h e^h \wh P(1/m,h)\wh f''(1/m,h)^2 \right] \ .
\eeq
Eq.~\eqref{exact_rel} with $y=1/m$ therefore implies that $\l_R=0$ everywhere in the fullRSB phase, 
which shows the marginal stability of the fullRSB solution. 
This result is important because we have shown in the previous sections that Eq.~\eqref{exact_rel} is the key ingredient to obtain
the critical exponents at jamming. This means that the marginal stability of the fullRSB solution plays a prominent role 
in characterizing the properties of the jamming transition.

\section{Comparison with results of molecular dynamics simulations in finite dimensions}
\label{sec:compnum}

\begin{figure}
\includegraphics[width=0.45\columnwidth]{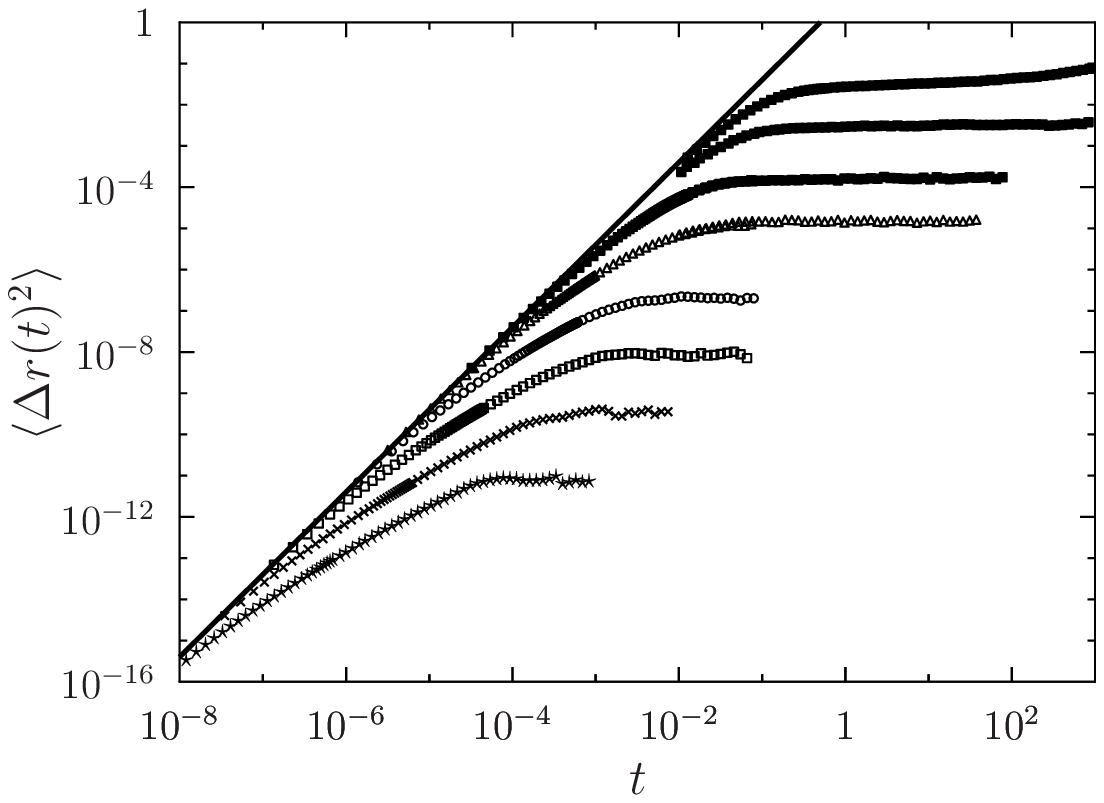}
\includegraphics[width=0.45\columnwidth]{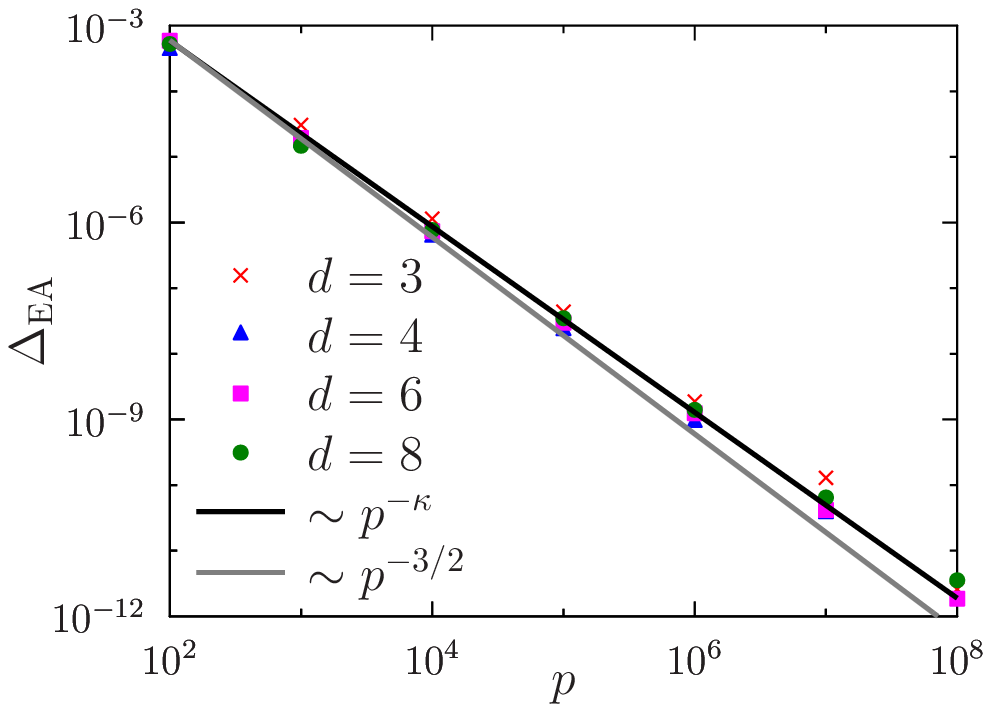}
\caption{(Left) Time evolution of the mean-square displacement in the equilibrated liquid at $\varphi=0.405$ as well as in the glass for different $p=10^{2}$--$10^{8}$ .
(Right) Pressure evolution of the Debye-Waller factor  $\D_{\rm EA}$ in 
$d=3,4,6,8$. The black line is a power-law with exponent $\k$ given by Eq.~\eqref{eq:exp_num}. 
For comparison, we also show a power-law $p^{-3/2}$ as a grey line.
}
\label{fig:4DMSD}
\end{figure}

The scaling of the fullRSB solution provides a number of predictions for the critical scaling at the jamming transition. It predicts that packings are isostatic with
average contact number $z=2d$, that the cage radius scales as $\D_{\rm EA} \sim p^{-\k}$, that the correlation function at jamming 
diverges on approaching contact as $(r-D)^{-\a}$, that the contact peak of the correlation function, on approaching jamming, 
has a scaling form given by Eq.~\eqref{eq:deltapeak} (see also~\cite{PZ10}) with a scaling function $\FF_\io(\l)\sim \l^{-2-\th}$ at large $\l$,
and that the force distribution is characterized by $P_\io(f) \sim f^\th$ for small $f$. The exponents
$\k,\a,\th$ are given in Eq.~\eqref{eq:exp_num}.

Some of these predictions have already been
verified in the past using molecular dynamics simulation. In particular isostaticity is a well-known
property of jammed packings (see~\cite{TS10,PZ10,He10} for reviews). Also, the exponent $\a$ in $g(r) \sim (r-D)^{-\a}$ has been independently 
measured with good precision by a number of groups~\cite{DTS05,SDST06,CCPZ12,LDW13,AST13} and its mostly accepted
value $\a \approx 0.42$ is perfectly compatible with
the prediction in Eq.~\eqref{eq:exp_num}. 

The scaling of $\D_{\rm EA} \sim p^{-\k}$ has been studied in~\cite{BW06,BW07,BW09b,IBB12} and the data were found to be compatible with $\k=3/2$, which
is slightly different from our prediction, and had been proposed in~\cite{WSNW05,BW09b} using a scaling argument based on marginal stability. 
To check whether the numerical
data are also compatible with our prediction for $\k$, we performed additional molecular dynamics simulations.
Hard-sphere systems in $d$=3, 4, 6, and 8 with $N$=8000 particles are simulated under periodic boundary conditions using a modified version of the event-driven molecular dynamics code described in 
Refs.~\cite{SDST06,CIPZ11,CIPZ12}. We consider monodisperse spheres with unit mass and unit diameter $D$ and unit mass $m$, hence time is expressed in 
units of $\sqrt{\beta m D^2}$ at fixed unit inverse temperature $\beta$. Hard sphere glasses are obtained using a Lubachevski-Stillinger 
algorithm initiated in the low-density fluid state with a slow growth rate $\dot{\gamma}=3\times 10^{-4}$. 
The fluid then falls out of equilibrium near the dynamical transition. 
Using these configurations, the mean-square displacement $\langle \Delta r^2(t)\rangle=\langle 1/N\sum_{i}[r_i(t)-r_i(0)]^2\rangle$ is obtained, 
and reported in Fig.~\ref{fig:4DMSD}. 
The rattlers, which are identified as the particles having fewer than $d+1$ contacts at $p=10^{10}$ (following Ref.~\cite{CCPZ12}), 
are removed when analyzing systems at $p\gtrsim 10^5$.
The long-time mean square displacement plateau provides $\langle \Delta r^2(t \to\io)\rangle = d \, \D_{\rm EA}$, where the Debye-Waller factor 
$\D_{\rm EA}$ is an estimate of the average cage size in the glass~\cite{CIPZ12}.
Using this approach, we find that in all dimensions the exponent $\k$ is close to $3/2$, but the data are better described by our
predicted value of $\k$, see Fig.~\ref{fig:4DMSD}. 
Note that the theoretical prediction is that $\D_{\rm EA} = \wh\D_{\rm EA} /d^2$ should decrease as $d^2$ (at fixed $\wh\f$), 
while in Fig.~\ref{fig:4DMSD} we see that $\D_{\rm EA}$ is
roughly independent of dimension in the range of $d$ we investigated.
This discrepancy is probably due to the fact that the numerically investigated dimensions 
are quite far from the asymptotic $d\to\io$ limit (as far as prefactors are concerned), 
as it has been already noted in~\cite{CIPZ11,CIPZ12}. 
An approximate analytical computation of the prefactor $\D_\io$ 
of the $p^{-\k}$ scaling in finite $d$ is certainly possible and would shed light on this issue.

The measure of the exponent $\theta$ is more problematic. 
In fact, previous attempts at measuring $\theta$ using different techniques reported 
results in the range $0.2 \div 0.45$~\cite{CCPZ12,LDW13}.
In~\cite{LDW13} it has been shown that
the behavior of the force distribution $P(f)$ at small forces is dominated by two different ways
in which the force network responds to an external perturbation:
extended modes and local buckling modes. According to the results of~\cite{LDW13,DLBW14},
extended modes give an exponent $\th_{\rm e} = 1/\a -2$, which perfectly agrees with our results
given in Eq.~\eqref{eq:exp_num}.
Buckling modes, instead, give an exponent $\th_{\rm b} = 1-2\a$ and using the value of 
$\a$ predicted by our theory in Eq.~\eqref{eq:exp_num}, one obtains $\th_{\rm b}= 0.17462$, which
agrees with the numerical value reported in~\cite{LDW13,DLBW14}. 
According to the analysis
of~\cite{LDW13}, in presence of both modes the force distribution is dominated by
the smallest exponent, hence $\th = \min\{\th_{\rm b}, \th_{\rm e}\}=0.17462$ 
is the value that enters in $P(f)$. 
However, in our approach
we do not see any trace of the buckling modes and $P(f)$ is characterized by the exponent
$\th$ associated with extended modes. This is probably due to the fact that buckling modes
disappear in large dimensions, similarly to what happens to rattlers~\cite{CCPZ12}. A careful
numerical investigation of this effect would be important.

\section{Conclusions}

We have derived the fullRSB equations that describe infinite-dimensional hard spheres (and, {\it en passant}, more general potentials). 
We have shown that a marginal fullRSB phase exists at high pressure
and that it correctly {\it predicts} isostaticity and the critical exponents $\k,\a,\th$ associated with the jamming transition, 
unlike the 1RSB solution. The predicted values of the exponents are given in Eq.~\eqref{eq:exp_num}.
These predictions have been reviewed and compared with numerical simulations in Sec.~\ref{sec:compnum}.

Of course, a lot of work is still needed to understand and characterize this phase. Let us summarize a certain number of research directions that should
be explored in the near future:
\begin{itemize}
\item
The exponents $\k$, $\a$ and $\th$ recently attracted a lot of attention because they are related to the marginal mechanical stability of the packing in real
space~\cite{WSNW05,Wy12,LDW13,KMT13,DLBW14}. Furthermore, 
the analysis of~\cite{Wy12,LDW13,DLBW14} predicts two scaling relations, $\a = 1/(2+\th)$ and $\k = 2 -2/(3+\th)$,
that are exactly verified by our predicted exponents, 
see Eq.~\eqref{eq:exp_num}.
Within the present approach, the critical regime around jamming 
emerges as a consequence of a marginal stability in phase space, a consequence
of the vanishing of the replicon eigenvalue of the fullRSB solution~\cite{DK83,MPV87}.
The scaling relations $\a = 1/(2+\th)$ and $\k = 2 -2/(3+\th)$ can be derived from Eq.~\eqref{eq:exprel}, 
using $b=(1+c)/2$ (which we proved) and $a+b=1$, a relation that we were not able to prove but 
holds within arbitrary numerical precision, 
see Eq.~\eqref{eq:exp_num}.
It seems therefore that our approach, based on phase-space marginality, and the approach of~\cite{Wy12,LDW13,DLBW14}, 
based on marginal mechanical stability, are intimately related, and making this connection more explicit would be extremely interesting.
\item
In~\cite{LDW13} it was claimed that another exponent $\th_{\rm b}$, associated to local buckling modes, 
controls the behavior of the small force distribution, 
and this exponent
is smaller than $\th$ and verifies a different scaling relation, see the discussion in Sec.~\ref{sec:compnum}. 
Yet we do not see any trace of the exponent $\th_{\rm b}$ in our $d=\io$ solution.
Clarifying this point is therefore of crucial importance. It might be possible that the modes leading to the exponent $\th_{\rm b}$ disappear
in the limit $d\to\io$. This could be checked through numerical simulations.
\item
An important technical issue is to perform a state following calculation~\cite{BFP97,KZ10}. Such a computation would allow one to compute the Gardner
transition and the fullRSB phase for a given glass phase, which would facilitate the comparison with numerical simulations.
\item
Following~\cite{YM10,Yo12,Yo12b}, one can hope to compute exactly the shear modulus, which is another important observable showing anomalous
scaling at the jamming transition~\cite{OLLN02,OT07}. This is particularly interesting in light of the recent numerical results of~\cite{OY13}, which show some
possible direct observations of fullRSB effects.
\item
It could be possible to compute the distribution of avalanche sizes, following~\cite{LMW10}, who performed a similar 
computation in the Sherrington-Kirkpatrick model.
\item
Although in this paper we focused only on hard spheres,
the computations can be easily extended to soft spheres, 
in order to study the complete scaling on both sides of the jamming transition~\cite{IBB12,BJZ11}.
A preliminary and incomplete account of this extension is reported in Appendix~\ref{app:A}.
It would also be interesting to check what is the temperature scale of the Gardner transition in thermal systems.
\item 
The results should be extended to finite dimensional systems (still within a mean-field approximation) 
by using the effective potential approximation scheme of~\cite{PZ10,BJZ11}.
This would be extremely important, as it
would allow one to compute precise numbers (e.g. for the distribution of mean-square displacements among different states)
to be compared with numerical simulations and experiments.
\item
It would be very important to understand how truly finite dimensional corrections around the mean field approximation,
i.e. critical fluctuations in a renormalization group approach, affect the scenario proposed in this paper. Some attemps
to study this problem have been made e.g. in~\cite{CDFMP10,CBTT11,YM12}.
\item
The most important point is, however, the study of the off-equilibrium 
dynamics, that is still poorly understood in presence of fullRSB effects even in spin glass models~\cite{MR03,MR04,Ri13}. 
In this paper, we assumed that, since fullRSB seems to characterize all glasses at high enough pressure, it will be present in whatever state is reached
by the off-equilibrium dynamics. However, how this is realized precisely, and which are the dynamical signatures of fullRSB in the context of structural
glasses, remains an open problem.
\end{itemize}

\acknowledgments

We are very indebted to Silvio Franz for an extremely useful discussion about the recursive procedure to write the replicated free energy in the Sherrington-Kirkpatrick model;
to Carolina Brito, Eric DeGiuli, Edan Lerner and Matthieu Wyart, for many important exchanges related to their work~\cite{LDW13,DLBW14}, and for sharing with us unpublished
material, that was particularly useful to detect an error in the original version of Appendix~\ref{app:A}; and to Florent Krzakala, Federico Ricci-Tersenghi, and Tommaso Rizzo for many
important discussions on the Gardner transition and fullRSB equations in the $p$-spin model, including their numerical resolution, that settled the basis for this work.
We also very warmly thank Ludovic Berthier, Giulio Biroli, and Hajime Yoshino for many interesting exchanges related to this work.

P.U. acknowledges the Physics Department of the University of Rome ``La Sapienza'' and the Laboratoire de Physique Th\'eorique et 
Mod\`eles Statistiques of the University of Paris-Sud 11 where part of this work has been done.

Financial support was provided by
the European Research Council through ERC grant agreement
no. 247328 and ERC grant NPRGGLASS.
PC acknowledges Sloan Foundation support.

\appendix
\section{Extension to soft spheres}
\label{app:A}

We consider a system of soft spheres interacting through the potential $v(r) = \ee (1 - |r|/D)^2 \th(1 - |r|/D)$.
Introducing $h = d (|r| - D)/D$ we have
\beq
e^{-\b v(r)} = e^{-\b \ee (1 - |r|/D)^2 \th(1 - |r|/D)} = e^{-(\b \ee/d^2) h^2 \th(-h)} = e^{- \wh\b h^2 \th(-h)} = e^{-\wh v(h)} \ ,
\eeq
where we introduced a scaled temperature $\wh\b = \b \ee/d^2$. Note that if we want to keep $\wh\b$ finite, we have
$\b\ee \sim d^2$ hence the soft sphere system is effectively at a very low temperature; for this reason the soft sphere
system is close to a hard sphere one and neglecting higher order virial diagrams is correct as in the case of hard spheres.
The equations for soft spheres have been derived in Sec.~\ref{sec:Gaussian}: they
are identical to Eq.~\eqref{eq:Sscaledfinal}, with only a difference in the initial condition for
$\wh f(y,h)$, which can be deduced from Eq.~\eqref{eq:GG1def} and
becomes\footnote{Remember that according to our definitions,
$m_k=1$, $\wh\D_k = m \wh\g_k$, $y_k=m_k/m=1/m$, and $\wh f(y_k,h) = y_k^{-1} \log g(m_k,h)$.}:
\beq
g(m_k,h) =  \g_{\wh\D_k} \star e^{-\wh v(h)} \ , 
\hskip30pt
\wh f(y_k,h) = m \log \g_{m \wh\g_k} \star e^{-\wh v(h)} \ .
\eeq
The effective potential in Eq.~\eqref{eq:phieff_dinf}, similarly to Eq.~\eqref{eq:phieffHS}, has the following expression:
\beq\label{eq:phieffSS}
\begin{split}
e^{-\phi_{\rm eff}(h)} &= e^{-\wh v(h)} \int_{-\io}^\io \de z \, e^{z-h} \, \g_{\wh\D_k}(h-z) \, \frac{e^{-z-\wh\D_1/2} P(m_k,z)}{g(m_k,z)}
=  e^{-\wh v(h)}  \int_{-\io}^\io \de z \,  \, \frac{\g_{\wh\D_k}(h + \wh\D_k-z)}{ \g_{m \wh\g_k} \star e^{-\wh v(z)}  }  \wh P(y_k,z) e^{\wh\D_k/2}
 \ .
\end{split}\eeq
In the case of soft spheres,
the jamming limit can be approached from above by an appropriate scaling of the parameters~\cite{BJZ11}.
First, the zero-temperature soft sphere limit is obtained by letting $m\to 0$ with $\wh T = 1/\wh \b = m \t$ and fixed $\t$.
After this limit has been taken, the jamming point is approached from above by letting $\t\to 0$.

\subsection{The zero temperature limit}

We now focus on the zero-temperature soft sphere limit. To take this limit we have to compute
\beq\label{eq:appeq}
\wh f(y_k,h) = m \log \g_{m \wh\g_k} \star e^{-\wh \b h^2 \th(-h)} 
\eeq
in the limit where $m\to 0$ and $\wh\b = 1/(m\t)$.
Note that it is natural to assume that $\wh\g_k$ remains finite in this limit: therefore, $\D_{\rm EA} = 
\wh\D_k \sim m \sim T$ vanishes
proportionally to temperature, as found numerically in~\cite{IBB12}.
We have
\beq
 \g_{m \wh\g_k} \star e^{-\wh \b h^2 \th(-h)} = \int_{-\io}^\io \de z \frac{e^{-\frac{(h-z)^2}{2m \wh\g_k}-\frac{1}{m\t} z^2 \th(-z)}}{\sqrt{2\pi m \wh\g_k}}
\eeq
which for $m\to 0$ can be evaluated by a saddle-point. For $h>0$, the saddle point is $z^*=h>0$, while for $h<0$ it is $z^*=h/(1+2\wh\g_k/\t) < 0$.
In both cases therefore the integral is strongly peaked around $z^*$. Because the integral is quadratic, computing the corrections to the saddle point
is equivalent to replacing $\th(-z)$ with $\th(-h)$, and we obtain
\beq\label{eq:ggSS}
 \g_{m \wh\g_k} \star e^{-\wh \b h^2 \th(-h)} \sim \int_{-\io}^\io \de z \frac{e^{-\frac{(h-z)^2}{2m \wh\g_k}-\frac{1}{m\t} z^2 \th(-h)}}{\sqrt{2\pi m \wh\g_k}}
 =
 \begin{cases}
 e^{-\frac{1}{m ( 2\wh\g_k + \t)} h^2} \left(1 + \frac{2\wh\g_k}\t \right)^{-1/2} & \text{for $h<0$} \\
 1 & \text{for $h>0$} 
 \end{cases}
\eeq
For $m\to 0$, one therefore obtains
\beq\label{eq:fSS}
\wh f(y_k,h) = - \frac{1}{2 \wh\g_k + \t} h^2 \th(-h) \ .
\eeq 
This is therefore the appropriate initial condition for $\wh f(y_k,h)$ for soft spheres at $T=0$.
Note that the asymptotic behavior is therefore modified with respect to the hard sphere case.
As in hard spheres,
inserting these asymptotes in the evolution equation for $\wh f(y_i,h)$, one can show that
Eq.~\eqref{eq:fasymptotes} becomes
\beq\label{eq:fasymptotesSS}
\wh f(y_i,h\to -\io) \sim - \frac{h^2}{2\wh\g_i + \t} \ .
\eeq
The correct definition of $\wh j(y_i,h)$ is therefore
\beq
\wh f(y_i,h) = -\frac{h^2 \th(-h)}{2\wh\g_i + \t} + \wh j(y_i,h) \ .
\eeq
It is also convenient to define $\wh\g_i^\t = \wh\g_i + \t/2$.
Then Eqs.~\eqref{eq:Sscaledfinal} become, recalling that $y_k = 1/m$ is formally infinite:
\beq\label{eq:SscaledfinalSS}
\begin{split}
\SS_{k{\rm RSB}} &=   \sum_{i=1}^k \left( \frac{1}{y_{i}} - \frac{1}{y_{i-1}}  \right) \log ( \wh \g_i^\t - \t/2 )  - \wh\f \, e^{-\wh\D_1/2}\int_{-\infty}^\infty \de h\, e^{h} 
  \left\{ 1- e^{ -\frac{h^2\th(-h)}{2\wh\g_1^\t} + \wh j(y_1,h) } \right\}  \ , \\
\wh\D_i &= \frac{\wh \g_i^\t}{y_i} + \sum_{j=i+1}^k \left(\frac1{y_j} - \frac1{y_{j-1}} \right) \wh \g_j^\t \ , \\
\wh j(y_k,h) & = 0  \ ,  \\
\wh j(y_i,h) &= \frac1{y_i} \log \left[ \int_{-\io}^\io \de z\, K_{\wh\g^\t_i, \wh\g^\t_{i+1}, y_i}(h,z) \,e^{y_i \wh j(y_{i+1},z)}  \right] \ , \hskip20pt i = 1 \cdots k-1 \ , \\
 \wh P(y_1,h) &= \,e^{-\wh\D_1/2  -\frac{h^2\th(-h)}{2\wh\g^\t_1}  + \wh j(y_1,h) } \ , \\
\wh P(y_i,h) &= \int \de z \, K_{\wh\g_{i-1}^\t, \wh\g_i^\t, y_{i-1}}(z,h) \,  \wh P(y_{i-1},z) \, e^{z-h} \, e^{ -y_{i-1} \wh j(y_{i-1},z)+y_{i-1} \wh j(y_i,h)} 
\hskip20pt i = 2 , \cdots , k
\ , \\
\wh\kappa_i & = \frac{ \wh\f}2  \int_{-\infty}^\infty \de h \, e^h \,  \wh P(y_i,h) \left(  -\frac{h \th(-h)}{\wh\g^\t_i}  +   \wh j'(y_i,h) \right)^2 \ , \\
 \frac{1}{\wh \g^\t_i - \t/2}  &=  y_{i-1} \wh\kappa_i -
\sum_{j=1}^{i-1} ( y_j - y_{j-1}) \wh\kappa_j  \ ,
\end{split}
\eeq
where the kernel $K$ is the same as the one for hard spheres given in Eq.~\eqref{eq:Kscaledfinal}. We notice that, in terms of $\wh\g_i^\t$, these
equations are almost identical to the ones of hard spheres, except for slight modifications to the first and last equations.
When $\t=0$, the equations correctly give back the HS equations for $m\to 0$~\cite{BJZ11,CCPZ12}.
In the continuum limit the equations above become
\beq\label{eq:Sscaledfinal_contSS}
\begin{split}
\SS_{\io{\rm RSB}} & = 
-   \int_1^{\io}\frac{\de y}{y^2}\log\left[ \g_\t(y)-\frac\t2 \right] 
-   \wh \f \,
e^{-\D(1)/2}\int_{-\infty}^\infty \de h\, e^{h} 
  [1- e^{-\frac{h^2\th(-h)}{2\g_\t(1)} + \wh j(1,h)}] \ , \\
  \D(y) &= \frac{\g_\t(y)}y - \int_y^{\io} \frac{\de z}{z^2} \g_\t(z) \ , \hskip20pt \Leftrightarrow \hskip20pt
\g_\t(y) =  y\D(y) + \int_y^{\io} \de z  \D(z) \ , \\
\lim_{y\to \io}\wh j(y,h) & = 0 \ ,  \\
  \frac{\partial \wh j(y,h)}{\partial y} &= 
    \frac 12 \frac{\dot\g_\t(y)}y \left[ - \frac{\th(-h)}{\g_\t(y)} +  \frac{\partial^2 \wh j(y, h)}{\partial h^2}
  - 2y   \frac{h \th(-h)}{\g_\t(y)}  \frac{\partial \wh j(y, h)}{\partial h}
  +y   \left(  \frac{\partial \wh j(y, h)}{\partial h}\right)^2    \right] \ , \\
 \wh P(1,h) &= \,e^{-\D(1)/2  -\frac{h^2\th(-h)}{2\g_\t(1)}  + \wh j(1,h) } \ , \\
 \frac{\partial \wh P(y,h)}{\partial y} &=-\frac12 \frac{\dot\g_\t(y)}y e^{-h} \left\{ 
\frac{\partial^2 [e^h \wh P(y, h)]}{\partial h^2} - 2y \frac{\partial}{\partial h} \left[ 
e^h \wh P(y,h)  \left(   - \frac{h \th(-h)}{\g_\t(y)}  + \frac{\partial \wh j(y, h)}{\partial h}\right)
\right]
\right\} \ , \\
\kappa(y) & = \frac{ \wh\f}2  \int_{-\infty}^\infty \de h \, e^h \,  \wh P(y,h) \left(  -\frac{h \th(-h)}{\g_\t(y)}  +   \wh j'(y,h) \right)^2 \ , \\
 \frac{1}{ \g_\t(y)-\frac \t2 }  &=  y \kappa(y) -
\int_1^{y} \de z \kappa(z)
\ .
\end{split}
\eeq
Using these equations it is possible to derive the marginal stability condition also in the soft sphere case.
We can derive the equation for $\k(y)$ in terms of $\wh P$ with respect to $y$ to get
\beq
\left(\frac{\g_\t(y)}{\g_\t(y)- \t/ 2}\right)^{2}=\frac{\wh \varphi}{2}\int_{-\io}^\io\de h\, e^h \wh P(y,h) \wt f''(y,h)^2 \ ,
\eeq
where $\wt f(y,h)=\g_\t(y)\wh f(y,h)$. Note that again, this equation holds only if there is a domain where $\g_\t(y)$ 
is not piecewise constant, namely where a fullRSB solution is present. By deriving again this equation with respect to $y$ we get
\beq
-\t \frac{\g^2_\t(y)}{\left(\g_\t(y)-\frac \t2\right)^3}=\frac{\wh \varphi}{2}\int \de h\, e^h \wh P(y,h) \left[ 2\left( \wt f''(y,h)^2+ \wt f''(y,h)^3\right)-\frac{\g_\t(y)}{y} \wt f'''(y,h)^2 \right] \ ,
\eeq
that can be rewritten as
\beq
y=\frac{1}{2}\g_\t(y)\frac{\int \de h e^h \wh P(y,h) \wt f'''(y,h)^2}{\frac{\t}{\wh \varphi}\frac{\g_\t^2(y)}{\left(\g_\t(y)-\frac{\t}{2}\right)^3}+ \int \de h\, e^h \wh P(y,h) \wt f''(y,h)^2 \left[1+\wt f''(y,h)\right]} \ .
\eeq
The above equation can be seen as an equation for the breaking points of the fullRSB profile of $\g_\t(y)$.

It is interesting to compute the effective potential~\eqref{eq:phieffSS} in the limit $m\to 0$. Using Eq.~\eqref{eq:ggSS}, and with similar
manipulations, we obtain
\beq\label{eq:effpotSS}
e^{-\phi_{\rm eff}(h)} =\begin{cases}
e^{-\frac{1}{m\t} h^2} \int_{-\io}^\io \de z \, 
\frac{
e^{ -\frac{(h-z+m \wh\g_k)^2}{2 m \wh\g_k} + \frac1m \frac{z^2}{2\wh\g_k + \t}}
}{
\sqrt{2\pi m \,  \t\wh\g_k /(\t + 2 \wh\g_k)}
}
\wh P(y_k,z) \sim \left(1+\frac{2\wh\g_k}\t\right) e^{\frac{2\wh\g_k}\t h} 
\wh P\left[y_k, h \left(1+\frac{ 2\wh\g_k}{\t} \right)\right]
&\text{for } h<0 \\
 \int_{-\io}^\io \de z \, 
\frac{
e^{ -\frac{(h-z+m \wh\g_k)^2}{2 m \wh\g_k}}
}{
\sqrt{2\pi m \wh\g_k}
}
\wh P(y_k,z) \sim
\wh P(y_k, h)
&\text{for } h>0
\end{cases}
\eeq
From this result, we can compute the number of contacts using Eq.~\eqref{eq:ZZdef}. For soft spheres,
all particles with $h<0$ are in contact, and we have
\beq\label{eq:SScontacts}
z = d \wh\f \int_{-\io}^0 \de h e^h e^{-\phi_{\rm eff}(h)} =
 d \wh\f 
\left(1+\frac{2\wh\g_k}\t\right) \int_{-\io}^0 \de h  e^{\left(1 + \frac{2\wh\g_k}\t \right) h} 
\wh P\left[y_k, h \left(1+\frac{ 2\wh\g_k}{\t} \right)\right]
= d \wh \f \int_0^\io \de t e^{-t} \wh P\left[ y_k, -t \right] \ .
\eeq

\subsection{The jamming limit and the force distribution}

It is natural to expect that away from jamming
(a small) $\t$ acts as a cutoff for the scalings of Sec.~\ref{sec:scaling}, like $m$ for hard spheres.
We assume therefore that all the scalings of Sec.~\ref{sec:scaling} hold, but with a cutoff at $y_{\rm max}(\t)$.
In the limit $\t\to 0$, the cutoff must diverge to recover the jamming physics.

For soft spheres, the inter-particle force is $f = 2 \frac{\ee}D (1-|r|/D) = - \frac{2 \ee}{d D} h$ for $h<0$. Let us rescale the forces
by a factor $\frac{d D}{2\ee}$ in such a way that $f =  -h$ exactly. 
In any case we are not interested in the prefactors in the overall scaling of forces.
Then the probability distribution of forces is, following~\cite{CCPZ12} and recalling that $f \geq 0$:
\beq
P(f) = \frac{e^{-f - \phi_{\rm eff}(-f)}}{\int_0^\io \de f' e^{-f' - \phi_{\rm eff}(-f')}} \ .
\eeq
Using Eq.~\eqref{eq:effpotSS}, we then obtain
\beq
P(f) \propto e^{-f \left(1+\frac{ 2\wh\g_k}{\t} \right)} \wh P\left[y_k, -f \left(1+\frac{ 2\wh\g_k}{\t} \right)\right] \ .
\eeq
In the jamming limit $\t\to 0$,
we expect that 
$\wh\g_k \sim \g_\io y_{\rm max}^{-c}$ and $\wh P(y_k,h) \propto y_{\rm max}^{c} p_0(h y_{\rm max}^{c})$. 
Note that we also have
\beq
\D_{\rm EA} = \wh \D_k = m \wh\g_k = \frac{T}\t \g_\io y_{\rm max}^{-c} \ .
\eeq
We also must assume that $\wh\g_k/\t \sim \g_\io y_{\rm max}(\t)^{-c}/\t$ diverges for $\t\to 0$,
because $\D_{\rm EA}/T$ must diverge on approaching jamming to match with the hard sphere regime.
Therefore
\beq
P(f) \propto e^{-f \frac{ 2\wh\g_k}{\t}} \wh P\left[y_k, -f \frac{ 2\wh\g_k}{\t} \right]
\propto e^{-f \frac{ 2\g_\io y_{\rm max}^{-c}}{\t}}  
p_0\left( -f \frac{ 2\g_\io y_{\rm max}^{-c}}{\t}y_{\rm max}^{c}  \right)
=  e^{-f \frac{ 2\g_\io y_{\rm max}^{-c}}{\t}}  
p_0\left( -f \frac{ 2\g_\io}{\t} \right)
\sim p_0\left( -f \frac{ 2\g_\io}{\t} \right)
\eeq
where the exponential factor can be neglected in the last step because
the natural scale of forces is $f\sim \t$. This is correct because the pressure scales as $\t$~\cite{BJZ11}.

It is customary in the literature to scale the forces in such a way that the average force is 1.
We have for the average force in the limit $\t \to 0$
\beq
\overline{f} = \frac{
\int_0^\io \de f f p_0\left( -f  \frac{ 2\g_\io }{\t} \right)
}{
\int_0^\io \de f p_0\left( -f  \frac{ 2\g_\io }{\t} \right)
}
=\frac{\t}{ 2\g_\io}
\frac{
\int_0^\io \de t \, t \, p_0\left( -t \right)
}{
\int_0^\io \de t p_0\left( -t \right)
} 
= \frac{\t}{ 2\g_\io} \overline{t}
\eeq
Defining therefore a scaled force $\hat f = f  \frac{ 2\g_\io}{\t \overline{t} }$, we have for the scaled distribution
\beq
P(\hat f) = \frac{p_0(- \hat f \overline{t})}{\int_0^\io \de \hat f p_0(- \hat f \overline{t})} 
=\frac{p_0(- \hat f \overline{t})}{\int_0^\io \de \hat f \, \hat f \, p_0(- \hat f \overline{t})} 
\ ,
\eeq
which confirms the validity of Eqs.~\eqref{eq:Pp0} and~\eqref{eq:deltapeak}.
Note that from Eq.~\eqref{eq:SScontacts} we find that also jammed soft spheres are isostatic.
In fact, for $\t\to 0$, 
\beq
z = d \wh \f \int_0^\io \de t e^{-t} \wh P\left[ y_k, -t \right] \sim 
d \wh \f \int_0^\io \de t e^{-t} y_{\rm max}^{-c} p_0(-t y_{\rm max}^{-c})
\sim d \wh\f \int_0^\io \de t p_0(-t) = 2 d \ .
\eeq
Actually, using Eq.~\eqref{exact_rel} and Eq.~\eqref{wtf2}, we can also compute the scaling of the corrections to this result:
\beq\begin{split}
\frac{z}{2d} &= \frac{ \wh \f}2 \int_0^\io \de t e^{-t} \wh P\left[ y_k, -t \right] =
\frac{ \wh \f}2 \int_{-\io}^\io \de h e^{h} \wh P\left[ y_k, h \right] [\th(-h) + \wt f''(y_k,h)^2 - \wt f''(y_k,h)^2 ]  \\
&= 1 + \frac{ \wh \f}2 \int_{-\io}^\io \de h e^{h} \wh P\left[ y_k, h \right] [\th(-h)  - \wt f''(y_k,h)^2 ] 
= 1 + \frac{ \wh \f}2 \int_{-\io}^\io \de h e^{h} \wh P\left[ y_k, h \right] [  -\frac{c^2}4 J''(h y_k^b)^2 - c J''(h y_k^b) \th(-h)     ] \\
&= 1 + \frac{ \wh \f}2 \int_{-\io}^\io \de h e^{h} y_k^a p_1(h y_k^b) [  -\frac{c^2}4 J''(h y_k^b)^2 - c J''(h y_k^b) \th(-h)     ] 
= 1 + y_k^{a-b} \frac{ \wh \f}2 \int_{-\io}^\io \de t p_1(t) [  -\frac{c^2}4 J''(t)^2 - c J''(t) \th(-t)     ]  \ .
\end{split}\eeq
Therefore, we conclude that $\d z = z- 2d \propto y_{\rm max}^{a-b} = y_{\rm max}^{-c}$, 
using the relations
$b=(1+c)/2$ and $a=1-b=(1-c)/2$.

From this analysis we conclude that the pressure $P \propto \t$, that the cutoff is
such that $y_{\rm max}^{-c} \propto \d z$, and that
\beq
\D_{\rm EA} \propto \frac{T}{P} \d z \ .
\eeq
Note that we expect
that pressure scales linearly in $\d\wh\f = \wh\f - \wh\f_j$, hence also
$\t \propto P \propto \d\wh\f$. The scaling of $y_{\rm max}(\t)$ remains however undetermined
from this analysis. If we assume that $y_{\rm max} \sim \t^{-\nu} \sim \d\wh\f^{-\nu}$, then 
we have $\d z = \d\wh\f^{c \nu}$ and $\D_{\rm EA} \sim T \d\wh\f^{-1 + c \nu}$. We note that the
choice $\nu c = 1/2$ allows us to recover the scaling of~\cite{DLBW14}, i.e.
 $\d z = \d\wh\f^{1/2}$ and $\D_{\rm EA} \sim T \d\wh\f^{-1/2}$.

\bibliographystyle{mioaps}
\bibliography{HS}

\end{document}